\documentclass{aa}  
\usepackage{graphicx}
\usepackage{txfonts}
\usepackage[draft]{hyperref}
\usepackage{xcolor}

\begin{document} 

  \title{Formation channels of slowly rotating early-type galaxies\thanks{Based on observations with the NASA/ESA Hubble Space Telescope, obtained at the Space Telescope Science Institute, which is operated by the Association of Universities for Research in Astronomy, Inc., under NASA contract NAS5-26555. These observations are associated with program 13324.}}

        \titlerunning{Formation channels of slow rotators}
        \authorrunning{Krajnovi\'c et al. }

   \author{Davor Krajnovi\'c,
                \inst{1} 
                Ugur Ural\inst{1},
                Harald Kuntschner\inst{2}
                Paul Goudfrooij\inst{3},
                Michael Wolfe\inst{3}
                Michele Cappellari\inst{4}, 
                Roger Davies\inst{4}, 
                P.T. de Zeeuw\inst{5,6},
                Pierre-Alain Duc\inst{7},
                Eric Emsellem\inst{2,8},
                Arna Karick\inst{4}, 
                Richard M. McDermid\inst{9},
                Simona Mei\inst{10,11,12},
                Thorsten Naab\inst{13} }


   \institute{$^{1}$Leibniz-Institut f\"ur Astrophysik Potsdam (AIP), An der Sternwarte 16, D-14482 Potsdam, Germany (\email{dkrajnovic@aip.de})\\
        $^{2}$European Southern Observatory, Karl-Schwarzschild-Str. 2, 85748 Garching, Germany\\
        $^{3}$Space Telescope Science Institute, 3700 San Martin Drive, Baltimore, MD 21218, USA\\
        $^{4}$Sub-department of Astrophysics, Department of Physics, University of Oxford, Denys Wilkinson Building, Keble Road, Oxford OX1 3RH, UK\\
        $^{5}$Sterrewacht Leiden, Leiden University, Postbus 9513, 2300 RA Leiden, the Netherlands\\
        $^{6}$Max-Planck-Institut fuer extraterrestrische Physik, Giessenbachstrasse, 85741 Garching, Germany\\
        $^{7}$Observatoire Astronomique, Universit\'e de Strasbourg, CNRS, 11, rue de l'Universit\'e. F-67000 Strasbourg, France\\
        $^{8}$Univ Lyon, Univ Lyon1, ENS de Lyon, CNRS, Centre de Recherche Astrophysique de Lyon UMR5574, F-69230 Saint-Genis-Laval France\\
        $^{9}$Department of Physics and Astronomy, Macquarie University, North Ryde, NSW 2109, Australia\\
        $^{10}$Universit\'{e} Paris Denis Diderot, Universit\'e Paris Sorbonne Cit\'e, 75205 Paris Cedex 13, France\\
        $^{11}$LERMA, Observatoire de Paris, PSL Research University, CNRS, Sorbonne Universit\'es, UPMC Univ. Paris 06, F-75014 Paris, France\\
        $^{12}$Jet Propulsion Laboratory and Cahill Center for Astronomy \& Astrophysics, California Institute of Technology, 4800 Oak Grove Drive, Pasadena, California 91011, USA\\
        $^{13}$Max-Planck-Institut f\"ur Astrophysik, Karl-Schwarzschild-Str. 1, 85741 Garching, Germany }


   \date{Received 1 November, 2019; accepted 17 January ,2020}  

 
  \abstract
{
We study the evidence for a diversity of formation processes in early-type galaxies by presenting the first complete volume-limited sample of slow rotators with both integral-field kinematics from the ATLAS$^{\rm3D}$ Project and high spatial resolution photometry from the Hubble Space Telescope. Analysing the nuclear surface brightness profiles of 12 newly imaged slow rotators, we classify their light profiles as core-less, and place an upper limit to the core size of about 10 pc. Considering the full magnitude and volume-limited ATLAS$^{\rm3D}$ sample, we correlate the presence or lack of cores with stellar kinematics, including the proxy for the stellar angular momentum ($\lambda_{Re}$) and the velocity dispersion within one half-light radius ($\sigma_e$), stellar mass, stellar age, $\alpha$-element abundance, and age and metallicity gradients. More than half of the slow rotators have core-less light profiles, and they are all less massive than $10^{11}$ M$_\odot$. Core-less slow rotators show evidence for counter-rotating flattened structures, have steeper metallicity gradients, and a larger dispersion of gradient values ($\overline{\Delta [Z/H]} = -0.42 \pm 0.18$) than core slow rotators ($\overline{\Delta [Z/H]} = -0.23 \pm 0.07$).  Our results suggest that core and core-less slow rotators have different assembly processes, where the former, as previously discussed, are the relics of massive dissipation-less merging in the presence of central supermassive black holes. Formation processes of core-less slow rotators are consistent with accretion of counter-rotating gas or gas-rich mergers of special orbital configurations, which lower the final net angular momentum of stars, but support star formation. We also highlight core fast rotators as galaxies that share properties of core slow rotators (i.e. cores, ages, $\sigma_e$, and population gradients) and core-less slow rotators (i.e. kinematics, $\lambda_{Re}$, mass, and larger spread in population gradients). Formation processes similar to those for core-less slow rotators can be invoked to explain the assembly of core fast rotators, with the distinction that these processes form or preserve cores.}

   \keywords{galaxies: elliptical and lenticular, cD -- galaxies: evolution -- galaxies: formation -- galaxies: kinematics and dynamics -- galaxies: nuclei -- galaxies: structure -- galaxies: stellar content}

   \maketitle
%

\section{Introduction}

Early-type galaxies (ETGs) are typically considered to be featureless compared with spirals. Nevertheless, they have complex surface brightness profiles that cannot be reproduced with a single fixed form, but require a smooth variation \citep[e.g.][]{1993MNRAS.265.1013C,1994MNRAS.271..523D,1996ApJ...465..534G,2001MNRAS.326..869T,2006ApJS..164..334F}, as well as multiple components \citep[e.g. ][]{2001AJ....121..820G,2009ApJS..182..216K,2010MNRAS.405.1089L}. Even before high spatial resolution imaging was available, the brightest and the most massive galaxies were known to have cores: regions where the surface brightness profile flattens to a profile that remains constant or slowly rises as the radius approaches zero \citep{1966ApJ...143.1002K,1985ApJ...292..104L, 1985ApJ...292L...9K,1991A&A...244L..37N}. High spatial resolution imaging of the Hubble Space Telescope (HST) brought unequivocal evidence that the nuclear regions of some ETGs have cores, but it also showed that the majority of luminous ETGs have continually rising cuspy profiles \citep{1993AJ....106.1371C, 1994AJ....108.1598F,1995AJ....110.2622L,1997AJ....114.1771F}. Subsequent studies enlarged the sample of galaxies with high-resolution imaging capable of distinguishing central cores from cusps of various shapes \citep[e.g.][]{2001AJ....121.2431R, 2001AJ....122..653R, 2003AJ....126.2717L, 2005AJ....129.2138L, 2006ApJS..164..334F,2009ApJS..182..216K, 2011MNRAS.415.2158R, 2012ApJ...755..163D, 2013ApJ...768...36D}. 

The classification of nuclear surface brightness profiles depends on the definition of what a core is, and on the functional form used to fit the light profiles (as can be seen from the discussions in the cited papers). We discuss these technical details further in Section~\ref{s:nsb}. For the moment, we need to bear in mind that cores typically exist in galaxies brighter than $M_V = -21$. Cores in fainter galaxies are known, but are also rare \citep{2007ApJ...662..808L}. Furthermore, cores have typical sizes of 20-500 pc \citep[e.g.][]{2007ApJ...662..808L, 2011MNRAS.415.2158R, 2013AJ....146..160R}, but even kiloparsec-scale cores are known \citep{2012ApJ...756..159P,2014ApJ...795L..31L, 2016ApJ...829...81B, 2017MNRAS.471.2321D}. Finally, core size positively correlates with the total luminosity, surface brightness, mass, and stellar velocity dispersion \citep{2007ApJ...662..808L,2014MNRAS.444.2700D}: the more massive the galaxy, the more extended its core and the larger the difference between the observed brightness level in the core and the expected brightness based on the extrapolation of the large-scale profile \citep[e.g.][]{2004ApJ...613L..33G, 2006ApJS..164..334F, 2009ApJS..182..216K,2009ApJ...691L.142K,2012ApJ...755..163D, 2013ApJ...768...36D, 2014MNRAS.444.2700D, 2013AJ....146..160R}. This is a key finding that delineates the formation of cores, and we review it in Section~\ref{ss:cores}.

Stellar kinematics represents a crucial diagnostic of the internal structure of galaxies as it relates the projected structure with the intrinsic shape of galaxies \citep[e.g.][]{1991ApJ...383..112F, 1991ARA&A..29..239D, 1994ApJ...425..458S,1994ApJ...425..481S,1994ApJ...425..500S}. Furthermore, the information in the mean velocity, $V$, and the velocity dispersion, $\sigma$, of the line-of-sight velocity distribution (LOSVD) can be used to distinguish between the dominance of the ordered and random kinetic energy, or in terms of the tensor virial theorem \citep[][p. 360]{2008gady.book.....B}, whether the flattening of the galaxy is due to its rotation or to anisotropy in the velocity dispersion vectors. This was pioneered by \citet{1978MNRAS.183..501B}, who introduced the anisotropy diagram relating $V/\sigma$ and the projected shape of galaxies, $\epsilon$. However, the long-slit data that revealed the kinematic properties of ETGs \citep[e.g.][]{1975ApJ...200..439B,1977ApJ...218L..43I, 1979ApJ...229..472S, 1980MNRAS.193..931E,1983ApJ...266...41D, 1983ApJ...266..516D, 1983ApJ...265..664D, 1988ApJ...330L..87J, 1989AJ.....98..147J,1989ApJ...344..613F, 1990A&A...239...97B, 1994MNRAS.269..785B} are insufficient for the rigorous interpretation of the tensor virial theorem \citep{2005MNRAS.363..937B}. 

The observations with SAURON, an integral field unit (IFU) \citep{2001MNRAS.326...23B} of nearby ETGs \citep{2002MNRAS.329..513D, 2011MNRAS.413..813C}, showed that  stellar velocity maps can be used to recognise discs \citep{2008MNRAS.390...93K, 2013MNRAS.432.1768K}, and to rigorously apply the tensor virial theorem and use the $V/\sigma - \epsilon$ diagram to estimate the anisotropy of galaxies \citep{2007MNRAS.379..418C}. Furthermore, the IFU data allow us to measure a more robust proxy, $\lambda_{Re}$, for the projected specific stellar angular momentum of ETGs \citep{2007MNRAS.379..401E, 2011MNRAS.414..888E}. The regular or non-regular appearance of the velocity field of ETGs \citep{2011MNRAS.414.2923K}, directly related to the existence or lack of (embedded) discs, can also be related to the measured angular momentum \citep{2011MNRAS.414..888E}. \citet{2007MNRAS.379..401E} defined two classes of ETGs, where fast rotators have regular kinematics, while slow rotators have irregular velocity maps \citep{2011MNRAS.414..888E}. The kinematic classification from IFU studies\footnote{$\lambda_{Re}$ measures the project and specific (weighted by a mass proxy) angular momentum within one half-radius. Several studies noted that assuming a different scale, the relative values of $\lambda_{Re}$ can change \citep{2014ApJ...791...80A, 2016MNRAS.457..147F, 2017MNRAS.467.4540B,2017ApJ...840...68G}. A scale is a natural requirement  for any classification. The one effective radius was driven by the size of the IFU, the distance of the galaxies, and their brightness. As we work on galaxies from \citet{2011MNRAS.414..888E}, we retain their definition.} has a strong resemblance to the structural classification of galaxies based on imaging, but it resolves a crucial problem of recognising stellar discs that are hidden due to (non-physical) projection effects or (physical) multiple structures (e.g. embedded in spheroids). Based on the structural classification of ETGs \citep[based on HyperLeda,][]{2011MNRAS.413..813C}, it is easy to recognise galaxies that are misclassified \citep{2011MNRAS.414..888E}, but also to associate slow rotators with bright ellipticals and fast rotators with discy ellipticals and S0 galaxies \citep[e.g.][]{2016ARA&A..54..597C}. 

Using the $V/\sigma$ metric to separate fast and slowly rotatating ETGs, \citet{1997AJ....114.1771F} noted that essentially all core galaxies have low $V/\sigma$. Their sample was limited and selected in a heterogenous way \citep{1995AJ....110.2622L}. Similar in size, the SAURON sample \citep{2002MNRAS.329..513D}, for which the first robust stellar angular momenta were derived, had a more systematic selection, but was still only a representative sample. \citet{2007MNRAS.379..401E} showed that while there is a strong trend between cores and slow rotators, there is no 1:1 relation (i.e. neither do all slow rotators have cores, nor are all cores found in slow rotators). This was also emphasised by \citet{2011ApJ...726...31G}, while \citet{2013ApJ...768...36D} presented a sample of S0 galaxies (and therefore almost certainly fast rotators) with cores. 

\citet{2012ApJ...759...64L} investigated a subset of galaxies with WFPC2 imaging from the volume- and magnitude-limited ATLAS$^{\rm 3D}$ sample \citep{2011MNRAS.413..813C}, and confirmed the lack of a 1:1 relation between cores and slow rotators. He argued that slow rotators could be defined as galaxies with low angular momentum and core surface brightness profiles, imposing a limit of $\lambda_{Re} < 0.25$. This would resolve the issue of previous studies that highlighted core galaxies in fast rotators. 

\citet{2013MNRAS.433.2812K} collected all published nuclear surface brightness profiles and also analysed all unpublished archival HST imaging of the ATLAS$^{\rm 3D}$ galaxies, increasing the \citet{2012ApJ...759...64L} sample from 63 to 135 galaxies, and demonstrated that the option of using $\lambda_{Re} < 0.25$ would also include a number of galaxies without cores into slow rotators. Furthermore, \citet{2013MNRAS.433.2812K} investigated the physical differences between fast and slow rotators with cores. The study concluded that {\it core fast rotators} are morphologically, kinematically, and dynamically different from {\it core slow rotators} and argued against a classification scheme that combines these objects. 

The mixing of fast and slow and core and no-core options remains a puzzle for a comprehensive picture of galaxy (or more precisely, ETG) formation. One of the problems was that only 135 of 260 ATLAS$^{\rm 3D}$ galaxies have HST imaging at sufficient resolution. Crucially, one-third of the slow rotators were among those without classified nuclear surface brightness profiles  \citep{2013MNRAS.433.2812K}. These galaxies are all of relatively low mass ($<10^{11}$ M$_\odot$) with indications of dynamically cold structures and exponential (i.e. low S\'ersic index) photometric components \citep{2013MNRAS.432.1768K}. Based also on the typical properties of core galaxies from the studies cited above, \citet{2013MNRAS.433.2812K} made a case for these galaxies being core-less\footnote{The term "cusps" is sometimes used as a description of steep central surface brightness profiles \citep[e.g.][p. 238]{2000gaun.book.....S}. However, some confusion may emerge when the  "cuspy cores" is used as in \citet[e.g.][]{2009ApJS..182..216K}. In order to avoid any ambiguity, we prefer to use core-less as a direct opposite to core. }. 

After obtaining HST imaging for all remaining slow rotators of the ATLAS$^{\rm 3D}$ survey, we are now in a position to address the connection between nuclear surface brightness and stellar angular momentum, and develop a comprehensive view of the diversity of slow rotators and the implications for their formation and evolution, as well as their distinctiveness. Furthermore, in contrast to previous studies \citep[with the exception of][]{2009ApJS..182..216K}, we also make use of the stellar population parameters that are now available for the ATLAS$^{\rm 3D}$ sample \citep{2015MNRAS.448.3484M}. As will become clear later, this information is crucial for separating the different assembly pathways among ETGs. 

This paper is organised as follows. In Section~\ref{s:data} we describe the derivation of the stellar population parameters, and present the new HST observations and their reduction. Section~\ref{s:nsb} presents the nuclear surface brightness profiles, which allows us to present the first volume-limited sample of slow rotators with both IFU data and HST imaging. Section~\ref{s:res} updates the results of \citet{2013MNRAS.433.2812K}, presents global stellar population properties based on SAURON observation, and discusses the metallicity gradients in the context of nuclear light profiles. The discussion in Section~\ref{s:disc} reviews theories of core formation and connects the results on the light profiles with results from the global IFU observations. It ends by discussing the different mass-assembly process of fast and slow rotators with and without cores and presents evidence for two separate channels of formation of slow rotators. The paper ends with a list of conclusions in Section~\ref{s:con}.

%
%

\section{Data analysis}
\label{s:data}

\subsection{Observations}
\label{sub:obs}

We used two sets of observations based on spectroscopy and imaging. The first set was obtained using the integral-field spectrograph SAURON \citep{2001MNRAS.326...23B} as part of the ATLAS$^{3D}$ survey \citep{2011MNRAS.413..813C} of nearby early-type galaxies. In particular, we here present and make available unpublished products of the stellar population analysis based on \citet{2015MNRAS.448.3484M}, pertaining to age, metallicity, $\alpha-$element abundances, and gradients of these quantities. 

The second set of data is based on new HST imaging. The ATLAS$^{\rm 3D}$ galaxies that were not analysed by \citet{2013MNRAS.433.2812K} lacked HST observations suitable for extracting nuclear surface brightness profiles. We selected all 12 remaining slow rotators with the aim to complete this class with space-based high-resolution imaging. The general properties of these galaxies are listed in Table~\ref{t:sample}.

The HST data for this project were obtained through \emph{HST} Program GO--13324. The primary pointings used the Wide Field Camera 3 (WFC3) with F475W and F814W filters, observed during one HST orbit. Next to the primary WFC3 observations, data were taken with the ACS \citep{1998SPIE.3356..234F} using the same two filters in coordinated parallel observations, yielding data 360\arcsec away from the galactic nuclei. We here present and analyse only the WFC3 data. 

The WFC3/F475W and WFC3/F814W data for each target galaxy were split into two
dithered exposures with total exposure times of 1050 s and 1110 s,
respectively. A short 35s exposure was added in F814W to mitigate
potential saturation of the galaxy nuclei. 

\begin{table*}
   \caption{ General properties of the observed galaxies.}
   \label{t:sample}
$$
  \begin{array}{l c r c c c c c}
    \hline
    \hline
    \noalign{\smallskip}

$name$ & $Dist$  & \sigma_e & \lambda_{Re} & \epsilon &R_e & M_K & $log(M$_{JAM}$)$\\
              & $Mpc$ & $km/s$        &             &        &  $kpc$      &     & M_{\odot}          \\
     (1)    &   (2)       &    (3)&  (4) &  (5)    &  (6)   &  (7)     & (8)     \\
   \noalign{\smallskip} \hline \noalign{\smallskip}

$NGC\,0661$ & 30.6 & 178 & 0.14 & 0.31 & 2.85 & -23.19 & 10.932\\
$NGC\,1222$ & 33.3 & 91 & 0.15 & 0.28 & 2.67 & -22.71 & 10.504\\
$NGC\,1289$ & 38.4 & 124 & 0.18 & 0.39 & 4.06 & -23.46 & 10.717\\
$NGC\,3522$ & 25.5 & 98 & 0.06 & 0.36 & 2.34 & -21.67 & 10.305\\
$NGC\,4191$ & 39.2 & 124 & 0.11 & 0.27 & 3.23 & -23.10 & 10.704\\
$NGC\,4690$ & 40.2 & 98 & 0.15 & 0.27 & 4.38 & -22.96 & 10.620\\
$NGC\,5481$ & 25.8 & 122 & 0.10 & 0.21 & 3.15 & -22.68 & 10.613\\
$NGC\,5631$ & 27.0 & 150 & 0.11 & 0.13 & 3.56 & -23.70 & 10.887\\
$NGC\,7454$ & 23.2 & 114 & 0.09 & 0.36 & 3.26 & -23.00 & 10.627\\
$PGC\,28887$ & 41.0 & 129 & 0.14 & 0.32 & 2.42 & -22.26 & 10.534\\
$PGC\,50395$ & 37.2 & 81 & 0.14 & 0.23 & 2.06 & -21.92 & 10.145\\
$UGC\,3960$ & 33.2 & 83 & 0.12 & 0.19 & 4.01 & -21.89 & 10.390\\
       \noalign{\smallskip}
    \hline
  \end{array}
$$ 
{Notes -- Column (1): Galaxy name. Column (2): Distance. Column (3): Effective velocity dispersion. Column (4): Specific stellar angular momentum within one effective radius. Column (5): Half-light radius. Columns (6): Apparent K-band magnitude. Column (7): Dynamical mass. Values in Cols. (2) and (7) are taken from \citet{2011MNRAS.413..813C}. Columns (3), (6), and (8) are taken from \citet{2013MNRAS.432.1709C}, where Col. (8) is obtained by multiplying the mass-to-light ratio with the luminosity of galaxies as specified in that work. Columns (4) and (5) are taken from \citet{2011MNRAS.414..888E}.}
\end{table*}

\subsection{SAURON stellar populations}
\label{ss:pops}

We made use of stellar population parameters extracted from the SAURON \citep{2001MNRAS.326...23B} data cubes obtained within the ATLAS$^{\rm 3D}$ project. The extraction of stellar population parameters, including the emission-line correction and the measurement of the absorption-line strengths and star formation histories, are described in detail in \citet{2015MNRAS.448.3484M}. We associate with this paper the maps of age, metallicity ([Z/H]), and $\alpha-$element abundance ([$\alpha$/Fe]) for all galaxies in the ATLAS3D$^{\rm 3D}$ sample. These were derived based on single stellar population (SSP) models as described in \citet{2015MNRAS.448.3484M}. Briefly, following \citet{2006MNRAS.373..906M} and \citet{2010MNRAS.408...97K}, \citet{2015MNRAS.448.3484M} used \citet{2007ApJS..171..146S} models, which predict the Lick indices for a grid of various ages, metallicities, and $\alpha-$elemement abundances. \citet{2015MNRAS.448.3484M} derived the SSP parameters for SAURON data using three indices that are measured across the field of view of SAURON cubes (H$\beta$, Fe5015, and Mg{\it b}). The stellar population parameters are found by means of $\chi^2$ fitting, where the best-fit SSP provides the closest model values to our observed indices in a grid of age, metallicity, and $\alpha-$elemement abundances. Original models are oversampled using linear interpolation, while the uncertainties are included as weights in the sum. The errors of the final parameters are calculated as dispersions of all points that differ by $\Delta \chi^2=1$. 

We caution that the derived maps have to be interpreted as SSP-equivalent because we cannot expect that all stars within a region covered by a SAURON bin have the same age, metallicity, or abundance ratio. As \citet{2007MNRAS.374..769S} have shown, the SSP-equivalent age is biased towards the young populations, while the SSP-equivalent chemical compositions is dominated by the old population. 

Maps of stellar ages and metallicities are pertinent to this paper, which were used to extract age and metallicity profiles. Metallicity profiles were derived in the same way as in \citet{2010MNRAS.408...97K} by averaging the values on stellar population maps along the lines of constant surface brightness. In this way, we ignored possible (and known) differences between the projected distribution (shape) of the stellar population parameters \citep[e.g. metallicity,][]{2010MNRAS.408...97K} and flux. Uncertainties at each radial point were derived as the standard deviation of all points at this ring after applying a $3\sigma$ clipping algorithm. 

Gradients of the metallicity, $\Delta [Z/H]$, and the age, $\Delta$\,Age, were obtained by performing straight line fits to the metallicity and age profiles (weighted by their errors). Following the definition in \citet{2010MNRAS.408...97K}, the metallicity gradient is then defined as 
\begin{equation}
\label{eq:metgrad}
\Delta [Z/H] = \frac{\delta [Z/H]}{\delta \log R/R_e},   \\
\end{equation}

\noindent and the age gradient as 
\begin{equation}
\label{eq:agegrad}
\Delta {\rm Age} = \frac{\delta \log({\rm Age})}{\delta \log R/R_e}.
\end{equation}

\noindent The fits were limited to a region between 2\arcsec\, and the half-light radius. The inner boundary was set to avoid seeing effects, while our data rarely reach far beyond the half-light radius. For massive slow rotators, SAURON observations do not cover the full R$_e$, and in these cases, the outer limit is set by the data. We were not able to extract stellar population profiles and gradients for two galaxies (NGC\,4268 and PGC\,170172), but these galaxies do not have HST data and therefore are not relevant for this study. Age and metallicity gradients are presented in Table~\ref{t:pop} and in the online version of the paper. Age, metallicity, and $\alpha-$element abundance profiles, as well as the SAURON maps of the SSP equivalent stellar age, metallicity, and alpha-element abundances can be obtained from the ATLAS$^{\rm 3D}$ 
Project website\footnote{\label{ft:a3d} \url{http://purl.org/atlas3d}}.

\subsection{HST image combination}
\label{sub:drizzle}

The data processing was performed at the Space Telescope Science Institute
(STScI). It involved dedicated wrapper scripts that use modules from the
\href{http://www.stsci.edu/hst/HST\_overview/drizzlepac}{DrizzlePac} software
  package, including {\tt AstroDrizzle} and {\tt TweakReg}.

First, we ran {\tt AstroDrizzle} on every set of associated images (i.e. all
images taken in one visit with the same filter) using the setting {\tt
  driz\_separate = True}. This created singly drizzled output images,
which are the individual exposures after correction for geometric distortion
using the World Coordinate System (WCS) keywords in the image header. The
software package SExtractor \citep{1996A&AS..117..393B} was then run on each
singly drizzled image, using a signal-to-noise ratio threshold of S/N = 10. The
resulting catalogues were trimmed using object size, location, and shape
parameters chosen to reject most cosmic rays, detector artefacts, diffuse
extended objects, and objects near the edges of the images. Using these
cleaned catalogues, residual shifts and rotations between the individual singly
drizzled images were then determined using {\tt TweakReg}. This yielded formal
alignment uncertainties below 0.1 pixel. The reference image was 
always taken to be the image that was observed first in the visit. The resulting shifts
and rotations were then implemented in a second run of {\tt AstroDrizzle} to
verify the alignment. 

Cosmic-ray rejection was performed within {\tt AstroDrizzle}, which uses a
process involving an image that contains the median values (or minimum
value, see below) of each pixel in the (geometrically corrected and aligned)
input images as well as its derivative (in which the value of each pixel
represents the largest gradient from the value of that pixel to those of its
direct neighbours; this image was used to avoid clipping bright point sources) to
simulate a clean version of the final output image. For this step, we used
the default cosmic-ray reduction settings in {\tt AstroDrizzle},  {\tt
  combine\_type = minmed}, which use the median value 
unless it is higher than the minimum value by a 4\,$\sigma$ threshold. Pixels that were saturated in the long F814W exposures were dealt with by flagging them as such in the data quality (DQ) extensions of the corresponding {\tt \_flc.fits} files. {\tt AstroDrizzle} then effectively replaced the saturated pixels by the corresponding pixels in the short F814W exposures. 

Sky subtraction was performed on each individual image prior to the final
image combination, using iterative sigma clipping in the region shared by all
images with a given filter. The resulting sky values were stored by {\tt
  AstroDrizzle} in the header keyword {\sc mdrizsky} of the individual
{\tt \_flt.fits} or {\tt \_flc.fits} images; our wrapper script then calculated
the average sky rate in e$^-$/s for that filter and stored it in the header
keyword {\sc mdrskyrt} of the final {\tt AstroDrizzle} output file
({\tt \_drz.fits} or {\tt \_drc.fits}), which also uses e$^-$/s units. This
was done to allow the sky level to be added back in before performing surface
photometry, which is needed to determine proper magnitude errors. 

The final run of {\tt AstroDrizzle} was then performed using the so-called
inverse variance map (IVM) mode. These weight maps contain all components of
noise in the images except for the Poisson noise associated with the
sources on the image, and they are constructed from the flatfield reference
file, the dark current reference file, and the read noise values listed in the
image header. For the final image combination by {\tt AstroDrizzle}, we used a
Gaussian drizzling kernel, and parameters {\tt final\_pixfrac} = 0.90 pixel and
{\tt final\_scale} 0.032 $''$/pixel. These parameter values were determined after
experimentation with appropriate ranges of values. The HST imaging used in this work as well as the reduced parallel fields were assigned a doi number (10.17876/data/2020\_1\footnote{Resolvable via \url{https://doi.org/10.17876/data/2020_1}}), and are made available at ATLAS$^{\rm 3D}$ Project website (see footnote~\ref{ft:a3d}).

\subsection{Extraction of surface brightness profiles.}
\label{s:ext}

We followed the common procedure, which starts with the extraction of the light profiles using the {\tt STSDAS IRAF} task {\tt ellipse}. This method, as described in \citet{1987MNRAS.226..747J}, is based on the harmonic analysis along ellipses, fitting for the centre of the ellipse, its ellipticity $\epsilon,$ and position angle $\Phi$. The best-fit parameters describing the ellipse (ellipse centre, $\Phi$, $\epsilon$ ) were determined by minimising the residuals between the data and the first two moments in the harmonic expansion. The semi-major axis was logarithmically increased as the fitting progresses. The values of the ellipse parameters are susceptible to the influence of foreground stars and dust patches, which effectively distort the shape of the isophotes. Therefore we created object masks prior to fitting. The whole procedure is similar to that used in \citet{2006ApJS..164..334F} and \citet{2013MNRAS.433.2812K}.

We converted the light profiles into surface brightness using the standard conversion formulae and the zero-points provided in the headers of the HST images. The light profile of one galaxy, NGC\,1222, was considered too uncertain for any subsequent analysis. The reason for this is the presence of complex filamentary dust structures, extending several kiloparsec from the nucleus, mostly along the minor axis of the galaxy, and bounded by two bright complexes of young and bright stars. The stunning appearance of NGC\,1222 was shown in the NASA/ESA Photo Release\footnote{\url{https://www.spacetelescope.org/images/potw1645a/}} using the HST observations presented here. NGC\,1222 is a recent merger remnant, and while it is clearly a slow rotator, it is somewhat special because it exhibits a prolate-like rotation (around the major axis). The galaxy is also rich in atomic gas, which has the same kinematic orientation as the stars and is distributed in the polar plane of the galaxy, making a clear link with the dusty central regions \citep{2018MNRAS.477.2741Y}. For our study, however, the most relevant is the extinction caused by the dust in the centre, which prohibits a robust extraction of the light profile. Therefore we excluded this galaxy from further analysis, but we highlight it in relevant figures.

\begin{figure*}
  \centering
\includegraphics[width=0.4\textwidth]{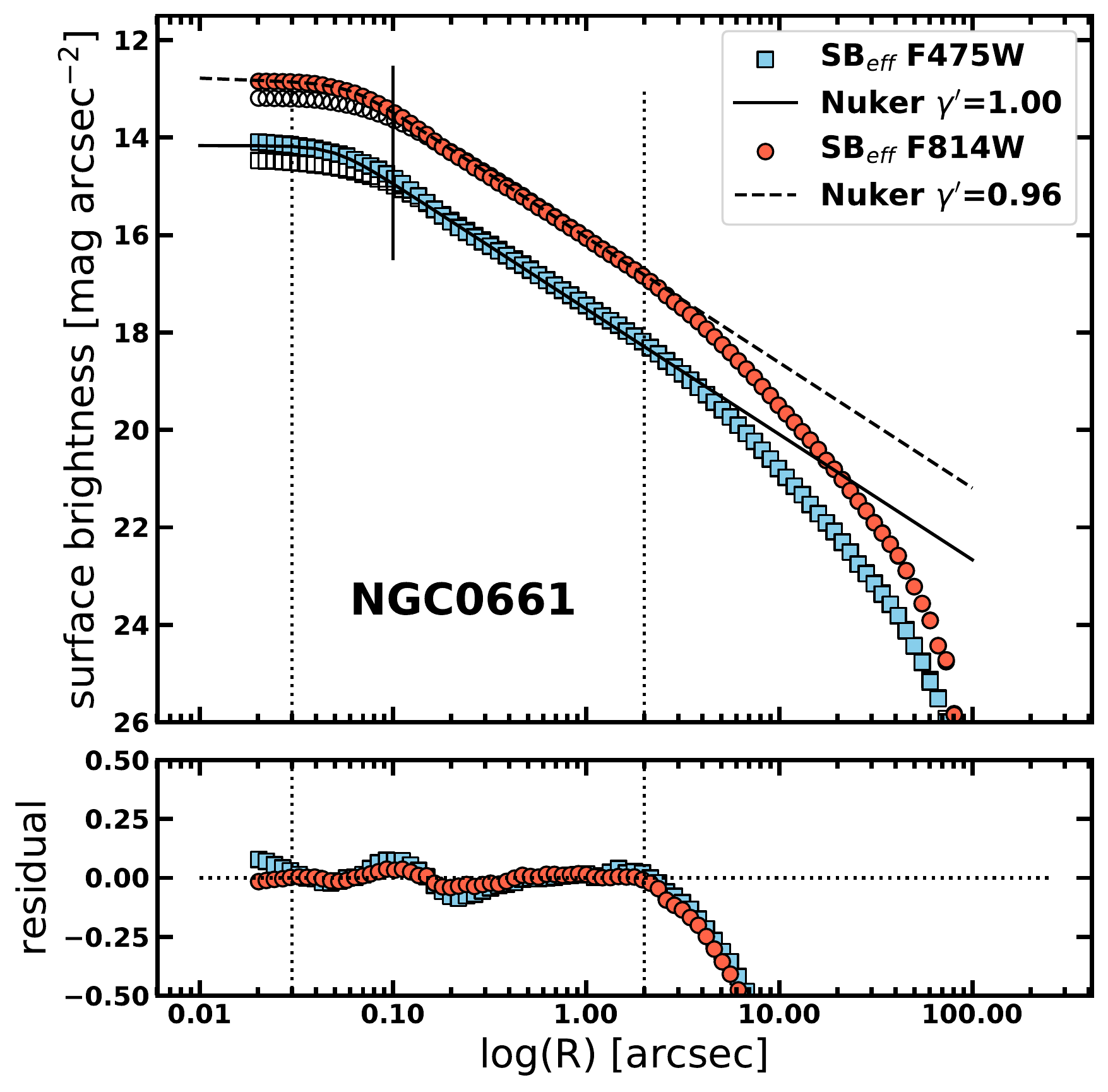}
\includegraphics[width=0.4\textwidth]{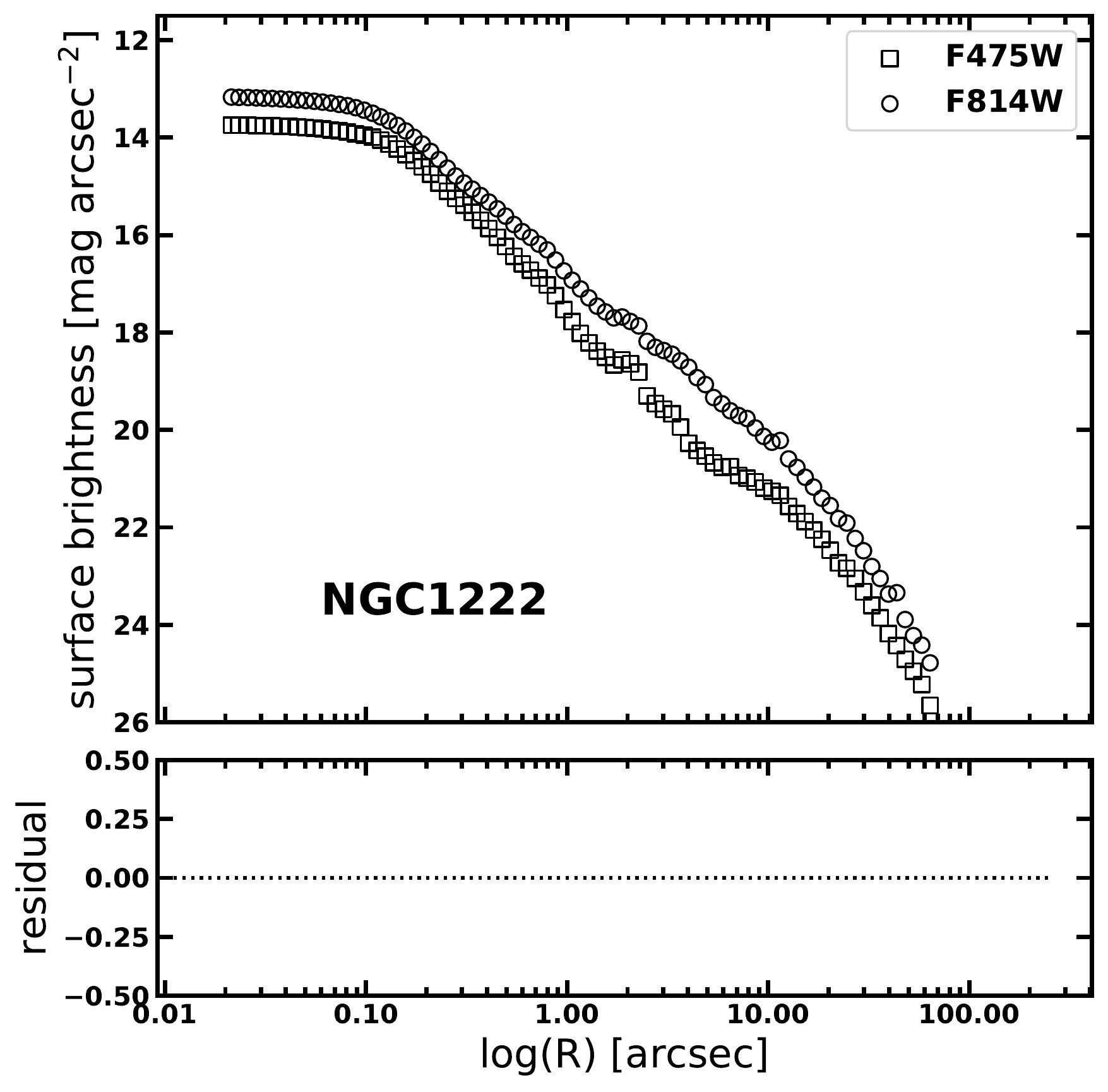}
\includegraphics[width=0.4\textwidth]{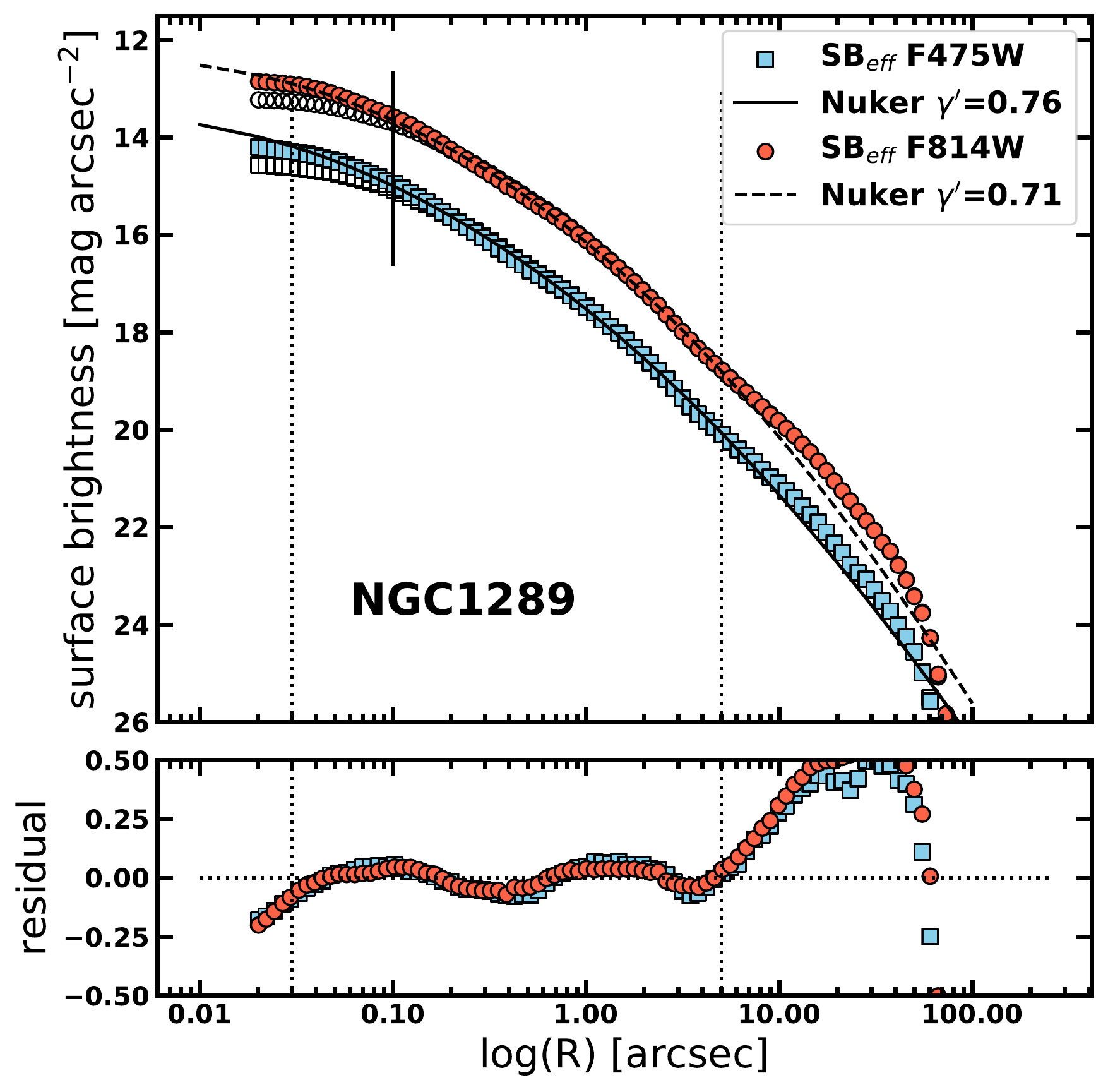}
\includegraphics[width=0.4\textwidth]{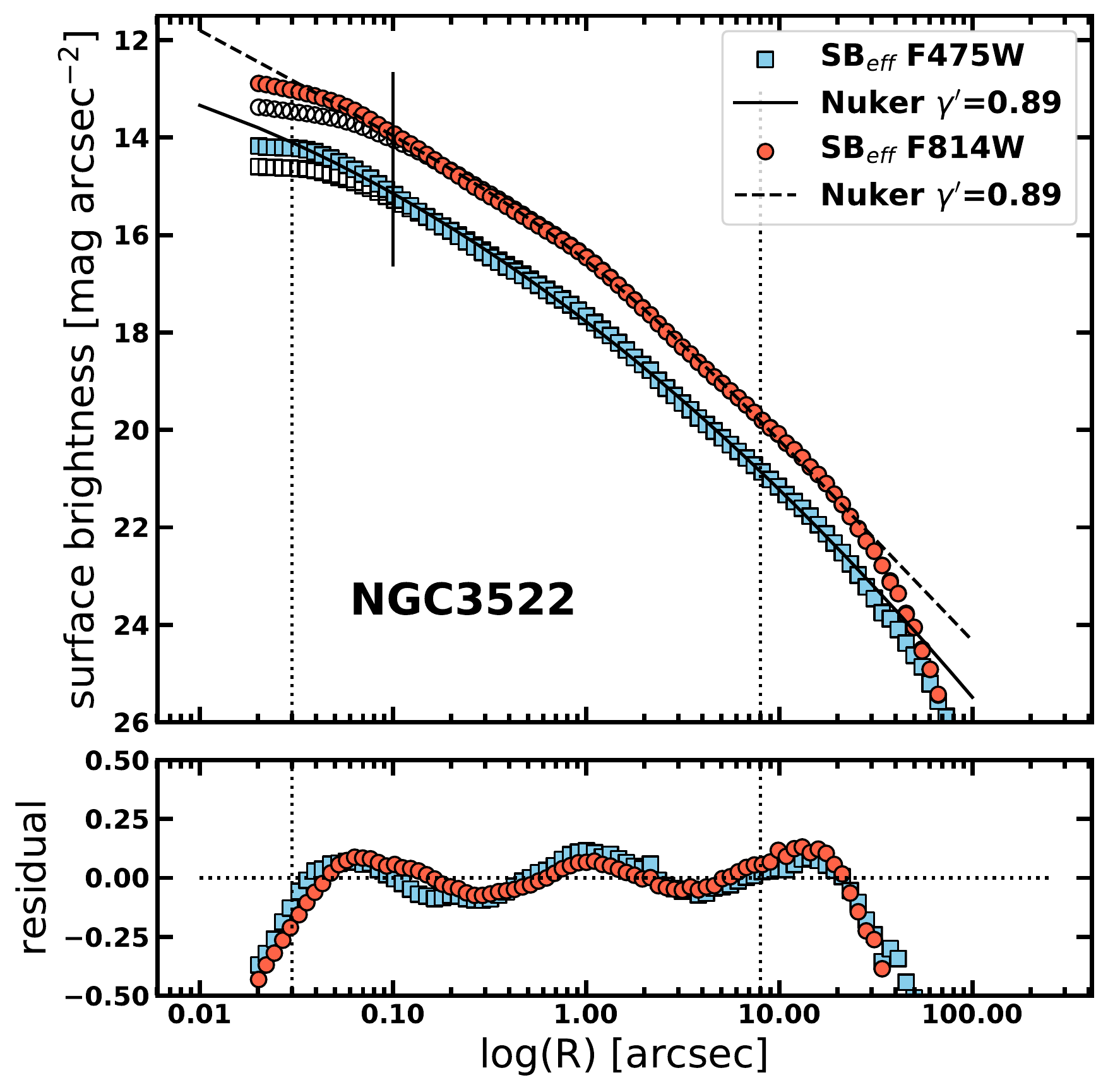}
\includegraphics[width=0.4\textwidth]{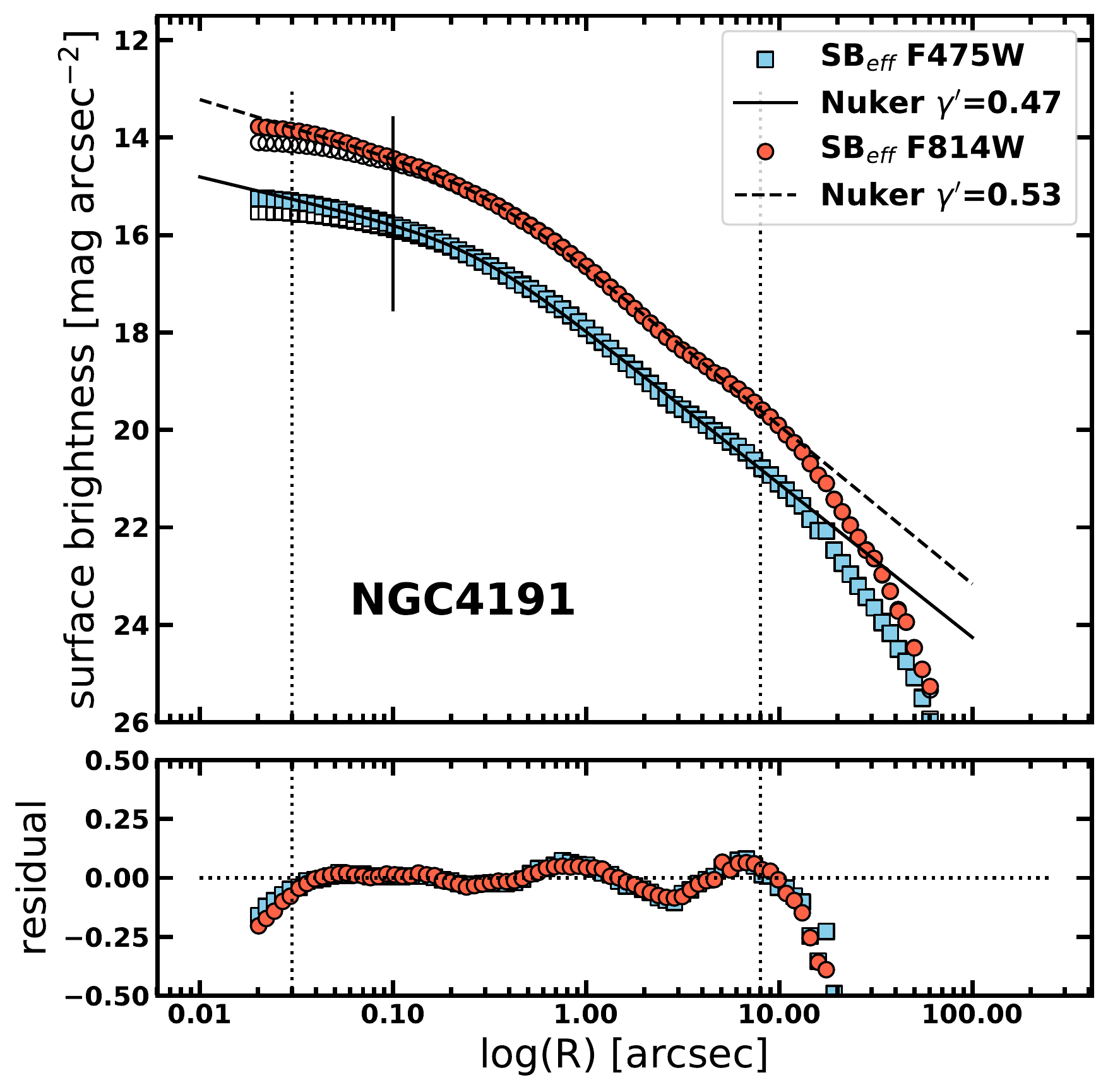}
\includegraphics[width=0.4\textwidth]{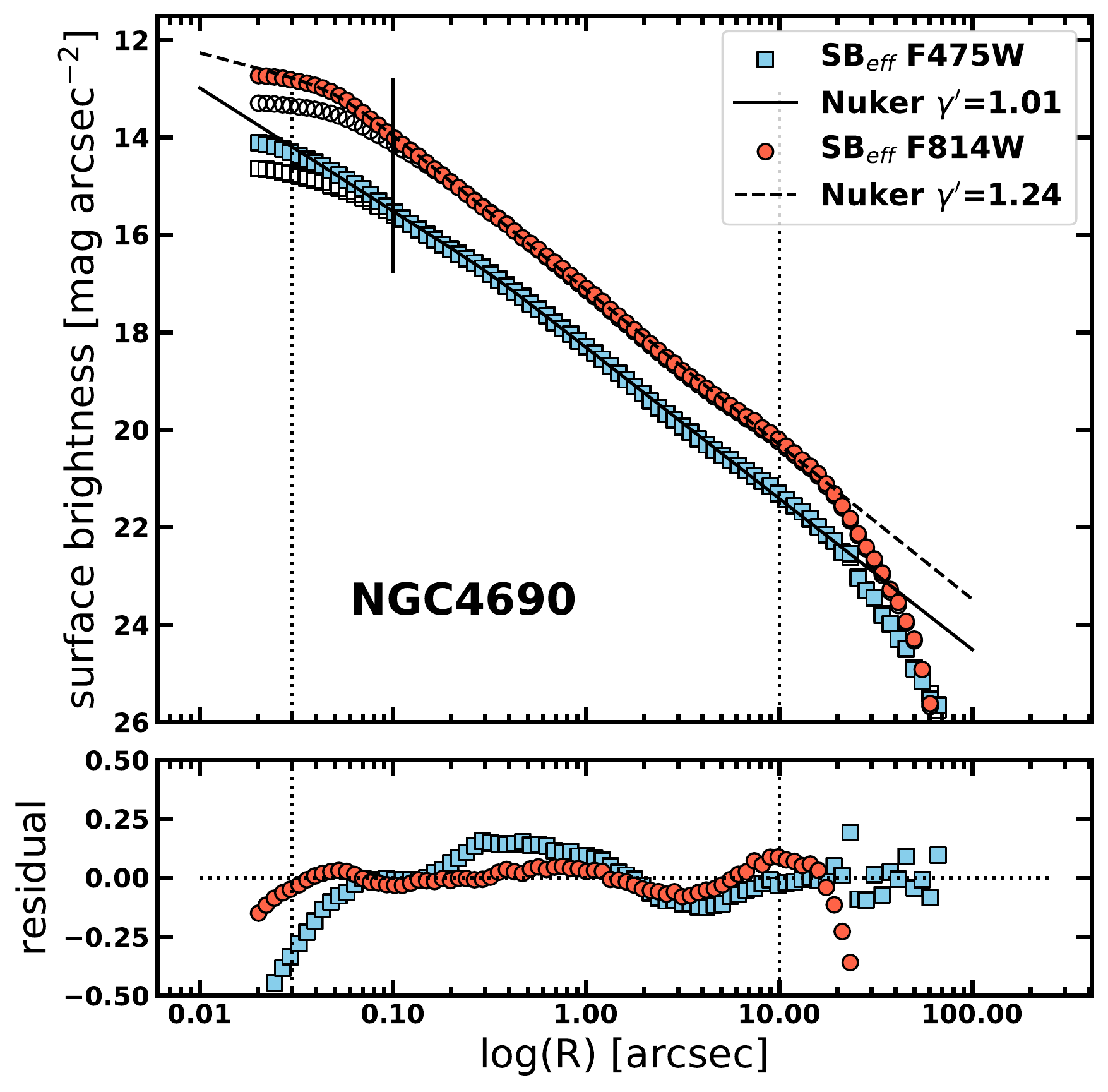}

\caption{Surface brightness profiles of the first six low-mass slow rotators from the ATLAS$^{\rm 3D}$ survey (ordered by name). Each galaxy is represented by two panels. The upper panel shows light profiles in F475W (light blue squares) and F814W (red circles) filters, and the smaller lower panel shows residuals from the fit. Open symbols are original (observed) light profiles, and filled symbols are effective (deconvolved) profiles used for the analysis. The sampling does not correspond to the pixels of the WFC3 camera, but it is defined by the tool for the isophote analysis. The Nuker fits to both filters are shown with solid (F475W) and dashed (F814W) lines. Two vertical dotted lines indicate the range used in the fit. The short vertical solid line indicates the location at which the $\gamma^\prime$ slope is measured. Our images have a pixel scale of 0\farcs032. For a comparison with Richardson-Lucy deconvolution results, see Appendix~\ref{a:deco_comp}.}
\label{f:sb1}
\end{figure*}

\begin{figure*}
  \centering
\includegraphics[width=0.4\textwidth]{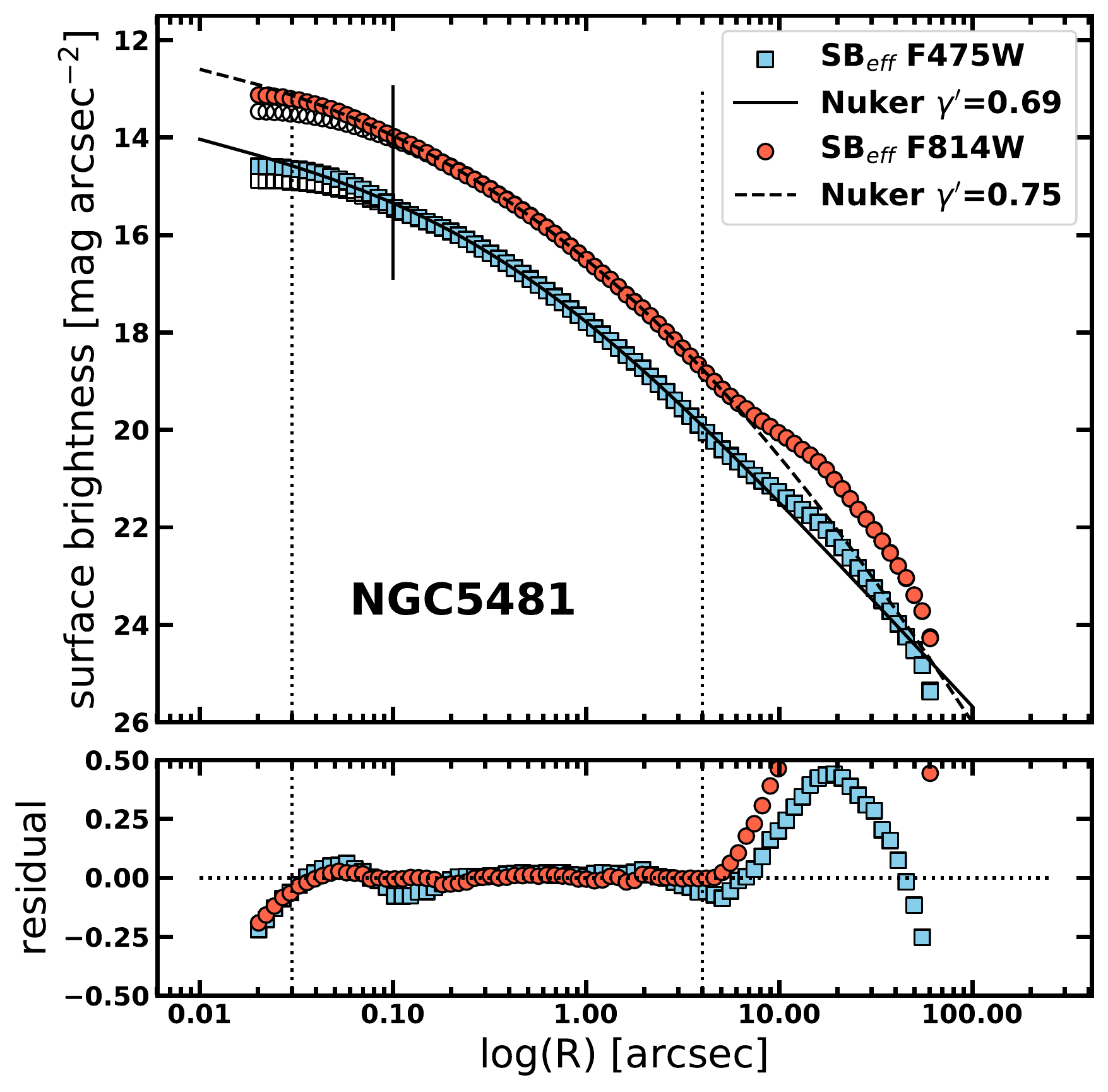}
\includegraphics[width=0.4\textwidth]{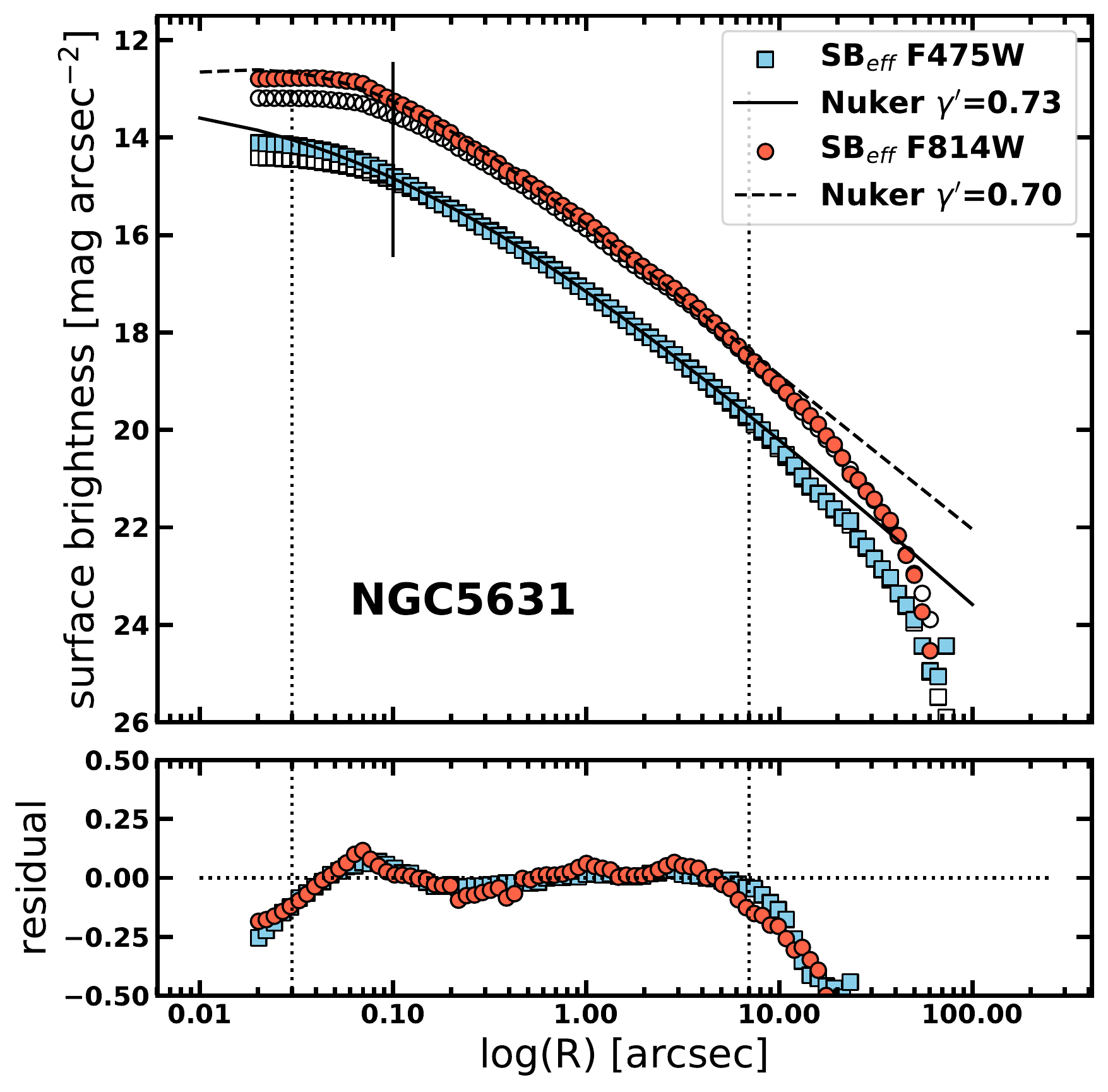}
\includegraphics[width=0.4\textwidth]{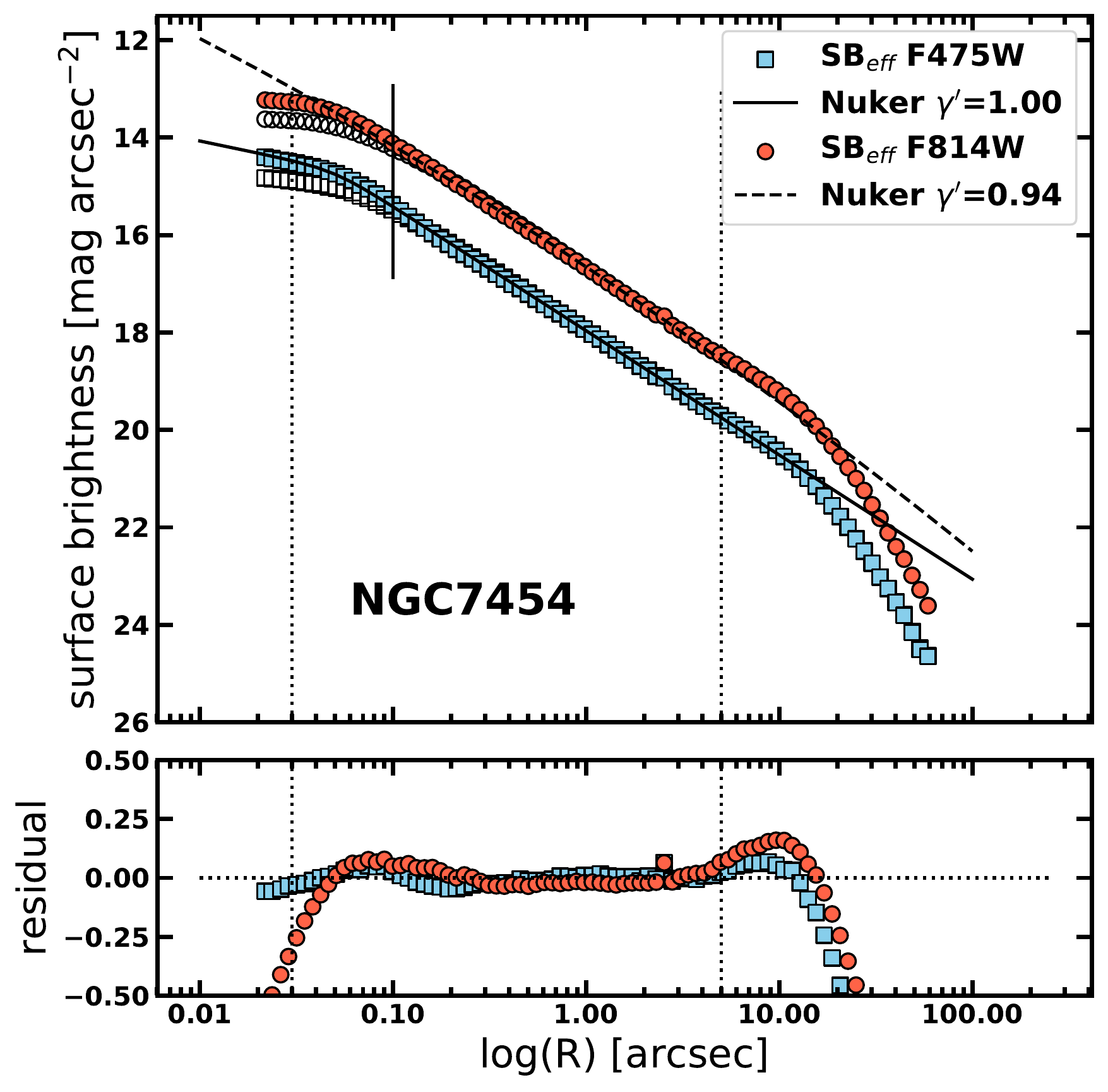}
\includegraphics[width=0.4\textwidth]{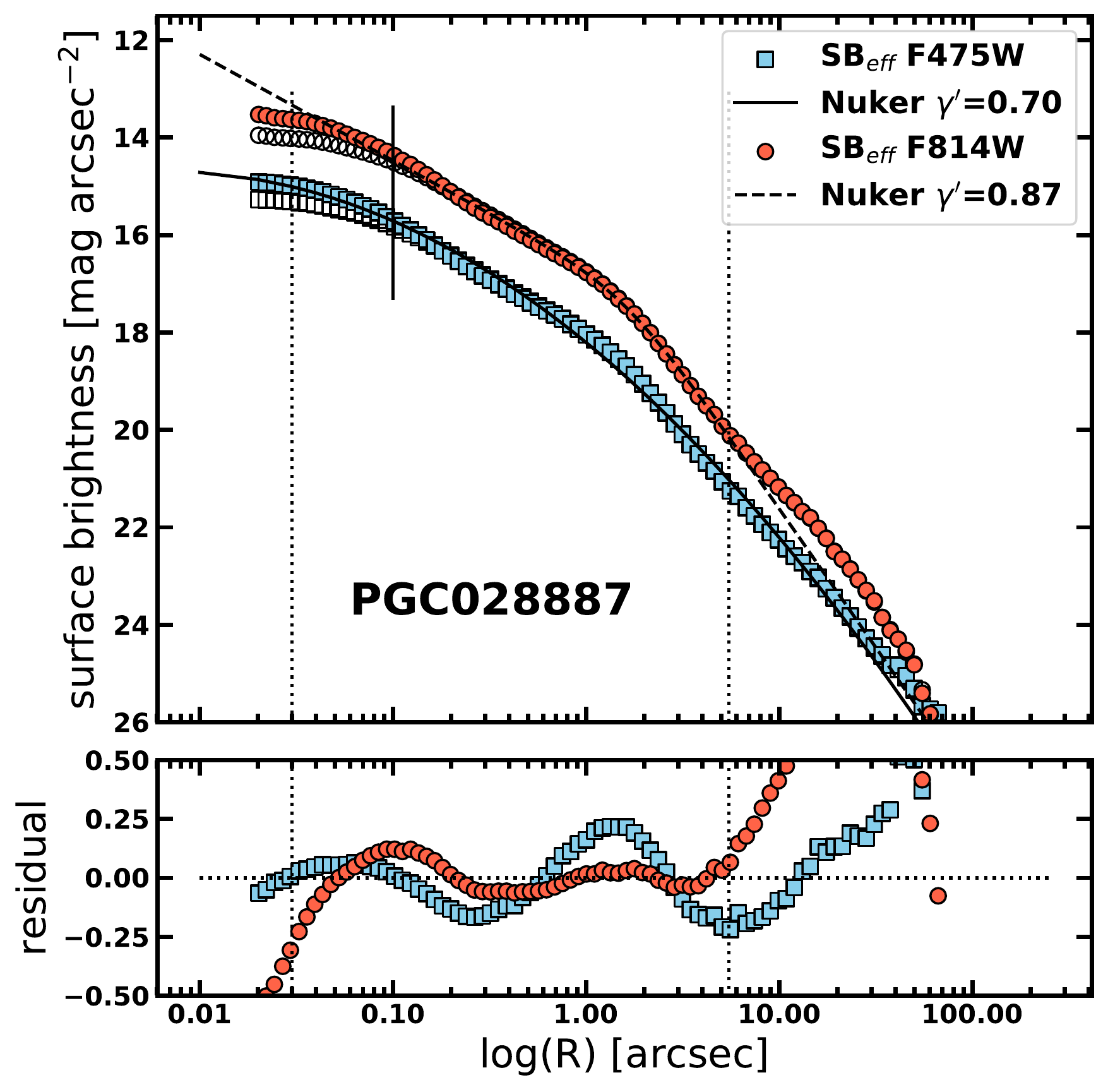}
\includegraphics[width=0.4\textwidth]{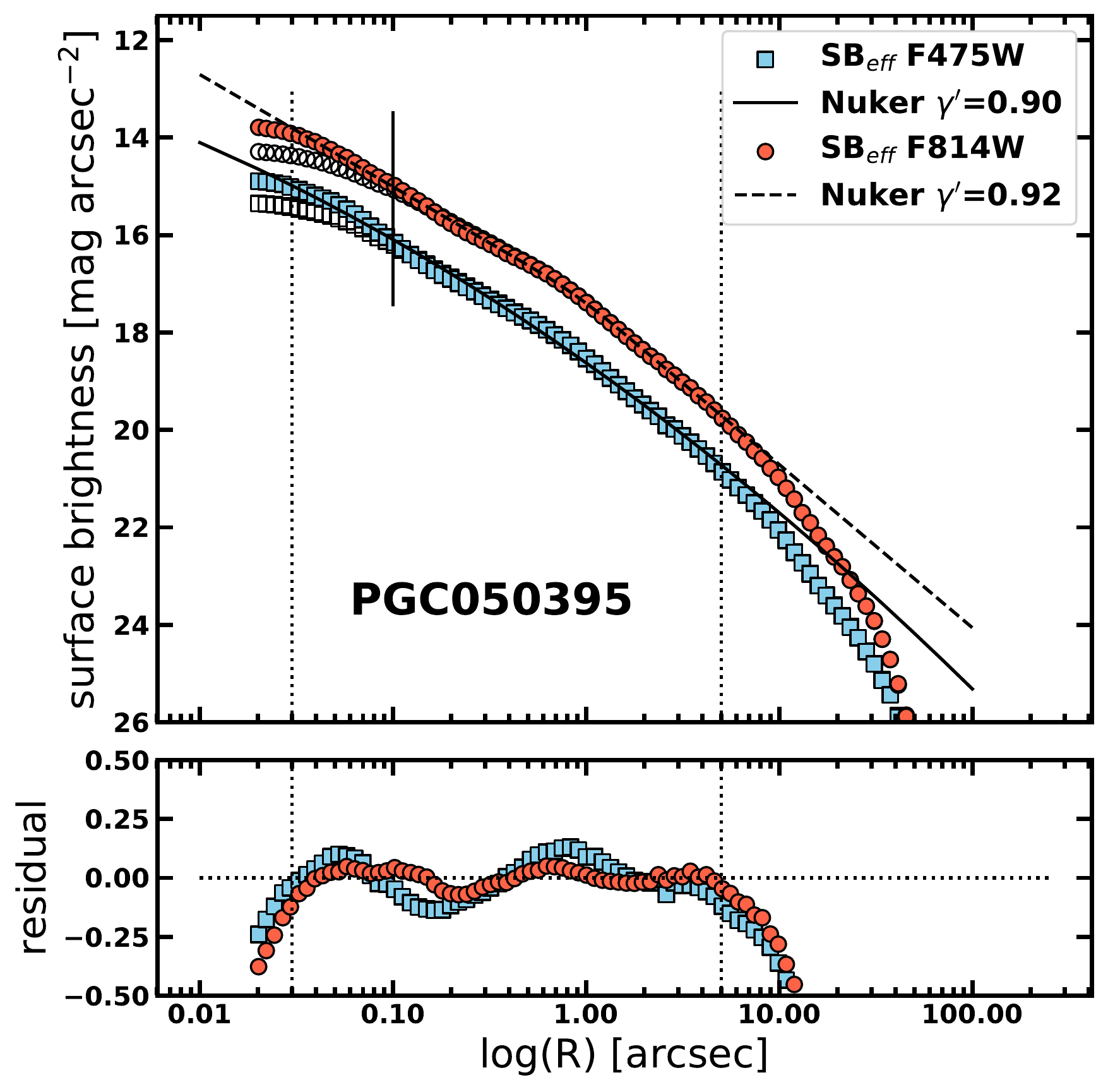}
\includegraphics[width=0.4\textwidth]{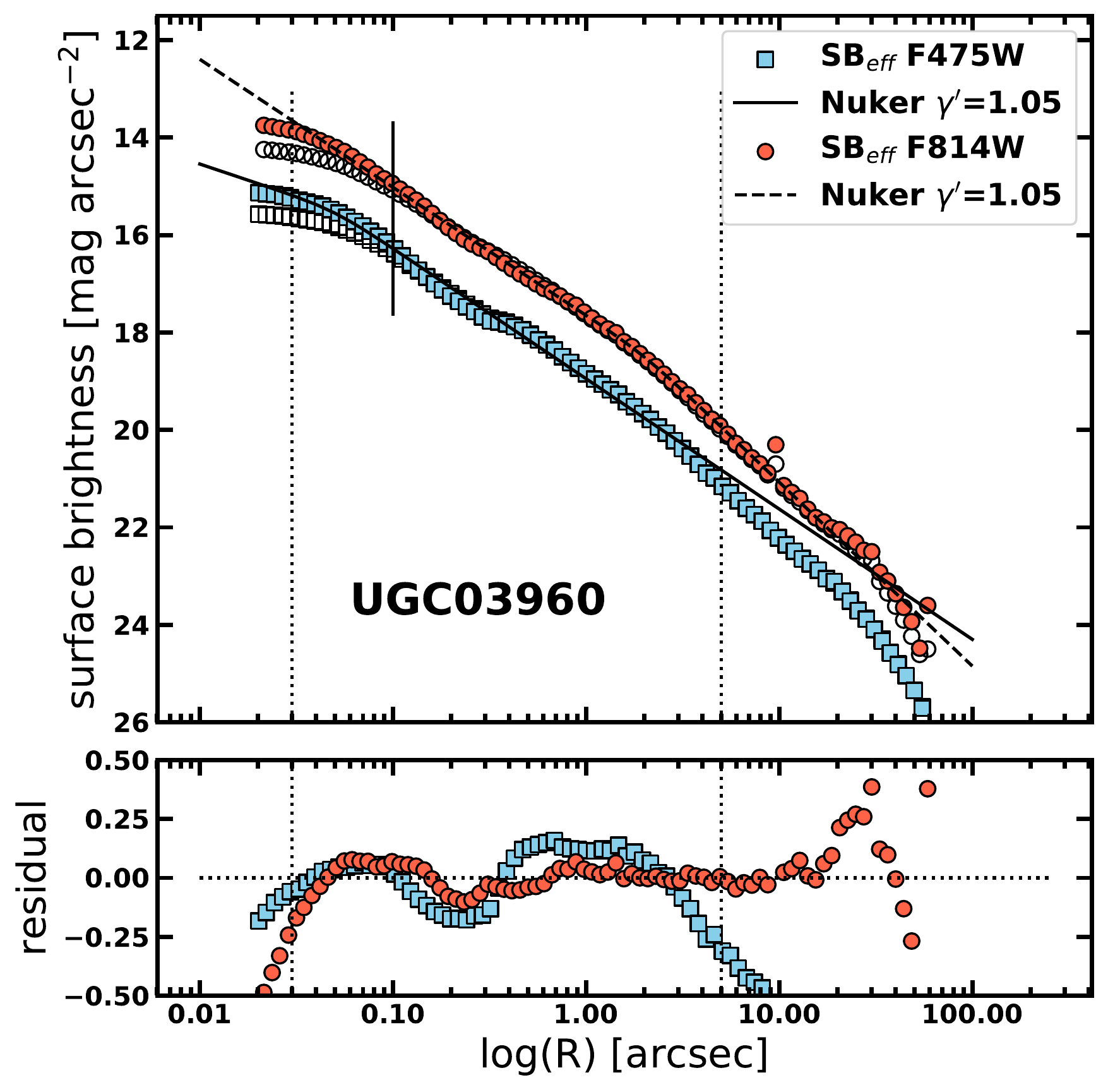}
\caption{Surface brightness profiles of the second six low-mass slow rotators from the ATLAS$^{\rm 3D}$ survey (ordered by name). Each galaxy is represented by two panels. The upper panel shows light profiles in F475W (light blue squares) and F814W (red circles) filters, and the smaller lower panel that shows residuals from the fit. Open symbols are original (observed) light profiles, and filled symbols are effective (deconvolved) profiles used for the analysis. The sampling does not correspond to the pixels of the WFC3 camera, but is defined by the tool for the isophote analysis.The Nuker fits to both filters are shown with solid (F475W) and dashed (F814W) lines. Two vertical dotted lines indicate the range used in the fit. The short vertical solid line indicates the location at which the $\gamma^\prime$ slope is measured. Our images have a pixel scale of 0\farcs032. For a comparison with Richardson-Lucy deconvolution results, see Appendix~\ref{a:deco_comp}.}
\label{f:sb2}
\end{figure*}

The final step in the preparation of the surface brightness profiles was to address the influence of the HST point spread function (PSF). We used the same method as in  \citet{2013MNRAS.433.2812K}: the single iteration of the Burger - van Cittert deconvolution \citep{1931vanCittert,1932Burger}. This method was shown to converge when $|1-{\bf A}(x)|<1$, where ${\bf A}$ is the PSF \citep{1954AuJPh...7..615B}, while \citet{1932Burger} advised that a single iteration is often sufficient. For each filter we constructed WFC3 PSF images using the code Tiny Tim \citep{2011SPIE.8127E..16K}. The original HST image was convolved with this PSF image using the {\tt STSDAS IRAF} task {\tt fconvolve} to create a smoothed image. This smoothed image was run through the {\tt ellipse} task, but now using the best-fit parameters (ellipse centre, $\Phi$, $\epsilon$) from the fit to the original HST image. This resulted in extraction of smoothed surface brightness profiles. We then approximated the effective profile as SB$_{eff} = 2\times $SB$_{orig}$ -- SB$_{conv}$, where SB$_{conv}$, SB$_{orig}$ , and SB$_{eff}$ are the convolved, original, and effective (deconvolved) surface brightness profiles, respectively. The errors on SB$_{eff}$ were estimated as a quadrature sum of the errors returned by the {\tt ellipse} task on SB$_{conv}$ and SB$_{orig}$. The original and the effective profiles in F475W and F814W filters are presented in Figs.~\ref{f:sb1} and \ref{f:sb2}, but the subsequent analysis was performed on the effective profiles only. In Appendix~\ref{a:deco_comp} we show through a comparison with the Richardson-Lucy deconvolution \citep{1972JOSA...62...55R,1974AJ.....79..745L} that the single iteration of the Burger-van Cittert deconvolution provides consistent surface brightness profiles down to the central WFC3 pixel. This means that the surface brightness profiles presented here have a spatial resolution comparable to those in the literature (and used in our previous work).

%
%

\section{Nuclear surface brightness profiles}
\label{s:nsb}

We are interested in characterising the surface brightness profiles as having or lacking a core. A core is a region interior to a certain break radius in which the surface brightness bends away from the steep outer profile to a shallower inner profile. The simplest functional form to describe such a surface brightness profile is a double power law \citep[e.g.][]{1992AJ....104..552L, 1994AJ....108.1598F}, which has subsequently been further developed into the Nuker profile \citep{1995AJ....110.2622L, 1996AJ....111.1889B}. 

An alternative parametrisation uses a combination of a power law and a \citet{1968adga.book.....S} function, the so-called core-S\'ersic model \citep{2003AJ....125.2951G}. Its main advantage is the fact that galaxy light profiles are typically well fitted with S\'ersic functions, and not with power laws. A core-S\'ersic model can, in principle, be expected to fit the surface brightness well across a wide range of radii \citep{2004AJ....127.1917T, 2006ApJS..164..334F}. Galaxies are, however, often made of multiple components, and as \citet{2009ApJS..182..216K} pointed out, instead of rigid analytic functions, it is necessary to use multiple (S\'ersic) functions and fit the light profiles piecewise.

Our main goal here is not to precisely fit the surface brightness profiles over the full range of the HST imaging. Instead, we wish to determine which galaxies can be classified as having cores, a task for which the double power-law model is sufficient. For consistency with \citet{2013MNRAS.433.2812K}, which was partly based on the literature values, we use the  Nuker profile. While the cores from the Nuker profile and the depleted cores from the core-S\'ersic model are not the same structures, in practice, there is only a small number of galaxies that are classified differently (e.g. where the Nuker profile and the core-S\'ersic model do not agree). Furthermore, it is often not clear if the differences are caused by the way the fit was  made (e.g. radial extent, as the core and outer profiles are interdependent), the difference in the data used, or the treatment of the PSF. We note, however, that the values of the parameters defining the core (e.g. the break radius, see Section~\ref{ss:nuk}) are not the same for both parametrisations, and they should be used with care. For more information, we refer to the discussion in \citet{2009ApJS..182..216K}, \citet{2012ApJ...755..163D,2014MNRAS.444.2700D}, and \citet{2013MNRAS.433.2812K}. 

\begin{table*}
   \caption{Parameters of the Nuker fits and kinematic structure of the analysed galaxies.}
   \label{t:fits}
$$
  \begin{array}{l c c r c c c c c c c c c c  }
    \hline \hline
    \noalign{\smallskip}

$name$ & $Filter$ &r_b & r_b & I_b & \alpha &\beta & \gamma & \gamma^\prime & $rms$ & r_\gamma & r_\gamma  & $Class$ & $Kinematic$ \\
              &              &      &        &      &           &         &               &                           &            &                  &                   &                & $structure$    \\
              &              &$(arcsec)$ & $(pc)$ &&  &&  & & & $(arcsec)$ & $(pc)$ & & \\
     (1)    &   (2)       &    (3)&  (4) &  (5)    &  (6)   &  (7)         &   (8)                   &    (9)    &     (10)       &    (11)         &    (12)      &   (13)        &(14)    \\
   \noalign{\smallskip} \hline \noalign{\smallskip}
$NGC\,0661$ & F475W & 0.05 & 7.42 & 14.32 & 5.00 & 1.03 & 0.00 & 1.00 & 0.04 & <0.10 & <14.84 & \setminus & $CRC$\\
$NGC\,0661$ & F814W & 0.06 & 8.90 & 13.04 & 5.00 & 1.03 & 0.06 & 0.96 & 0.02 & <0.10 & <14.84 & \setminus & $CRC$\\
$NGC\,1289$ & F475W & 0.40 & 74.47 & 16.36 & 0.11 & 5.13 & -3.00 & 0.76 & 0.05 & <0.10 & <18.62 & \setminus & $CRC$\\
$NGC\,1289$ & F814W & 2.00 & 372.34 & 17.19 & 0.20 & 4.11 & -1.15 & 0.71 & 0.03 & <0.10 & <18.62& \setminus & $CRC$ \\
$NGC\,3522$ & F475W & 0.40 & 49.45 & 16.63 & 0.07 & 5.17 & -3.00 & 0.89 & 0.06 & <0.10 & <12.36 & \setminus & $KDC$\\
$NGC\,3522$ & F814W & 1.03 & 127.34 & 16.56 & 1.25 & 1.66 & 0.85 & 0.89 & 0.05 & <0.10 & <12.36 & \setminus & $KDC$\\
$NGC\,4191$ & F475W & 0.27 & 51.31 & 16.47 & 2.18 & 1.26 & 0.38 & 0.47 & 0.04 & 0.06 & 11.55 &  \wedge & 2\sigma\\
$NGC\,4191$ & F814W & 0.31 & 58.91 & 15.27 & 2.34 & 1.30 & 0.48 & 0.53 & 0.04 & <0.10 & <19.00 & \setminus & 2\sigma\\
$NGC\,4690$ & F475W & 0.33 & 64.32 & 16.85 & 5.00 & 1.24 & 1.01 & 1.01 & 0.04 & <0.10 & <19.49 & \setminus & $NRR$\\
$NGC\,4690$ & F814W & 0.05 & 9.74 & 13.12 & 5.00 & 1.27 & 0.42 & 1.24 & 0.04 & <0.10 & <19.49 & \setminus & $NRR$\\
$NGC\,5481$ & F475W & 0.40 & 50.03 & 16.63 & 0.80 & 1.72 & 0.35 & 0.69 & 0.03 & <0.10 & <12.51 & \setminus & $KDC$\\
$NGC\,5481$ & F814W & 2.00 & 250.16 & 17.55 & 0.36 & 3.01 & -0.02 & 0.75 & 0.01 & <0.10 & <12.51 & \setminus & $KDC$\\
$NGC\,5631$ & F475W & 0.05 & 6.54 & 14.34 & 0.49 & 1.42 & -0.25 & 0.73 & 0.03 & <0.10 & <13.09 & \setminus & $KDC$\\
$NGC\,5631$ & F814W & 0.05 & 6.54 & 12.83 & 1.01 & 1.27 & -0.44 & 0.70 & 0.05 & <0.10 & <13.09 & \setminus & $KDC$\\
$NGC\,7454$ & F475W & 0.05 & 5.62 & 14.75 & 5.00 & 1.02 & 0.33 & 1.00 & 0.02 & <0.10 & <11.25 & \setminus & $NRR$\\
$NGC\,7454$ & F814W & 0.26 & 29.24 & 15.16 & 0.03 & 4.32 & -2.35 & 0.94 & 0.05 & <0.10 & <11.25 & \setminus & $NRR$\\
$PGC\,028887$ & F475W & 0.40 & 79.51 & 17.03 & 0.13 & 5.12 & -3.00 & 0.70 & 0.12 & <0.10 & <19.88 & \setminus & $KDC$\\
$PGC\,028887$ & F814W & 1.61 & 320.03 & 17.48 & 2.72 & 2.28 & 0.87 & 0.87 & 0.07 & <0.10 & <19.88 & \setminus & $KDC$\\
$PGC\,050395$ & F475W & 0.40 & 72.14 & 17.55 & 0.05 & 4.82 & -2.75 & 0.90 & 0.08 & <0.10 & <18.04 & \setminus &$CRC$\\
$PGC\,050395$ & F814W & 0.89 & 160.51 & 17.25 & 5.00 & 1.34 & 0.92 & 0.92 & 0.03 & <0.10 & <18.04 & \setminus &$CRC$\\
$UGC\,03960$ & F475W & 0.05 & 8.05 & 15.55 & 5.00 & 1.07 & 0.53 & 1.05 & 0.12 & <0.10 & <16.10 & \setminus &$NRR$\\
$UGC\,03960$ & F814W & 2.00 & 321.92 & 18.50 & 5.00 & 1.51 & 1.05 & 1.05 & 0.05 & <0.10 & <16.10 & \setminus &$NRR$\\

       \noalign{\smallskip}
    \hline
  \end{array}
$$ 
{Notes -- Column (1): Name of the galaxy. Column (2): HST WFC3 filter.  Column (3-8): Parameter of the Nuker fit as defined in Eq.~(\ref{eq:nuk}). Column (4) repeats the values of Col. (3) in the physical units. Column (9): Gradient of the luminosity profile evaluated at the limit of 0\farcs1 (assuming the HST resolution limit of 0\farcs05, the classification remains the same, except for NGC\,0661 and NGC\,5631, which then have intermediate profiles, but only in the F814W filter.). Column (10): Root mean square of the Nuker fit residuals. Columns (11 - 12): Cusp radius as defined by Eq.~(\ref{eq:cusp}) in arcseconds and parsec, respectively. Column (13): Classification into intermediate ($\wedge$) and power law ($\setminus$), where intermediate and power law are grouped as {\it core-less} in the text (see Section~\ref{s:nsb}). A {\it core} ($\cap$) does not occur in this sample. Column (14): Kinematic structure in the velocity maps taken from \citet{2011MNRAS.414.2923K}, where CRC means counter-rotating core, KDC means kinematically distinct core, $2\sigma$  is the double peak in the velocity dispersion map indicating counter-rotating discs, and NRR are non-regular velocity maps. NGC\,1222 is not included because we did not analyse this galaxy because its nuclear dust content is high. 
}
\end{table*}

\subsection{Nuker profiles}
\label{ss:nuk}

We used a double power-law function of the following form \citep{1995AJ....110.2622L}:
\begin{equation}
\label{eq:nuk}
I (r)  = 2^{(\beta - \gamma)/\alpha}  I_b \left(\frac{r_b}{r}\right)^\gamma \left[ 1 + \left(\frac{r}{r_b}\right)^{\alpha}\right]^{(\gamma- \beta)/\alpha},\\
\end{equation}

\noindent where $\gamma$ is the inner cusp slope as $r$ approaches 0. Galaxies with cores are marked with $r_b$, the radius at which a break in the light profiles occurs, and $I_b$ is the brightness at the break. Whether a light profile has a {\it core} or is {\it core-less} is  not parametrised by $\gamma$, but by the local (logarithmic) gradient $\gamma^\prime$ of the luminosity profile, evaluated at the HST angular resolution limit, $r^\prime$.  The definition \citep{2001AJ....121.2431R,2004AJ....127.1917T} of $\gamma^\prime$ is given by
\begin{equation}
\label{eq:gammaprime}
\gamma^\prime \equiv - \frac{d~log~I}{d~log~r} \bigg|_{r = r^\prime}   = - \frac{\gamma + \beta(r^\prime/r_b)^\alpha}{1 + (r^\prime/r_b)^\alpha}, \\
\end{equation}

\noindent where we adopted for $r^\prime$ = 0.1\arcsec as a measure of the HST resolution. In Appendix~\ref{a:deco_comp} we show that the deconvolution method we used can be trusted to about 0\farcs04 - 0\farcs05 for WFC3 data. Therefore we could also have selected a smaller radius to derive $\gamma^\prime$, for instance the resolution limit of 0.05. The main reason for not doing this is that part of the literature data based on Nuker profiles, including \citet{2013MNRAS.433.2812K} with values for the rest of the ATLAS$^{\rm 3D}$ galaxies, have used $r^\prime$ = 0.1\arcsec. Using a smaller radius does not change our conclusion here, as we discuss in more detail below. 

Following \citet{1995AJ....110.2622L}, {\it core} galaxies are defined to have $\gamma^\prime\le 0.3$, while power-law galaxies are defined to have profiles steeper than $\gamma^\prime > 0.5$. The values of $0.3 < \gamma^\prime < 0.5$ are nominally denoted as intermediate \citep{2001AJ....121.2431R}. In practice, we considered all galaxies with $\gamma^\prime > 0.3$ not to have resolved cores, and we refer to them as {\it core-less}. 

We also defined the cusp radius as the radius at which $\gamma^\prime=0.5$ \citep{1997ApJ...481..710C},
\begin{equation}
\label{eq:cusp}
r_\gamma \equiv r_b \left(  \frac{0.5 - \gamma^\prime}{\beta - 0.5}\right)^{1/\alpha}.\\
\end{equation}

\noindent This radius is considered to be a more robust core scale radius, which reasonably approximates the core radius obtained from core-S\'ersic fits \citep{2012ApJ...755..163D}. For galaxies that have $\gamma^\prime\ge0.5$, we adopt $r_\gamma<0.1$\arcsec. 

The fits to the deconvolved surface brightness profiles in both filters are shown in Figs.~\ref{f:sb1} and \ref{f:sb2} and their parameters are presented in Table~\ref{t:fits}. The fitting was performed using a least-squares minimisation routine based on {\tt MPFIT} \citep{2009ASPC..411..251M}, an implementation of the MINPACK algorithm \citep{1980ANL.....80.74M}. The deconvolved profiles in both filters were fitted between the inner radius of 0.03\arcsec\, and an outer radius that was chosen for each galaxy, limiting the spatial range in which Eq.~(\ref{eq:nuk}) was used. Reasonable fits were obtained when the outer radius was generally close to 10\arcsec, although in a few cases, they were considerably smaller (e.g. 2\arcsec\, for NGC\,0661). This is a typical range for fits with the Nuker law \citep{1995AJ....110.2622L,2001AJ....121.2431R, 2005AJ....129.2138L,2013MNRAS.433.2812K}. 

In a few cases, the residual plots in Figs.~\ref{f:sb1}  and ~\ref{f:sb2} show significant deviations between the Nuker model and the data within the fitted region (e.g. NGC\,3522, NGC\,4191, PGC\,028887, PGC\,050395, and UGC\,03960). The deviations in the inner parts (within the fitting region) arise partially because galaxy light does not follow a power-law profile. The Nuker fit therefore needs to be limited to different regions for different galaxies. In addition, some of our galaxies likely contain multiple light components that are most appropriately decomposed with S\'ersic profiles. At large scales ($>2.5\arcsec$), light profiles of our galaxies are well fitted by a S\'ersic profile, while PGC\,028887 and UGC\,03960 are better fit with a double S\'ersic model \citep{2013MNRAS.432.1768K}. The HST data show that additional (S\'ersic) components are also necessary within the central few arcseconds to reproduce the profiles well. We did not attempt a full decomposition of the radial profiles because we are only interested in the existence (or lack) of cores.

\subsection{Are cores of our galaxies beyond the HST resolution limit?}
\label{ss:detect}

Our galaxies are at distances of between 25 and 40 Mpc, and it is not obvious that even with the HST resolution we would be able to resolve or even detect their cores. Based on our choice for $r^\prime=0.1\arcsec$ and a limiting distance of 40Mpc, the lower limit to the size of cores that we can detect is about 19 pc, while sizes a factor of two smaller would be detectable for  $r^\prime=0.05\arcsec$. This means that we cannot expect to detect any core with a physical radius smaller than $\sim10$pc. Cores detected in previous works have characteristic sizes typically larger than 20 pc \citep[e.g.][]{2007ApJ...662..808L, 2013MNRAS.433.2812K, 2014MNRAS.444.2700D}.

As defined in the previous section, two relevant radii are related to the core size within the Nuker profile: $r_b$ and $r_\gamma$. The former is the radius at which the Nuker profile has the maximum curvature, or the location of the transition between the two power laws of Eq.~(\ref{eq:nuk}). The latter is a characterisation of the physical size of the core, defined as the location at which the logarithmic slope of the galaxy surface brightness reaches a given value of $\gamma^\prime$, as shown in Eq.~(\ref{eq:cusp}). The choice of $\gamma^\prime$ is somewhat arbitrary, but as \citet{1997ApJ...481..710C} and \citet{2007ApJ...662..808L} showed, $\gamma^\prime=0.5$ is a natural way to separate cores from core-less galaxies, and it provides tighter relations with other galaxy parameters. However, $r_\gamma$ does not specify the actual size of the core, but should be considered, in the words of \citet{2007ApJ...662..808L}, as  ``just a convenient representative scale". 

Our galaxies have $\gamma^\prime>0.5$ (all except NGC\,4191 in F475W filter), and therefore the cusp radius is not well defined for the combination of the $\alpha$ and $\beta$ parameters. As is the custom for such galaxies, we placed an upper limit on the core size of $r_\gamma<0.1\arcsec$ (see Table~\ref{t:fits} for values in parsec). Given the distances to our galaxies, this places a limit to the core scales of $<10-20$ pc, as expected from the resolution arguments. The possibility remains that our galaxies harbour smaller cores.   

Known galaxies with cores are all massive, bright, and have large velocity dispersions, therefore it might be an issue that the current scaling relations, such as $r_\gamma - \sigma$ or the $r_\gamma - $ M$_V$ relations, are not representative of  our galaxies. In order to use them for our galaxies, they need to be extrapolated to $\sigma \sim100$ km/s or M$_V<-20$, while they are currently confined to $\sigma>150$ and M$_V>-21$ \citep[Figs. 4 and 5 in ][]{2007ApJ...662..808L}. When we apply these relations to estimate the sizes, the potential cores in our galaxies would have $r_\gamma<5$ pc (for $r_\gamma - \sigma$) and $r_\gamma<10$ pc (for $r_\gamma - $ M$_V$, assuming V-K = 3 colour for our galaxies and using the absolute K-band magnitudes from Table~\ref{t:sample}). Even though $r_\gamma < r_b$, and not the physical size of the core, it is likely that we would not be able to detect such small cores. The same conclusion remains valid for most galaxies when we use assume $r^\prime=0.05$ (and $r_\gamma <0.05\arcsec$).

\begin{figure}
\includegraphics[width=\columnwidth]{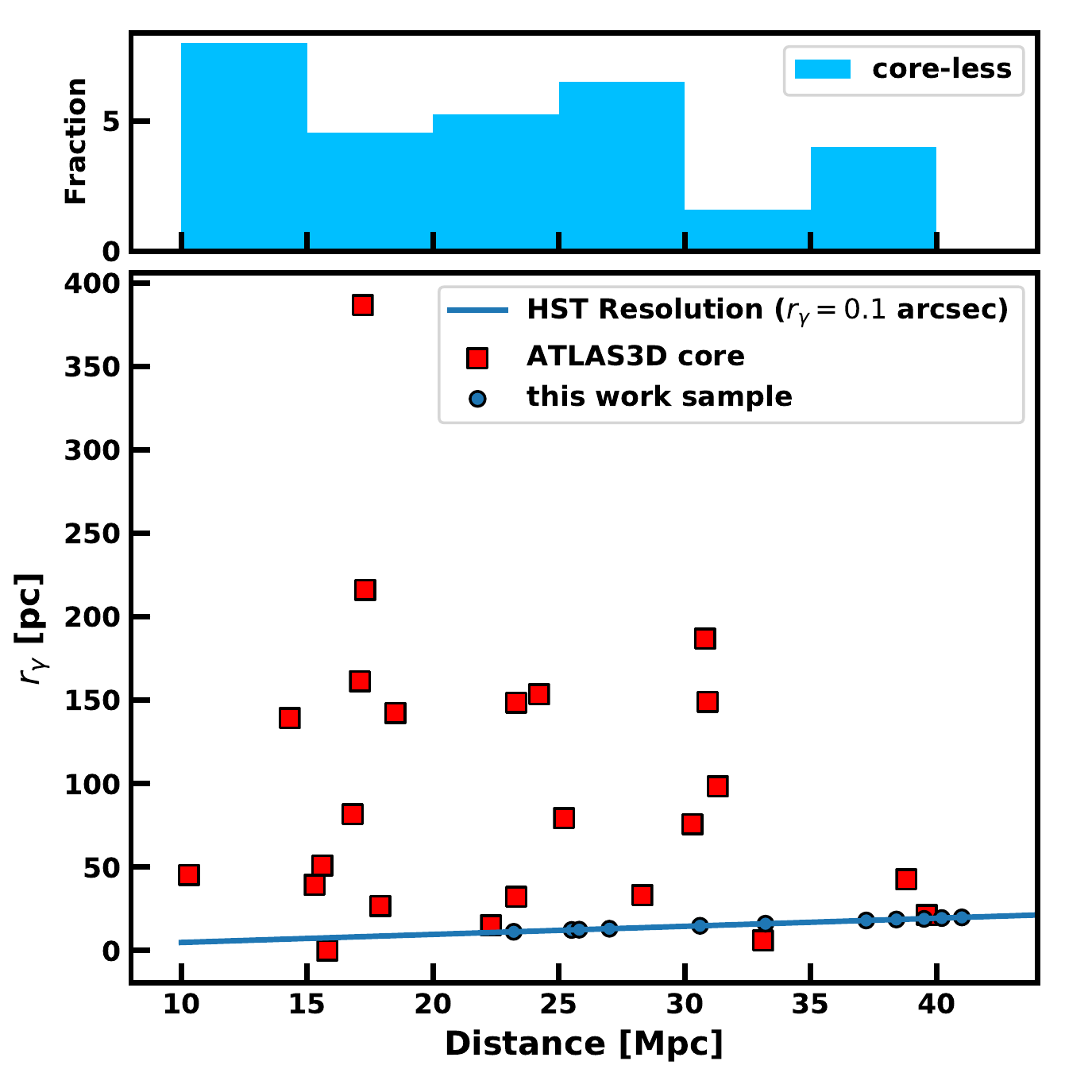}
\caption{Distribution of sizes ($r_\gamma$) and distances to core ATLAS$^{\rm 3D}$ galaxies (red squares). The blue solid line shows the HST limit of $r_\gamma<0.1$\arcsec\, assumed for all galaxies that have $\gamma^\prime >0.5$. Below this line, cores cannot formally be detected. Circles on the line are the upper limits on possible core sizes for galaxies presented here with values from Table~\ref{t:fits}. The histogram at the top shows the fraction of core-less galaxies in bins of distance. There is no evidence for an increase of the fraction of core-less galaxies as a result of resolution effects. }
\label{f:rg}
\end{figure}

An alternative is to use relations from \citet{2014MNRAS.444.2700D}, such as their $r_b - \sigma$ or $r_b - \mu_0$ (Fig. 5 in that paper). These relations are made for core-S\'ersic fits and are based on a smaller sample than the \citet{2007ApJ...662..808L} relations. They also need to be extrapolated because the velocity dispersion is limited to $\sigma>200$ km/s, while $\mu_0>14$ (V-band surface brightness). Furthermore, we recall that $r_b$ (core-S\'ersic) $\approx 1/2 r_b$ (Nuker) \citep{2014MNRAS.444.2700D}, but $r_\gamma \sim r_b$ (core-S\'ersic)  \citep{2012ApJ...755..163D}. Nevertheless, using the $r_b - \mu_0$ relation on our galaxies with typical surface brightness in F475W filter between 14 and 15 mag/\arcsec$^2$, we might expect core sizes of 20 -- 40 pc, parametrised as $r_b$ (core-S\'ersic), while the $r_b - \sigma$ relation predicts for most of galaxies $r_b<5$ pc. These values show a considerable spread in the estimated sizes, and point to a general problem of predicting the (relevant) sizes of cores: they are highly uncertain. 

In Fig.~\ref{f:rg} we compare the sizes of cores ($r_\gamma$) and distances to galaxies in the ATLAS$^{\rm 3D}$ sample. We over-plot the effect of the resolution of the HST and the upper limits for our galaxies from Table~\ref{t:fits}. They show that our observations are able to detect cores with sizes typical for ETGs, even at the distance limit of the sample. We also show a histogram with the fraction of core-less galaxies as a function of distance. The purpose is to demonstrate that there is no sudden increase in the fraction of core-less galaxies with distance, which might be the case if we were missing cores because of the resolution effects. There is a lack of galaxies with large cores at distances beyond $\sim35$ Mpc, but this is a feature of the local universe and the ATLAS$^{\rm 3D}$ sample, which contains no massive galaxies at these distances.

Table~\ref{t:fits} shows that none of our galaxies have a core larger than 10-20pc. They could have undetected cores from sub-parsec up to a few parsec in size, but these are obviously very different from cores in other slow rotators. As we discuss later, the galaxies presented here differ from other slow rotators, most notably in mass and stellar population parameters. It is crucial that the galaxies we investigated here are slow rotators because this information was absent from all previous samples that were used to investigate nuclear surface brightness profiles. We showed that among slow rotators are galaxies that have large cores, tens to hundreds of parsec in size (\citealt{2013MNRAS.433.2812K}, and see also \citealt{2012ApJ...759...64L}), and we here address slow rotators that in the most extreme case cannot have cores larger than a few parsec.

For the rest of the paper we assume that the galaxies analysed here are all core-less.  When we assume that all galaxies with non-regular kinematics have similar formation histories and that slow rotators should be core galaxies \citep[e.g.][]{2012ApJ...759...64L}, then this is already an unexpected result. 

%
%

\section{Results: not all slow rotators have cores}
\label{s:res}

In this section, we first update \citet{2013MNRAS.433.2812K} with information on the surface brightness profiles for all ATLAS$^{\rm 3D}$ slow rotators. We then extend the analysis using the information on the stellar populations, in particular the metallicity and age gradients.

\subsection{Cores versus rotation}
\label{ss:nocores}

\subsubsection{Global kinematic parameters}
\label{sss:glob}

\begin{figure}
\includegraphics[width=\columnwidth]{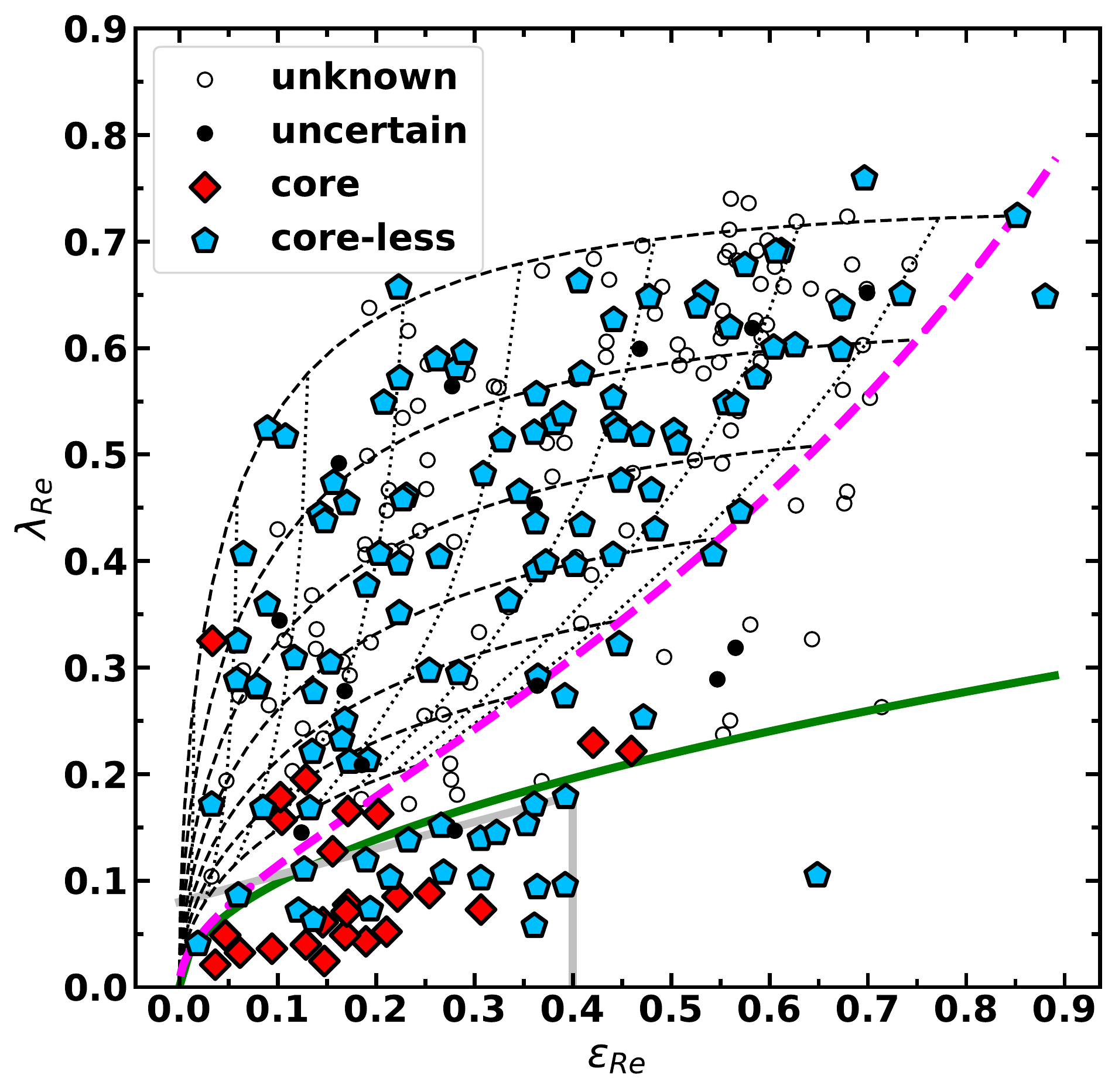}
\caption{Specific stellar angular momentum vs. the observed ellipticity of the ATLAS$^{\rm 3D}$ galaxies. Small open circles are galaxies with no available HST observations, and small filled circles are galaxies for which the central surface brightness profile is uncertain, mostly because of a dusty nucleus. Red diamonds are core galaxies ($\gamma^{\prime} \le 0.3$), and blue pentagons are core-less galaxies ($\gamma^{\prime}> 0.3$). The green solid line separates fast from slow rotators following \citet{2011MNRAS.414..888E}, and the grey solid line is the \citet{2016ARA&A..54..597C} alternative. The dashed magenta line shows the edge-on view for spheroidal galaxies integrated up to infinity with $\beta=0.7\times \epsilon_{intr}$, as in \citet{2007MNRAS.379..418C}. Other dashed lines show the same relation projected at inclinations of 80\degr, 70\degr, 60\degr, 50\degr, 40\degr, 30\degr, 20\degr , and 10\degr \ (from right to left). The dotted lines show the change in location for galaxies of intrinsic $\epsilon_{intr}=0.85, 0.75, 0.65, 0.55, 0.45, 0.35,$ and 0.25 (from top to bottom). This plot differs from Fig.4 (left) of \citet{2013MNRAS.433.2812K} in that all slow rotator galaxies now have nuclear surface brightness characterisation, but the number of slow rotators with cores did not increase. NGC\,1222 is the small black symbol on the grey line. }
\label{f:lamR}
\end{figure}

\begin{figure}
\includegraphics[width=\columnwidth]{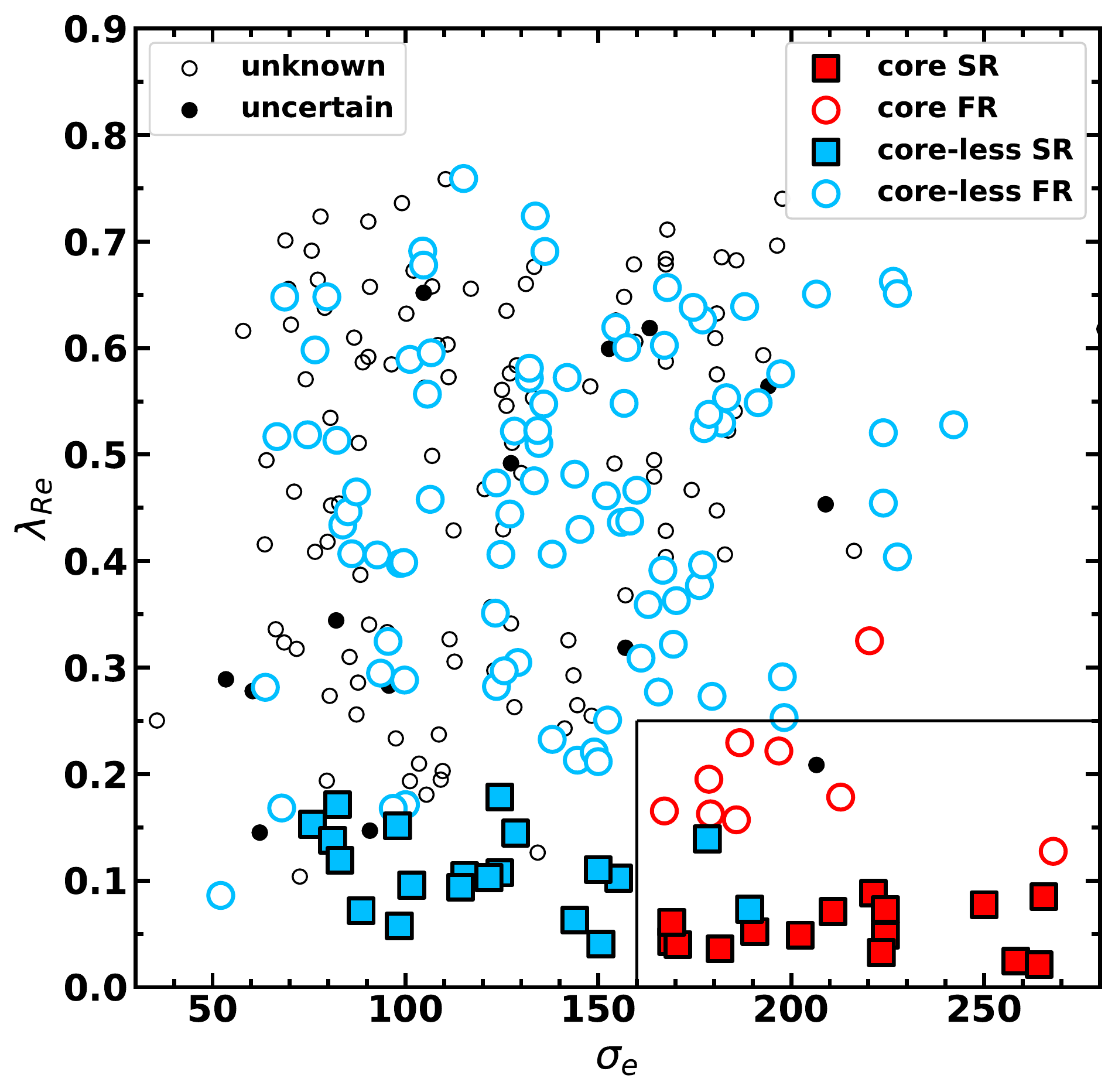}
\caption{Specific stellar angular momentum vs. stellar velocity dispersion within the half-light radius for the ATLAS$^{\rm 3D}$ sample. Galaxies are separated into {\it core slow rotators}, {\it core fast rotators}, {\it core-less slow rotators,} and {\it core-less fast rotators,} as shown on the legend. Galaxies with uncertain profiles and galaxies with no HST data are shown with small solid and open symbols, respectively. The box delineated by solid black lines is from Fig. 7 of \citet{2013MNRAS.433.2812K}, where mostly cores occur. Data presented here show that outside the box there are no new core galaxies, while the one galaxy that was previously unclassified (NGC\,0661, the upper blue symbol within the box) also has a core-less surface brightness profile. The galaxy with an uncertain surface brightness profile in the box is NGC\,3607, while NGC\,1222 is the small solid circle with $\sigma \sim90$ km/s and $\lambda_R\sim0.15$.}
\label{f:sig}
\end{figure}

Figs.~\ref{f:lamR} and \ref{f:sig} present the specific angular momentum ($\lambda_{Re}$) versus the observed elliptiticy and the velocity dispersion within one half-light radius for ATLAS$^{\rm 3D}$ galaxies, respectively. The data are the same as in Figs. 4 (left) and 7 of \citet{2013MNRAS.433.2812K}, where we now complete the information on nuclear light profiles for all remaining slow rotators. Two results are evident. Stellar angular momentum alone remains an ambiguous predictor of the presence of cores (Fig~\ref{f:lamR}). The transition region between fast and slow rotators ($0.1<\lambda_{Re}<0.2$) contains galaxies with both types of nuclear surface brightness profiles. Only galaxies with the lowest measured values for $\lambda_{Re}$ are more likely to have cores.  Notably, core-less light profiles seem to be more likely associated with flatter slow rotators, dominating the distribution for $\epsilon_{Re}>0.25$. 

\begin{table*}
   \caption{Incidence of kinematic and photometric features among ATLAS$^{\rm 3D}$ slow rotators.}
   \label{t:stat}
$$
  \begin{array}{c | c c c c c | c c | c c c c c }
    \hline \hline
    \noalign{\smallskip}
    
$class$      & $KDC$ & $CRC$ & 2\sigma & $LV$ & $NRR$ & $Total$ & $Fraction$ & $f(KDC)$ & $f(CRC)$ & $f($2\sigma$)$ & $f(LV)$ & $f(NRR)$\\
(1)      & (2) & (3) & (4) & (5) & (6) & (7) & (8)  & (9) & (10) & (11) & (12) & (13) \\
   \noalign{\smallskip} \hline \noalign{\smallskip}
$core-less$   &   6        &   4        &   4          &   1      & 5       &   20       &0.57              &  0.30     & 0.20    & 0.20    & 0.05   & 0.25\\ 
$core$          &    4       &   3        &    0          &   4      &  4      &   15       &0.43              &  0.27   & 0.20    & 0.00   & 0.27  & 0.27\\
       \noalign{\smallskip}
    \hline
  \end{array}
$$ 
{Notes -- Column (1): Classification of the surface brightness profiles based on this work and Table C1 from \citet{2013MNRAS.433.2812K}. Columns (2-6): Kinematic classes from Table D1 of \citet{2011MNRAS.414.2923K}: KDC is the kinematically distinct core, CRC is the counter-rotating core, $2\sigma$ is the double velocity dispersion peak (indicative of counter-rotation), LV is the low-level velocity (no net rotation), and NRR is the non-regular rotation (but without special features, except for a possible kinematic twists). Columns (7) and (8): Total number and fraction of slow rotators with different types of surface brightness profiles (NGC\,1222 is excluded), respectively.  Columns (9-13): Fraction of galaxies with KDC, CRC, $2\sigma$, LV, and NRR kinematic features.} 
\end{table*}

On the other hand, a combination of the effective velocity dispersion, $\sigma_e$, and $\lambda_{Re}$, remains the best predictor of the nuclear light profile structure (Fig.~\ref{f:sig}), as noted in \cite{2013MNRAS.433.2812K}. All slow rotators with a velocity dispersion lower than about 160 km/s have core-less light profiles. Within the rectangle, defined as $\sigma_e > 160$ km/s and $\lambda_R < 0.25$, where essentially all core galaxies are found, there are now two core-less galaxies, and one with an uncertain profile. This means that only about 10\%\  of the galaxies within the box are likely to have core-less profiles. When we restrict this to $\sigma_e > 200$ km/s and $\lambda_{Re} < 0.25$, essentially all galaxies have cores. This provides an interpretation for studies of large samples, such as the one of \citet{2018MNRAS.477.4711G}, who investigated the kinematics of MANGA galaxies for which high-resolution nuclear stellar profiles are not available. Assuming that our low number statistics can be taken as an indicator, about 15 per cent of galaxies with $160<\sigma_e<200$ km/s and $\lambda_{Re}<0.25$ in the MANGA sample could be core-less. Nuclear surface brightness profiles of all MANGA galaxies with $\sigma_e>200$ km/s and $\lambda_{Re}<0.25$ most likely exhibit cores. 

Galaxies in the boxed region of Fig.~\ref{f:sig} are separated into two groups with a jump in $\lambda_{Re}$ for about 0.05-0.1. The group of the higher $\lambda_{Re}$ (and $\sigma_e<220$ km/s) corresponds to the group of fast rotators with cores in Fig.~\ref{f:lamR}. Slow rotators with cores are confined to the lowest values of $\lambda_{Re}$, but extend to the highest velocity dispersions. 

Slow rotators are heterogeneous in terms of the mass (spanning almost two orders of magnitude in the ATLAS$^{\rm 3D}$ sample), environment, and kinematics \citep{2011MNRAS.414..888E, 2016ARA&A..54..597C}. Their velocity maps exhibit no net rotation, various types of KDCs, as well as velocity maps that show rotation, but it is irregular and with twists \citep{2008MNRAS.390...93K,2011MNRAS.414.2923K}. The velocity maps of the slow rotator sub-sample presented here are as diverse. Notably, 7 of 11 galaxies (we also excluded NGC\,1222 from the kinematic analysis of the sample) have KDCs or counter-rotating cores (CRC), one galaxy is classified as a $2\sigma$ (it contains a counter-rotating disc, recognisable with two peaks in the velocity dispersion maps), with the remaining three having non-regular rotation (NRR)\footnote{Velocity maps characterised as NRR by \citet{2011MNRAS.414.2923K} have a relatively low  rotation amplitude, an irregular appearance, and possible kinematic twists. These all result in deviations from the regular rotation (as found in fast rotators) that is described by a simple disc-like velocity model.} velocity maps. In the last column of Table~\ref{t:fits} we copy the kinematic structure of these galaxies from \citet{2011MNRAS.414.2923K}.

The high incidence of KDC/CRCs among {\it core-less slow rotators} is worth a closer look, especially when we consider that CRCs are a sub-class of KDCs in which the rotation of the KDC is opposite to the orientation of the main body (the angle difference is $\sim180$\degr). In Table~\ref{t:stat} we combine the information from this work (Table~\ref{t:fits}), the surface brightness profile classification from table~C1 of \citet{2013MNRAS.433.2812K}, and the kinematic structures from table~D1 of \citet{2011MNRAS.414.2923K}. We removed NGC\,1222 from the total of 36 ATLAS$^{\rm 3D}$ slow rotators, and show that only 43\%\  of the remaining 35 slow rotators have cores. The relative fraction of KDCs or CRCs is similar between core and core-less galaxies, however, and the same is true for galaxies with NRR velocity maps. The clear difference in the kinematics is visible in the remaining two kinematic classes. Low-velocity (LV) features are almost entirely found among {\it core slow rotators}, while $2\sigma$ features are found only among {\it core-less slow rotators}. 

An exception to the rule is NGC\,6703, classified as LV and a core-less slow rotators, but its non-rotation arises because this galaxy is seen almost face-on, as has been suggested by \citet{2011MNRAS.414..888E} and confirmed by dynamical modelling of \citet{2013MNRAS.432.1709C}. We also highlight the case of the core galaxy NGC\,5813, the first galaxy that was recognised as having a KDC \citep{1980MNRAS.193..931E,1982MNRAS.201..975E}. Recent high-quality MUSE data also showed a $2\sigma$ feature in their velocity dispersion map \citep{2015MNRAS.452....2K}. This galaxy is unusual because it apparently does not have two counter rotating discs, but the MUSE data can be reproduced with a dynamical model constructed of two counter-rotating short axis tube orbital families. Strictly speaking, NGC\,5813 could be included in Table~\ref{t:stat} as the only core $2\sigma$, but we prefer to keep it as a KDC, reserving the $2\sigma$ class for galaxies that are made of counter-rotating discs. Nevertheless, NGC\,5813 is an important case because it shows that counter-rotation does not need to be solely associated with discs, but likely comes in a spectrum of possible orbital  structures. 

On the other hand, NGC\,0661 (a CRC) and NGC\,7454 (a NRR) could also be considered $2\sigma$ galaxies. For these two galaxies, \citet[][fig.~12]{2016ARA&A..54..597C} constructed successful dynamical models made of two counter-rotating discs. If we assumed that NGC\,0661 and NGC\,7454 were such objects, the fraction of various kinematics classes of {\it core-less slow rotators }would change: f(CRC)=0.15, f($2\sigma$)=0.3, and f(NRR)=0.2, but the overall conclusions remain the same. A significant fraction of {\it core-less slow rotators} are dominated by counter-rotation, which decreases the net angular momentum.

\begin{figure}
\includegraphics[width=\columnwidth]{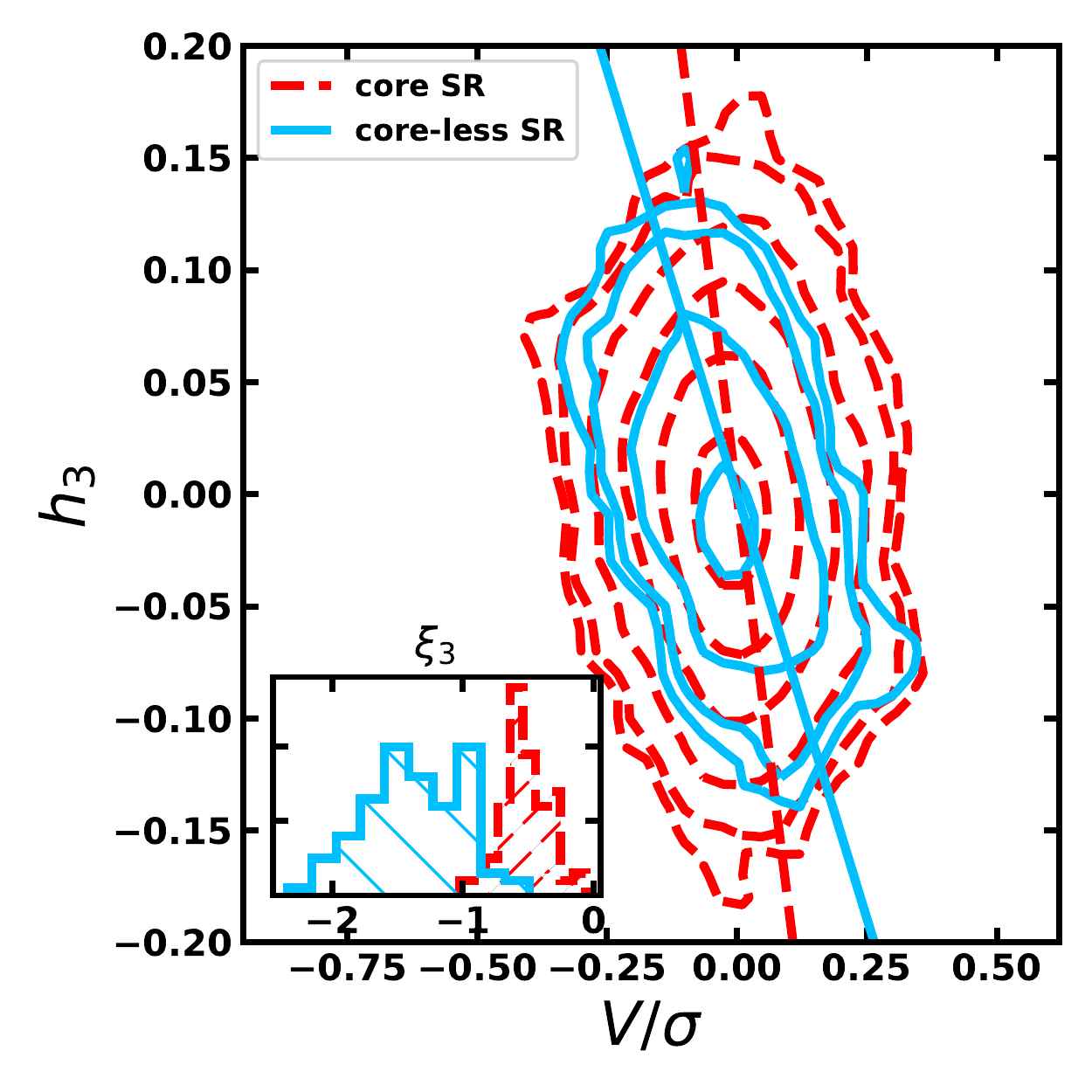}
\caption{Local $V/\sigma - h_3$ diagram for slow rotators separated into core (dashed red line) and core-less (solid blue line). Contours are based on logarithmic number counts starting from 0.25 and 0.5 and then increase with a step of 0.5 until 2.5 for {\it core slow rotators} and 1.5 for {\it core-less slow rotators}, respectively. Only bins with an uncertainty $\delta h_3<0.05$ and $\sigma_e>120$ km/s are plotted. Straight lines show the slope of the distributions as measured by the $\xi_3$ parameter (see text for details). The inset histogram shows the distribution of the $\xi_3$ obtained using the jackknife method, where from the original distributions for {\it core slow rotators} and {\it core-less slow rotators} one galaxy (in each sub-sample) was randomly removed, and the $\xi_3$ remeasured. The histograms are notably different, with {\it core-less slow rotators} having steeper slopes ($\xi_3<1$). This confirms the robustness of the weak anti-correlation between the $V/\sigma$ and $h_3$ distributions of {\it core-less slow rotators}.   }
\label{f:vsh3}
\end{figure}

\subsubsection{Local kinematic parameters}
\label{sss:loc}

The difference in the kinematics between {\it core slow rotators} and {\it core-less slow rotators} must originate in their formation. The statistics in Table~\ref{t:stat} suggests that the main difference is the existence of hidden disc-like structures, which by virtue of the counter-rotation leads to a low angular momentum. To investigate this conjecture further, we also considered the higher order moments of the LOSVD, in particular, the $h_3$ Gauss-Hermite moment, as defined by \citet{1993ApJ...407..525V}. Figure~\ref{f:vsh3} shows that {\it core slow rotators} and {\it core-less slow rotators} have marginally different distributions in the $V/\sigma - h_3$ plane: that there is an anti-correlation between $V/\sigma$ and $h_3$ for {\it core-less slow rotators}. This anti-correlation is not strong, but the distribution of the blue contours ({\it core-less slow rotators}) is clearly skewed with respect to the symmetric distribution of red contours ({\it core slow rotators}). We quantify the difference between the distributions in Fig.~\ref{f:vsh3} using 
\begin{equation}
\label{eq:zeta3}
\xi_3 = \frac{\sum_i F_i h_{3,i}(V_i/\sigma_i)}{\sum_i F_i h_{3,i}^2}\end{equation}

\noindent from \citet{2019MNRAS.489.2702F}, where for each spatial bin $i$, there is the local flux $F_i$, the mean velocity $V_i$, the velocity dispersion $\sigma_i$, and the skewness parameter $h_{3,i}$ of the LOSVD. This global parameter measures the slope of the distribution of points in the $h_3 - V/\sigma$ plane, as shown by straight lines in Fig.~\ref{f:vsh3}. It is tuned such that when $h_3$ and $V/\sigma$ are fully (anti-)correlated, the correlation is given by $h_3 = (1/\xi_3)V/\sigma$. \citet{2019MNRAS.489.2702F} showed examples of various $h_3$ and $V/\sigma$ distributions with and without correlations, and corresponding $\xi_3$ parameters. Fast-rotating galaxies have $\xi_3 <-4$, while slow rotators are expected to have $\xi_3$ close to 0. Positive $\xi_3$ are also possible and often found in barred systems. In our case, as shown in Fig.~\ref{f:vsh3}, {\it core-less slow rotators} combined have $\xi_3=-1.3$, and {\it core slow rotators} combined have $\xi_3 =-0.5$. The difference between the two distributions is small because all galaxies are slow rotators, but it is significant. 

We tested the significance of the difference between the two distributions by randomly removing one {\it core-less slow rotator} and one {\it core slow rotator} from the distributions and remeasuring the slope of the distribution through the $\xi_3$ parameter. The aim was to show how the distributions of points in the $h_3 - V/\sigma$ are dependent on individual galaxies, that is, whether the distributions are skewed by, for example, a single galaxy. The resulting histograms of a jackknife sequence of 100 such samples are shown in the inset panel of Fig.~\ref{f:vsh3}. The difference in the two $\xi_3$ distributions is clearly visible, where {\it core slow rotators} show a relatively narrow distribution that peaks at low $\xi_3$ values compared to {\it core-less slow rotators}. As expected from the original sample, the distribution of $\xi_3$ values for {\it core-less slow rotators} is centred on a value indicating a stronger anti-correlation between $V/\sigma$ and $h_3$. However, the distribution is wide and it also has multiple peaks that result from large variations between individual galaxies. The small overlap between the two histograms indicates a clear difference of the orientations in the $h_3 - V/\sigma$ plane, and a stronger anti-correlation between $h_3$ and $V/\sigma$ for {\it core-less slow rotators}.

The anti-correlation between $V/\sigma$ and $h_3$ is one of the crucial differences between galaxies with and without discs \citep{1994MNRAS.269..785B, 2011MNRAS.414.2923K,2013MNRAS.432.1768K,2017ApJ...835..104V}. These anti-correlations are typically found in remnants of gas-rich mergers \citep{2000MNRAS.316..315B, 2005MNRAS.360.1185J, 2006MNRAS.372L..78G, 2006MNRAS.372..839N,2007ApJ...658..710N,2007MNRAS.376..997J, 2009ApJ...705..920H,2014MNRAS.445.1065R} or in simulations of objects that did not have a strong feedback mechanism turned on \citep[e.g. no AGN feedback][]{2016MNRAS.463.3948D,2019MNRAS.489.2702F}. We therefore conclude that it is likely that core-less slow rotators originate from dissipative processes and contain embedded discs or disc-like structures.

\begin{figure}
\includegraphics[width=\columnwidth]{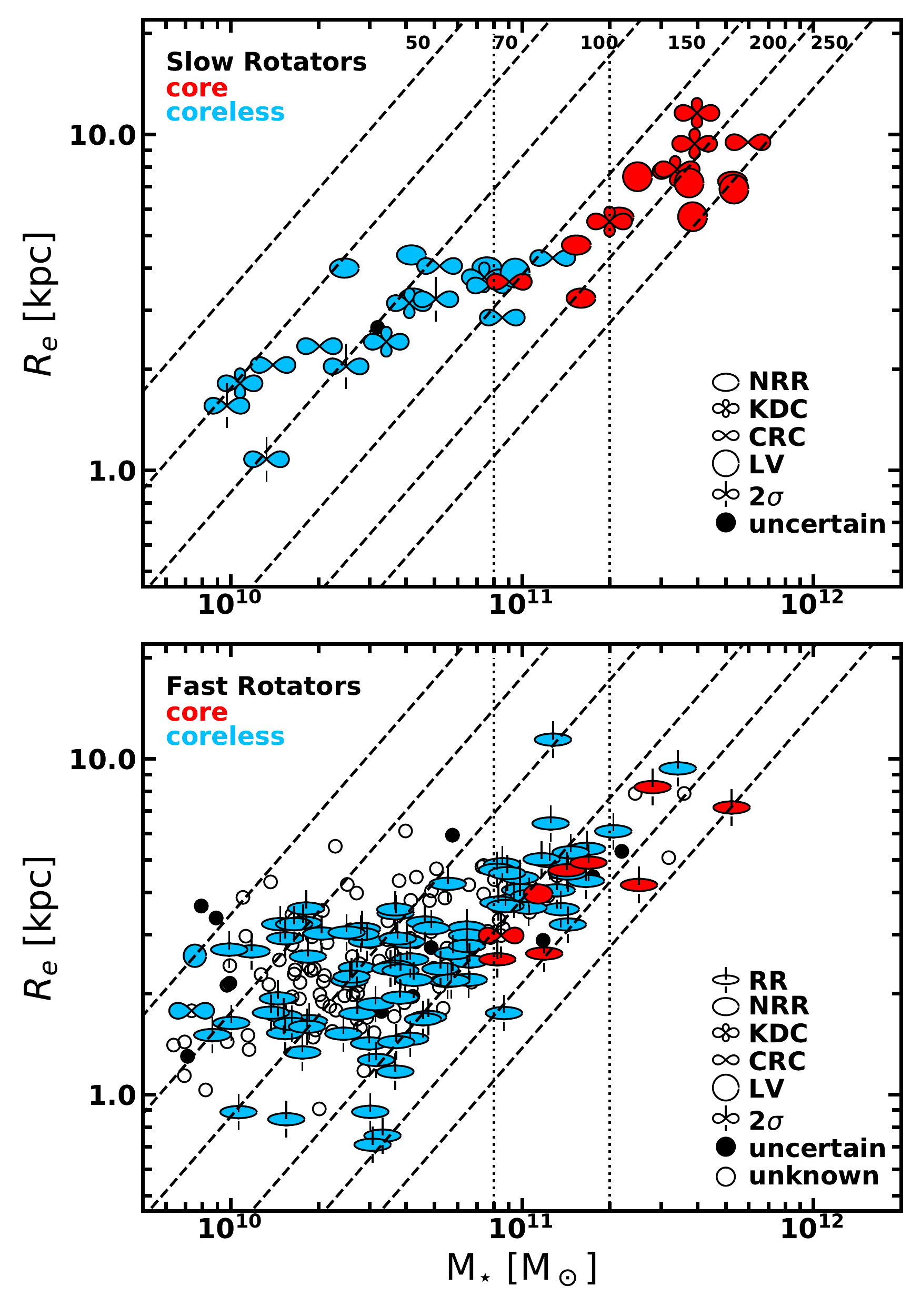}
\caption{Mass -- size relations for ATLAS$^{\rm 3D}$ fast (bottom) and slow (top) rotators. In both panels, small open symbols show galaxies with no HST imaging, and small filled symbols represent galaxies with uncertain light profiles. The colour specifies core (red) and core-less galaxies (light blue). Symbols refer to kinematic classes defined in \citet[][see also Table~\ref{t:stat}]{2011MNRAS.414.2923K}, including KDC, CRC, $2\sigma$ , LV, NRR,  and RR. Vertical lines are drawn at characteristic masses of 0.8 and $2\times 10^{11}$ M$_\odot$. Constant velocity dispersions are shown by dashed lines. Compared to Fig.~6 of \citet{2013MNRAS.433.2812K}, the new HST data show that core galaxies do not appear in galaxies less massive than $0.8\times10^{11}$ M$_\odot$, and that all slow rotators more massive than $10^{11}$ M$_\odot$ have cores. The black symbol in the top panel is NGC\,1222.
}
\label{f:ms}
\end{figure}

\subsection{Cores in the (M, R$_e$) diagram}
\label{ss:ms}

The difference between {\it core slow rotators} and {\it core-less slow rotators} is also well illustrated in the mass - size relation. Fig.~\ref{f:ms} is an update of Fig. 6 from \citet{2013MNRAS.433.2812K} and shows the mass - size relation for slow rotators (top) and fast rotators (bottom) in the ATLAS$^{\rm 3D}$ sample, with masses and sizes from \citet{2013MNRAS.432.1709C}. Again there is a confirmation of the expectation that all low-mass ETGs ($<0.8\times10^{11}$ M$_\odot$) are core-less \citep{1997AJ....114.1771F,2009ApJS..182..216K}, but the size, the velocity dispersion, or the mass are not decisive parameters for finding cores. Mass seems to be a robust discriminator between core and core-less galaxies only for slow rotators; there are no {\it core-less slow rotators} above $\sim10^{11}$ M$_\odot$. It should be noted that ATLAS$^{\rm 3D}$ does not probe galaxies more massive than $8\times10^{11}$ M$_\odot$ , and because beyond this mass fast rotators become very rare \citep[e.g.][]{2017MNRAS.464..356V}, it could be that beyond some high value, the galaxy mass remains the only parameter separating core from core-less galaxies, regardless of their stellar angular momentum content. To settle this issue, more observations of most massive galaxies are required because there are BCGs that have core-less profiles \citep{2003AJ....126.2717L}. Their absolute magnitude is typically less bright than -23 mag in V band, which limits their mass to about $10^{12}$ M$_\odot$. Nevertheless, it would be interesting to see if these galaxies are fast or slow rotators because not all central galaxies, or BCGs in particular, are found to be slow rotators \citep{2011MNRAS.414L..80B, 2013ApJ...778..171J,2017AJ....153...89O}.

Fig.~\ref{f:ms} also shows the kinematic type of galaxies. As shown before, complex kinematic features, which include KDC, CRC, LV, NRR, and $2\sigma$, are found in slow rotators \citep{2011MNRAS.414.2923K,2011MNRAS.414..888E}. Among slow-rotators there is a weak trend that KDC are found in more massive galaxies than CRC and $2\sigma$ features, but exceptions exist. Much more robust is the fact that {\it core-less slow rotators} overlap with (core-less) fast rotators in the mass, size, and velocity dispersion. Conversely, {\it core slow rotators} occupy a special place in the mass - size diagram, being both the most massive and the largest galaxies and having the highest velocity dispersions. They extend beyond the location of fast rotators and spiral galaxies \citep[e.g.][]{2011MNRAS.413..813C,2013MNRAS.432.1862C} and form a progressively thin distribution clustering close to the zone of exclusion \citep{2018MNRAS.477.5327K}.

The most conspicuous difference between {\it core-less slow rotators} and {\it core slow rotators} is their masses. The high-mass {\it core slow rotators} are found in dense regions, such as clusters and groups of galaxies, and they often are the central galaxies in such environments \citep[see review by][]{2016ARA&A..54..597C}. Low-mass {\it core-less slow rotators} are found in various environments from clusters to the field \citep{2011MNRAS.416.1680C}. Our sample is too small to distinguish between mass or environmental effects as the driver for the kinematic differences \citep[e.g.][]{2017ApJ...844...59B, 2017ApJ...851L..33G,2019arXiv191005139G}. Nevertheless, as the galaxy mass increases, core-less galaxies give way to core galaxies. According to the currently favoured core formation scenario (see Section~\ref{s:disc}), cold gas needs to be absent for making cores. The transition between {\it core-less slow rotators} and {\it core slow rotators} visible in the top panel of Fig.~\ref{f:ms} therefore may be the result of a decreasing role for the nuclear cold gas in the mass assembly. 

\begin{figure}
\includegraphics[width=\columnwidth]{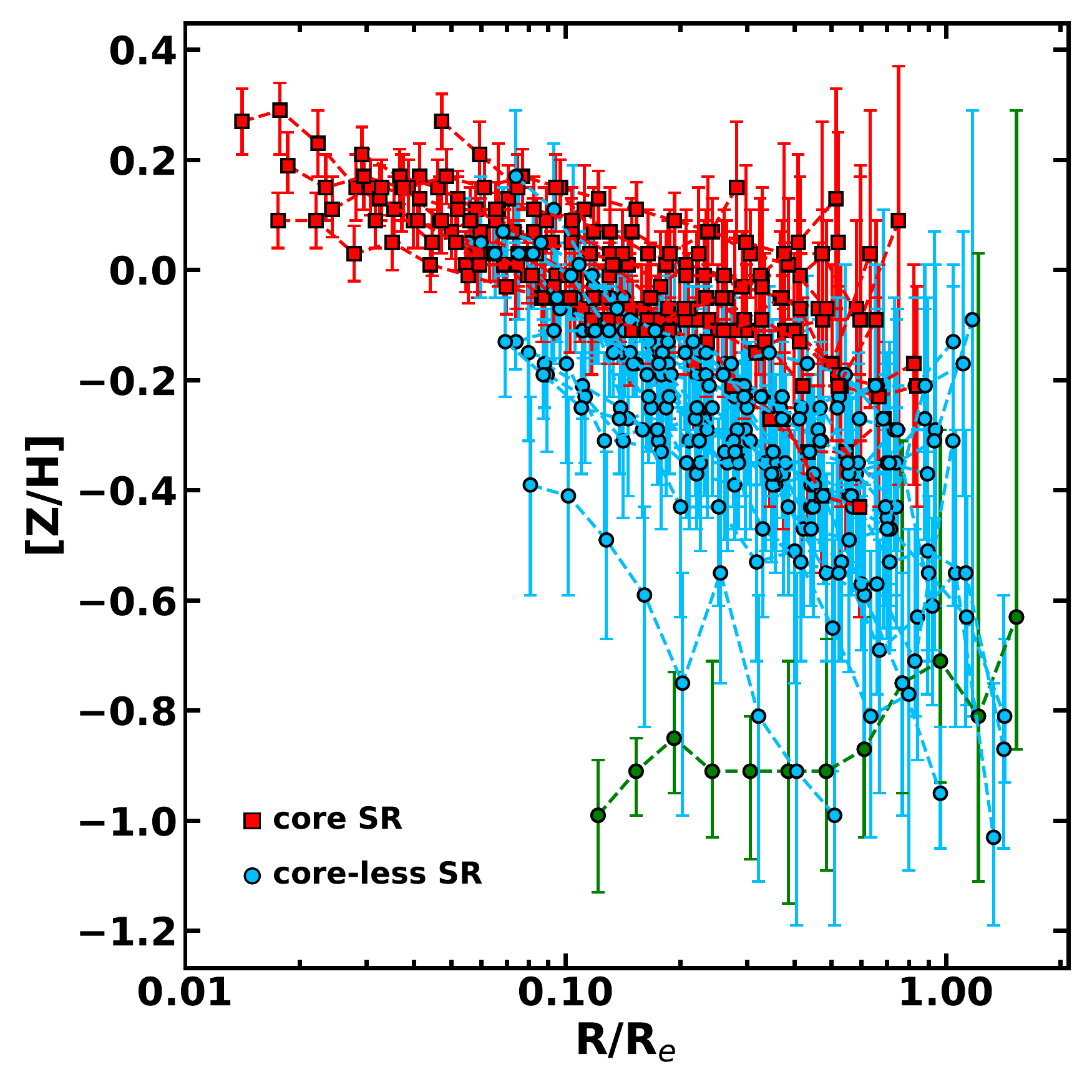}
\caption{Radial variation of metallicity profiles averaged along isophotes for all ATLAS$^{\rm 3D}$ slow rotators and normalised by the half-light radii. Blue circles show metallicity profiles of slow rotators with core-less surface brightness profiles, and red squares show profiles of slow rotators with cores. Green circles belong to NGC\,1222, which was not classified because of the dust. As seen in Fig.~\ref{f:ms}, core galaxies are more massive slow rotators, and the offset between the metallicity profiles of core and core-less galaxies is explained as the mass trend. Core galaxies are also larger than core-less galaxies, which explains the offsets along the horizontal axis between the two types of galaxies.  The profiles differ also in their slopes, however. }
\label{f:ZHprof}
\end{figure}

Cores also exist in fast rotators. They are rare (8\%\ of fast rotators with HST imaging, compared e.g. to 57\%\ of core-less slow rotators), and their hosts are kinematically different from {\it core slow rotators} \citep{2013MNRAS.433.2812K}. Compared to the rest of fast rotators, {\it core fast rotators} are typically more massive, have a higher effective velocity dispersion, and a lower stellar angular momentum (e.g. Figs.~\ref{f:lamR} and~\ref{f:ms}). They occupy the same regions in the mass - size space as slow rotators with cores, except that they do not extend as high in mass and size. These galaxies are further discussed in Section~\ref{ss:frcore}.

\subsection{Metallicity gradients: evidence for different assembly processes of core and core-less galaxies}
\label{ss:metG}

Mass assembly can also be traced by stellar population properties. In this respect, gradients of stellar populations, in particular, their metallicity gradients, are heralded as discriminators between various formation models \citep{1980MNRAS.191P...1W}. We investigate this prediction by first showing the radial metallicity profiles of all ATLAS$^{\rm 3D}$ slow rotators in Fig.~\ref{f:ZHprof}. We consider metallicity profiles in galaxies with young stellar populations as not reliable (for our sample, this means anything younger than 5 Gyr). The slow rotator with the youngest stellar populations is NGC\,1222, which we exclude from this analysis and highlight in the figures. We highlight in passing also NGC\,4191 and NGC\,4690 because their stellar ages fall below the 5 Gyr limit at some radial bins. The luminosity-weighted stellar ages within the half-light radius for these galaxies are 6 and 4 Gyr, respectively \citep{2015MNRAS.448.3484M}. For this reason, and because their metallicity radial profiles do not look different from the rest of the slow rotators, we kept them for further analysis. 

\begin{figure*}
\includegraphics[width=\textwidth]{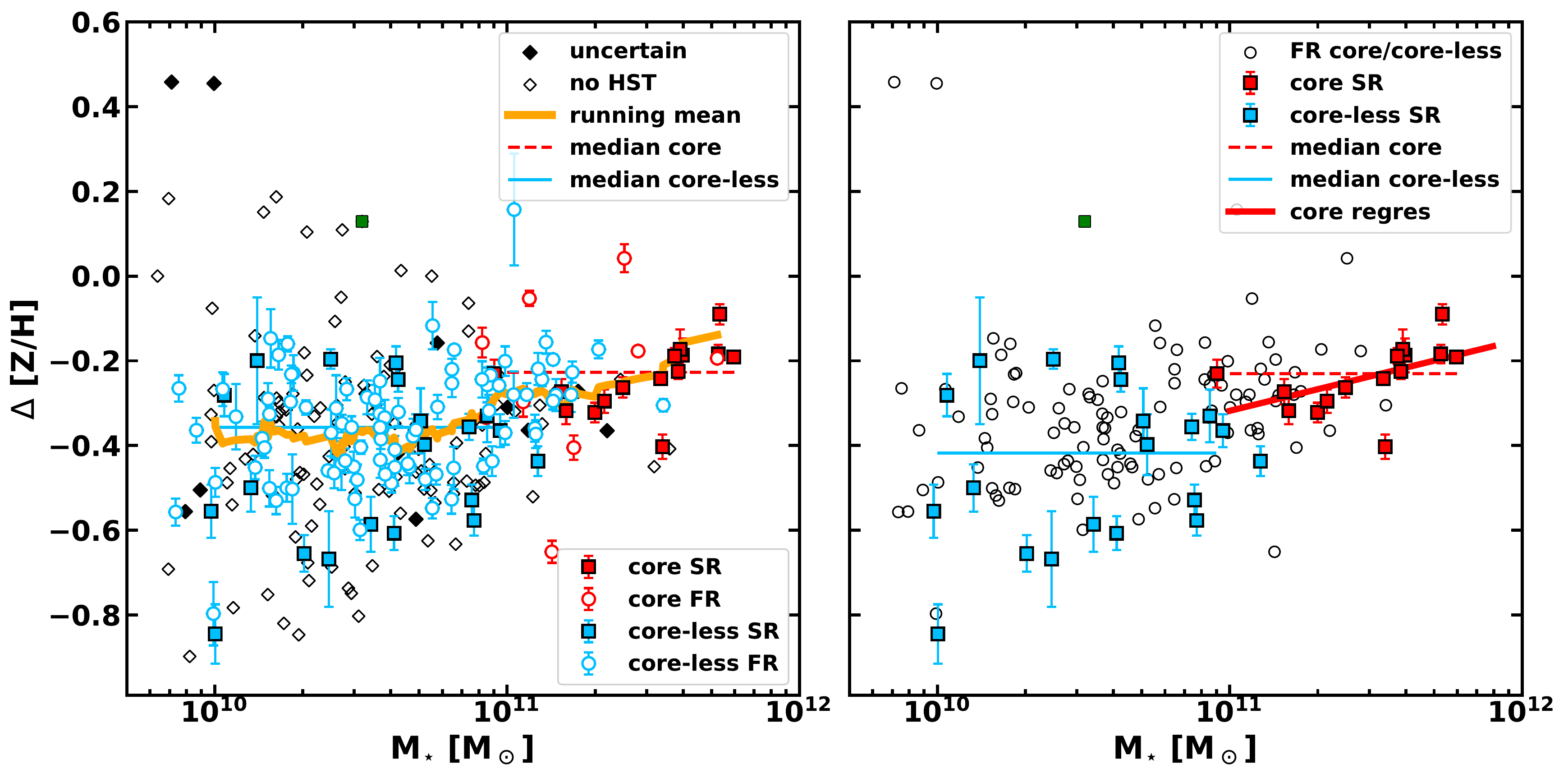}
\caption{Metallicity gradients vs. the stellar mass of the ATLAS$^{\rm 3D}$ galaxies. The {\it left-hand} panel shows all ATLAS$^{\rm3D}$ galaxies. Galaxies with cores are shown with red symbols, and core-less galaxies are represented in blue. The shape of the symbols indicates whether the galaxy is a slow (square) or a fast (circle) rotator. Diamonds show galaxies without HST imaging (empty) or with uncertain profiles (filled). NGC\,1222 is shown as a green square. Red (dashed) and blue (solid) straight lines indicate the median values for core and core-less galaxies, respectively. The lengths of these lines are arbitrary and separated at mass of $10^{11}$ M$_\odot$. The thick orange line is the running mean of $\Delta [Z/H]$. The {\it right-hand} panel focuses on the ATLAS$^{\rm 3D}$ galaxies with HST imaging only, where coloured symbols are slow rotators, either with cores (red) or core-less (blue). Open symbols are fast rotators. Dashed lines indicate the median values for core and core-less galaxies, respectively. The thick red line is the best-fit relation for {\it core slow rotators} (red squares). Galaxies with cores tend to have shallow gradients, which flatten with the increase in mass, while core-less galaxies show a large spread. }
\label{f:ZHgrad}
\end{figure*}

The metallicity profiles in Fig.~\ref{f:ZHprof} are normalised to the effective radius of galaxies \citep[from][]{2013MNRAS.432.1709C} to remove the size dependence. The mass trend, in which more massive {\it core slow rotators} have higher mean absolute values of the metallicity profiles, is visible in the figure. The size difference between {\it core slow rotators} and {\it core-less slow rotators} is also highlighted by the fact that in {\it core slow rotators}, which are typically larger galaxies, we probe smaller relative radii. Furthermore, there seems to be a global difference in the slope of the metallicity profiles between {\it core slow rotators} and {\it core-less slow rotators}, and we quantify this by considering metallicity gradients. 

Fig.~\ref{f:ZHgrad} shows metallicity gradients for all ATLAS$^{\rm 3D}$ galaxies (left-hand panel) and ATLAS$^{\rm 3D}$ galaxies with HST imaging, highlighting the slow rotators (right-hand panel). In both panels we highlight core and core-less galaxies and their correlation with the fast or slow rotators. A similar figure showing a mass dependence on the metallicity gradient was also presented in \citet{2015IAUS..311...53K}. The trend in the figure is consistent with trends in the literature when the specific sample selections are taken into account \citep{2009ApJ...691L.138S, 2010MNRAS.408..272S,2010MNRAS.401..852R,2011MNRAS.417.1643K,2012MNRAS.426.2300L,2018MNRAS.476.1765L}. Most ATLAS$^{\rm 3D}$ galaxies have negative gradients, implying higher values in the centre, while many of the galaxies with positive gradients have younger stellar populations. Recent IFU surveys of late- and early-type galaxies showed that metallicity gradients typically steepen as the mass increases \citep{2015A&A...581A.103G,2017MNRAS.466.4731G,2018MNRAS.476.1765L}. Figure~\ref{f:ZHgrad} shows, however, that for the most massive galaxies, there is a reverse trend where the gradients become shallower. An indication of this trend is also visible in Fig.~13 of \citet{2015A&A...581A.103G}, where for masses higher than $10^{11}$ M$_\odot$ there is a turnover and metallicity gradients become less steep. More recently, \citet{2018MNRAS.476.1765L} analysed more than 2000 spirals and ETGs and showed that galaxies above $2\times10^{11}$ M$_\odot$ have flatter metallicity gradients than the rest of the galaxies. In this respect, our smaller sample is consistent, but additionally provides the information on the shape of the surface brightness profiles.

The shallower gradients for more massive galaxies can be demonstrated by the running mean plotted in Fig.~\ref{f:ZHgrad}. The mean value of the gradient $\overline{\Delta [Z/H]}$ does not change for masses $<10^{11}$ M$_\odot$, beyond which there is a gradual increase for about 0.1 dex, with a tendency for further increase. When we divide the sub-sample with HST imaging into galaxies that are less and more massive than $10^{11}$ M$_\odot$, measure the median gradient and its standard deviation, we obtain the following values: the high-mass sub-sample has a median $\overline{\Delta [Z/H]} = -0.27 \pm 0.13$, while the low-mass sub-sample has $\overline{\Delta [Z/H]} = -0.36 \pm 0.19$.  This is in line with predictions from numerical simulations that more massive galaxies should have flatter profiles. A very similar results is achieved when we calculate the median and the standard deviation of the gradients for core ($\overline{\Delta [Z/H]} = -0.23 \pm 0.13$) and core-less galaxies ($\overline{\Delta [Z/H]} = -0.36 \pm 0.15$), strengthening an assembly connection between the mass and the nuclear light structures. 

\begin{table}
   \caption{Median values and scatter of the metallicity gradients for the galaxies in the ATLAS$^{\rm 3D}$ sample.}
   \label{t:grad}
$$
  \begin{array}{l| c c r}
    \hline \hline
    \noalign{\smallskip}

$ type $ & \overline{\Delta[Z/H]} & \delta (\Delta [Z/H] )   & $No.$ \\
(1) & (2) & (3)   & (4) \\
 \noalign{\smallskip} \hline  \noalign{\smallskip}
 $all ATLAS$^{\rm3D}$ $ &   &   & \\
 \hline 
M>10^{11} M_\odot  & -0.28  & 0.13   & 211\\
M<10^{11} M_\odot  &   -0.37& 0.21  & 49 \\
 \noalign{\smallskip} \hline \noalign{\smallskip}
$ATLAS$^{\rm 3D}$ with HST$ &&&\\
\hline 
M>10^{11} M_\odot &   -0.27& 0.13  & 41\\
M<10^{11} M_\odot &  -0.36 & 0.19  & 106\\
$core$ &  -0.23 & 0.13  & 24 \\
$core SR$ &  -0.23 & 0.07  & 15\\
$core FR$ &  -0.19 & 0.19  & 9\\
$core-less$ &  -0.36 & 0.15  & 109\\
$core-less SR$ & -0.42  & 0.18  & 20\\
$core-less FR$  & -0.35 & 0.13  & 89\\
       \noalign{\smallskip}
    \hline
  \end{array}
$$ 
{Notes -- Column (1): Sub-sample type. Column (2): Median metallicity gradient values for various sub-samples of the ATLAS$^{\rm 3D}$ galaxies. Column (3): Standard deviations of the metallicity gradients. Column (4): Number of galaxies in each sub-sample. }
\end{table}

Dividing galaxies according to their nuclear profiles and angular momentum adds important information (Table ~\ref{t:grad}). As expected, both {\it core slow rotators} and {\it core fast rotators} are characterised by flatter metallicity gradients (close to $-0.2$), while {\it core-less slow rotators} and {\it core-less fast rotators} have steeper gradients ($>-0.35$). Furthermore, the standard deviations of the metallicity gradients of {\it core fast rotators}, {\it core-less fast rotators,} and {\it core-less slow rotators} are similar among each other,  $\sim0.13-0.19,$ and to the values reported above. Significantly, the standard deviation of $\Delta [Z/H]$ for {\it core slow rotators} is only 0.07, at least a factor of 2 smaller. This tightening of the spread in metallicity gradients among {\it core slow rotators} is not visible when a selection in mass alone is considered, and we discuss this further. 

The right-hand panel of Fig.~\ref{f:ZHgrad} focuses on slow rotators. Here we again divided the {\it core slow rotators} and {\it core-less slow rotators} and plot in the background all other galaxies with HST imaging (fast rotators). This plot visualises the strong difference between {\it core slow rotators} and {\it core-less slow rotators} in terms of the dispersion of their $\Delta [Z/H]$ values. {\it Core-less slow rotators} can essentially have any value of $\Delta [Z/H]$ typical for the underlying fast rotators. {\it Core slow rotators} are located in a much more limited space of $\Delta$[Z/H]. To quantify these differences in metallicity gradients, we performed a Kolmogorov-Smirnov test. The hypothesis that {\it core slow rotators} and {\it core-less slow rotators} are drawn from the same continuous distribution can be rejected because its probability is 0.0003. Similarly, the probability that metallicity gradients for {\it core} and {\it core-less} ATLAS$^{\rm 3D}$ galaxies are drawn from the same distribution is only 0.001.  A Kolmogorov-Smirnov test, however, cannot reject the hypothesis that {\it core-less slow rotators} and fast rotators in general are drawn from the same distribution (the rejection probability is 0.11).

Next to the conclusion that {\it core slow rotators} and {\it core-less slow rotators} (and galaxies in general) have different metallicity gradients, we see in the right-hand panel of Fig.~\ref{f:ZHgrad} another interesting feature: there seems to be a correlation of the metallicity gradient of {\it core slow rotators} with their mass. We fitted a linear regression and found a relation $\Delta [Z/H] = 0.16 \log(M_\star) - 2.18$, with correlation coefficient of 0.51. The correlation is not limited to the 15 {\it core slow rotators} in the ATLAS$^{\rm 3D}$ sample because adding {\it core fast rotators} (9 galaxies) does not change its shape by much ($\Delta [Z/H] = 0.14 \log(M_\star) - 1.87$). The correlation coefficient drops to 0.28, however, as expected because the dispersion of $\Delta [Z/H] $ of {\it core fast rotators} is significantly larger. 

Fig.~\ref{f:ZHgrad} and Table~\ref{t:grad} provide evidence for different formation scenarios between {\it core slow rotators} and other ETGs. The trend of flattening metallicity gradients with increasing mass is the dominant effect, seen both globally (in our full sample) and locally (among {\it core slow rotators}). Higher mass galaxies also have a lower dispersion of metallicity gradients, but when the selection is made for only {\it core slow rotators}, the spread in $\Delta [Z/H]$ is significantly minimised. The consequence of this small dispersion is that these galaxies must follow very similar formation scenarios, whereas the flat gradients suggest a lack of star formation in the assembly events. As a contrast, the steepness of the metallicity gradients of {\it core-less galaxies} are indicative of an inside-out formation, while the larger spread of the gradient values is indicative of more varied star-formation histories. {\it Core fast rotators} are somewhere in between the two extremes, having similar mean metallicity gradients like {\it core slow rotators}, but the dispersion of the gradients is more similar to core-less galaxies. The latter suggests that there are multiple ways of forming this class of galaxies. 

We add two caveats pertinent to our sample. Firstly, there are only nine {\it core slow rotators} in the ATLAS$^{\rm 3D}$ sample, and the results are susceptible to low number statistics. Secondly, above a mass of $2-3 \times 10^{11}$ M$_\odot$ there are no more fast rotators, and only {\it core slow rotators} remain. Selecting this mass cut would reproduce the same result as by selecting on {\it core slow rotators}, but for a small number of galaxies. Although we cannot fully separate the effects of galaxy mass, it plays a pivotal role. In Section~\ref{s:disc} we discuss the influence of the mass on formation of flat metallicity gradients and cores in more detail.

\subsection{Stellar age, age gradients, and star formation histories}
\label{ss:age}

\begin{figure}
\includegraphics[width=\columnwidth]{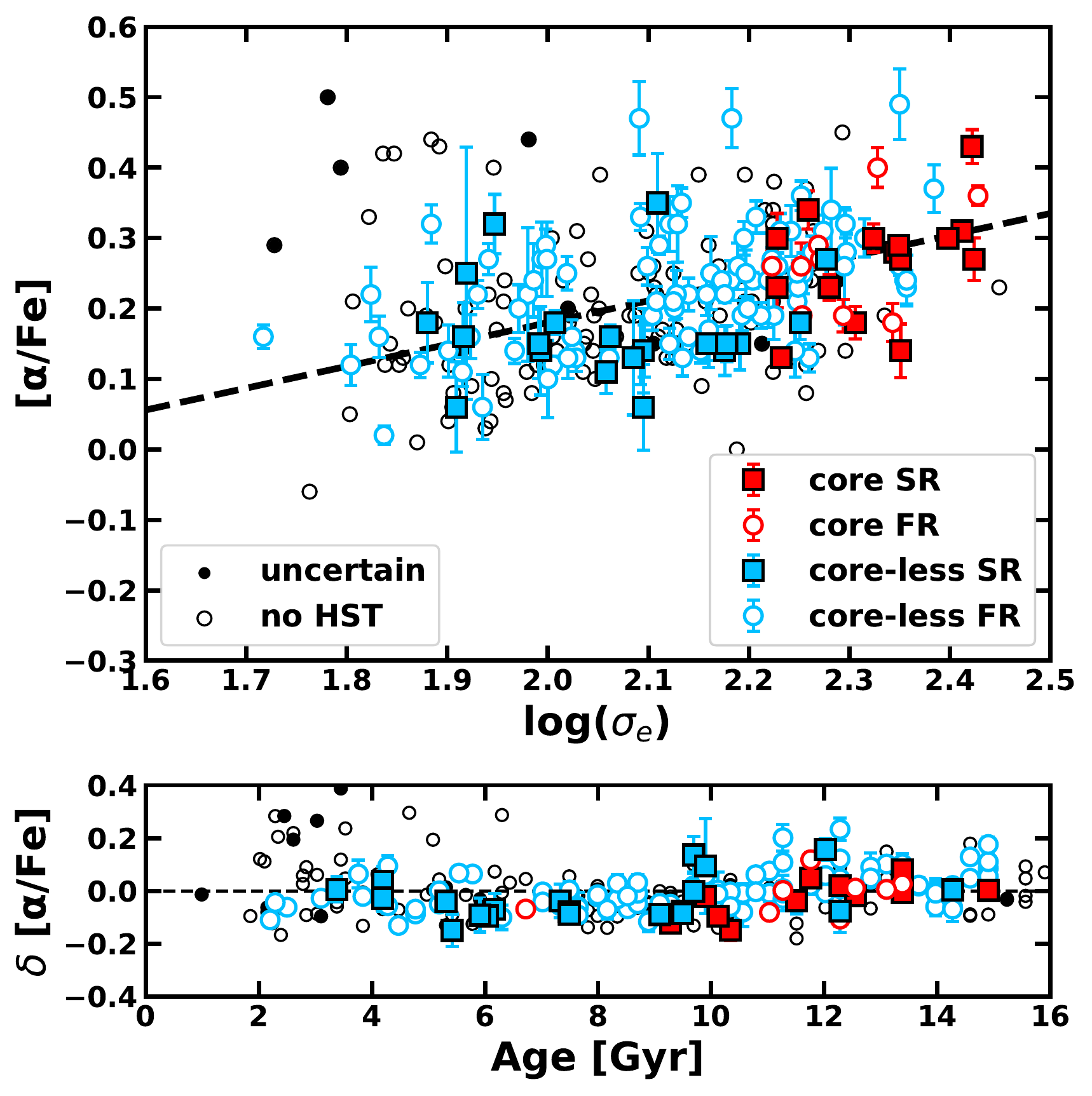}
\caption{ {\it Top:} Luminosity-weighted SSP $\alpha$-element abundance vs. effective velocity dispersion of the ATLAS$^{\rm 3D}$ galaxies. The dashed black line is the best-fit relation from \citet[][see their Table~5, relation {\it i}]{2015MNRAS.448.3484M}. {\it Bottom:}  $\alpha$-element abundance corrected for the $\sigma_e$ dependance (by subtracting the values as given by the dashed line in the top panel) vs. the SSP age. Both SSP ages and [$\alpha$/Fe] values are taken from \citet{2015MNRAS.448.3484M}, which also show that very old SSP ages are consistent with the fiducial age of the Universe \citep{2016A&A...594A..13P} when measurement errors and SSP mode uncertainties are taken into account. In both panels the symbols are the same and as described in the legend.  {\it Core-less slow rotators} have a range of SSP ages, while {\it core slow rotators} as well as all but one {\it core fast rotators} are older than 10 Gyr. }
\label{f:age}
\end{figure}

We conclude our presentation of results by addressing the relation between the kinematics, surface brightness profiles, and age properties of stellar populations. For this purpose, we use the results of \citet{2015MNRAS.448.3484M}, specifically their SSP-based ages and $\alpha$-element abundances within one effective radius (their Table~3). In Fig.~\ref{f:age} we focus on [$\alpha$/Fe] abundances as a measure of the star formation timescales. Similar to \citet{2009ApJS..182..216K}, we plot it against the effective velocity dispersion \citep[as in Fig.~11 of][]{2015MNRAS.448.3484M}, and after removing the best-fit relation, against the SSP ages. As before, we highlight the core and core-less galaxies as well as fast and slow rotators. 

The $\sigma_e$, or the mass, dependance of this relation is well known \citep[e.g.][]{2005ApJ...621..673T}, while \citet{2009ApJS..182..216K} also noted that core-less galaxies have lower $\alpha$-elements abundances than core galaxies. This implies that the stars in core galaxies formed over a shorter timescale than stars in core-less galaxies. The ATLAS$^{\rm 3D}$ sample shows a relatively large scatter in [$\alpha$/Fe] - $\sigma_e$ relation, where some of the largest [$\alpha$/Fe], and therefore shortest star formation histories, are found among {\it core-less fast rotators}. When we focus only on slow rotators, however, they alone are consistent with the global relation \citep[we take the best fit from][as given in their Table~5, where the slope and intercept are $0.31\pm 0.03$ and $-0.44\pm 0.05$, respectively]{2015MNRAS.448.3484M}. {\it Core-less slow rotators} typically have lower [$\alpha$/Fe]. This seems to be a purely $\sigma_e$ driven effect because when the $\sigma_e$ dependance is removed from the relation (bottom panel), there are no significant differences in relative abundances. In Section~\ref{ss:form} we return to this point, but as gas-free merging cannot increase the value of [$\alpha$/Fe], {\it core-less slow rotators} cannot be progenitors of {\it core slow rotators}, unlike some {\it core-less fast rotators}. 

The bottom panel of Fig.~\ref{f:age} shows the distribution of light-weighted SSP ages. As already presented in \citet{2015MNRAS.448.3484M}, galaxies with complex kinematics (i.e. slow rotators) can have a range of ages. When {\it core slow rotators} and {\it core-less slow rotators} are separated, it becomes clear that slow rotators with the youngest stellar ages are found among core-less galaxies. Stars in {\it core slow rotators} are on average older than 10 Gyr, while {\it core-less slow rotators} can be as old as any ATLAS$^{\rm 3D}$ galaxy. Notably, cores are found in galaxies with old stellar populations (with the exception of NGC\,4382), regardless of whether they are fast or slow rotators. As the core formation occurred at or after the starburst (see Section~\ref{ss:cores}), the formation redshift of cores is lower than 2-3.

\begin{figure}
\includegraphics[width=\columnwidth]{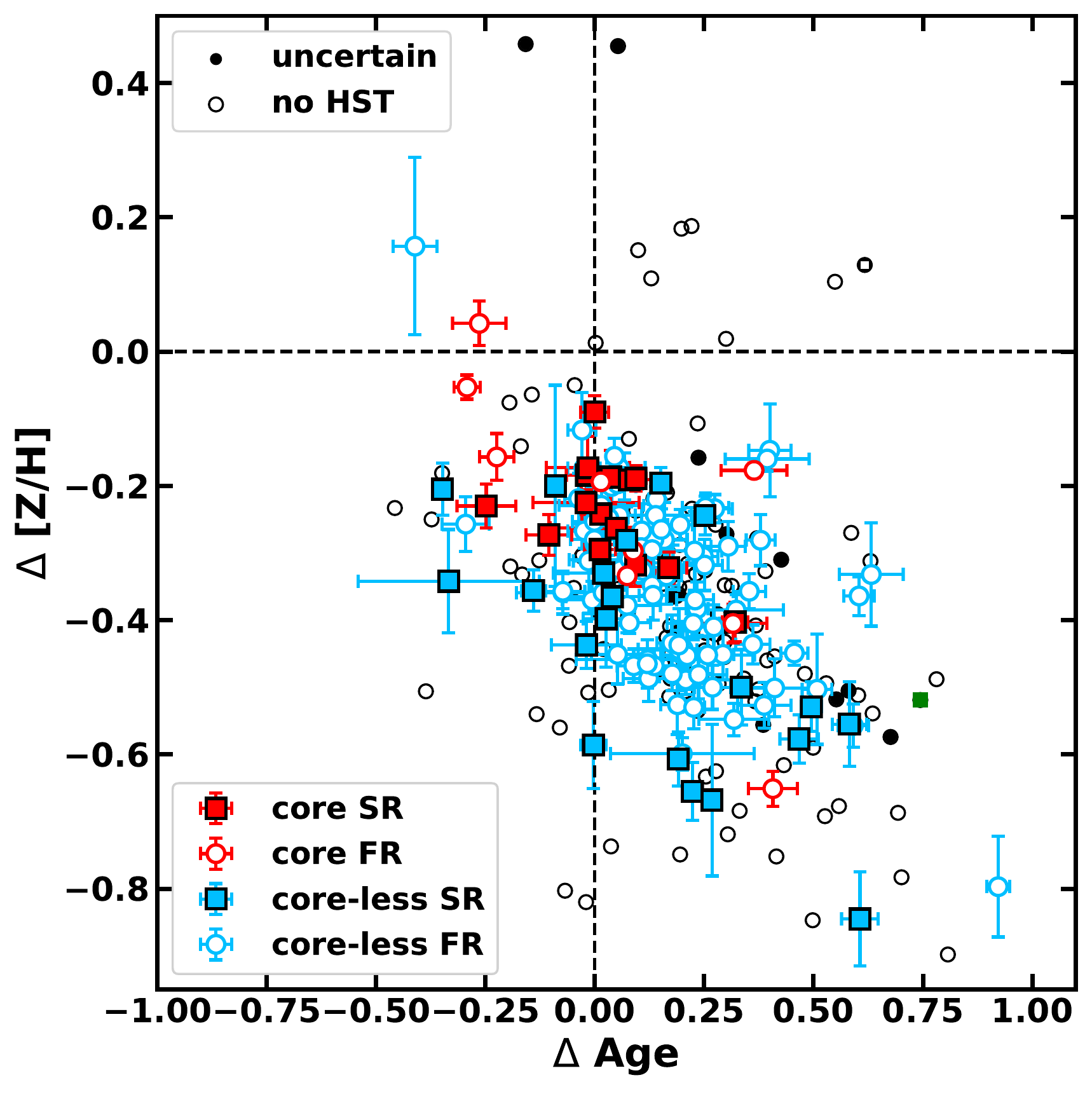}
\caption{ Age gradients vs. metallicity gradient. Galaxies with cores are shown with red symbols, and core-less galaxies are represented in blue. The shape of the symbols indicates whether the galaxy is a slow (square) or a fast (circle) rotator. All galaxies follow an anti-correlation trend between the metallicity and age gradients. {\it Core slow rotators} are, however, found to have close to zero age gradients, while {\it core-less slow rotators}  can have extreme positive and negative age gradients. Similar behaviour is observed for {\it core fast rotator} galaxies. }
\label{f:ageGrad}
\end{figure}

In Fig.~\ref{f:ageGrad} we present the anti-correlation between the age and metallicity gradients. This is an expanded version of the plot presented in \citet{2015IAUS..311...53K}. This time we add the information on the kinematics and the shape of the surface brightness profiles. Next to the strong anti-correlation between the age and metallicity gradients, consistent with other studies \citep[e.g.][]{2010MNRAS.401..852R,2011MNRAS.417.1643K}, this figure reveals a remarkable location of the core and core-less galaxies. Before we remark on them, we note that positive age gradients are expected to be produced by nuclear starbursts, which also enrich the medium and are responsible for the steepening of the negative metallicity gradients \citep{1994ApJ...437L..47M,2004MNRAS.347..740K,2009ApJS..181..135H}. The overall anti-correlation in Fig.~\ref{f:ageGrad} is therefore as expected and consistent with other stellar population relations \citep{2015IAUS..311...53K,2015MNRAS.448.3484M}. 

Fig.~\ref{f:ageGrad} shows that {\it core slow rotators} have shallow age gradients (close to zero), while {\it core-less slow rotators} can essentially have any age gradient. They are found among galaxies with the most negative as well as most positive age gradients. This supports the conjecture where {\it core-less slow rotators} can be produced through a large variety of formation scenarios, while the formation of {\it core slow rotators} is much more restricted to a specific formation channel. 

The location of {\it core fast rotators} is also noteworthy. These galaxies are distributed similarly to {\it core-less slow rotators} such as that there are some with negative age and nearly positive metallicity gradients, some with flat but negative metallicity and zero age gradients, and some with positive age and negative metallicity gradients. The implication is again that {\it core fast rotators} can be made through processes involving different levels of star formation and subsequent gas-poor merging that tend to flatten gradients. 

%
%

\section{Discussion}
\label{s:disc}

\begin{figure*}
  \centering
\includegraphics[width=0.45\textwidth]{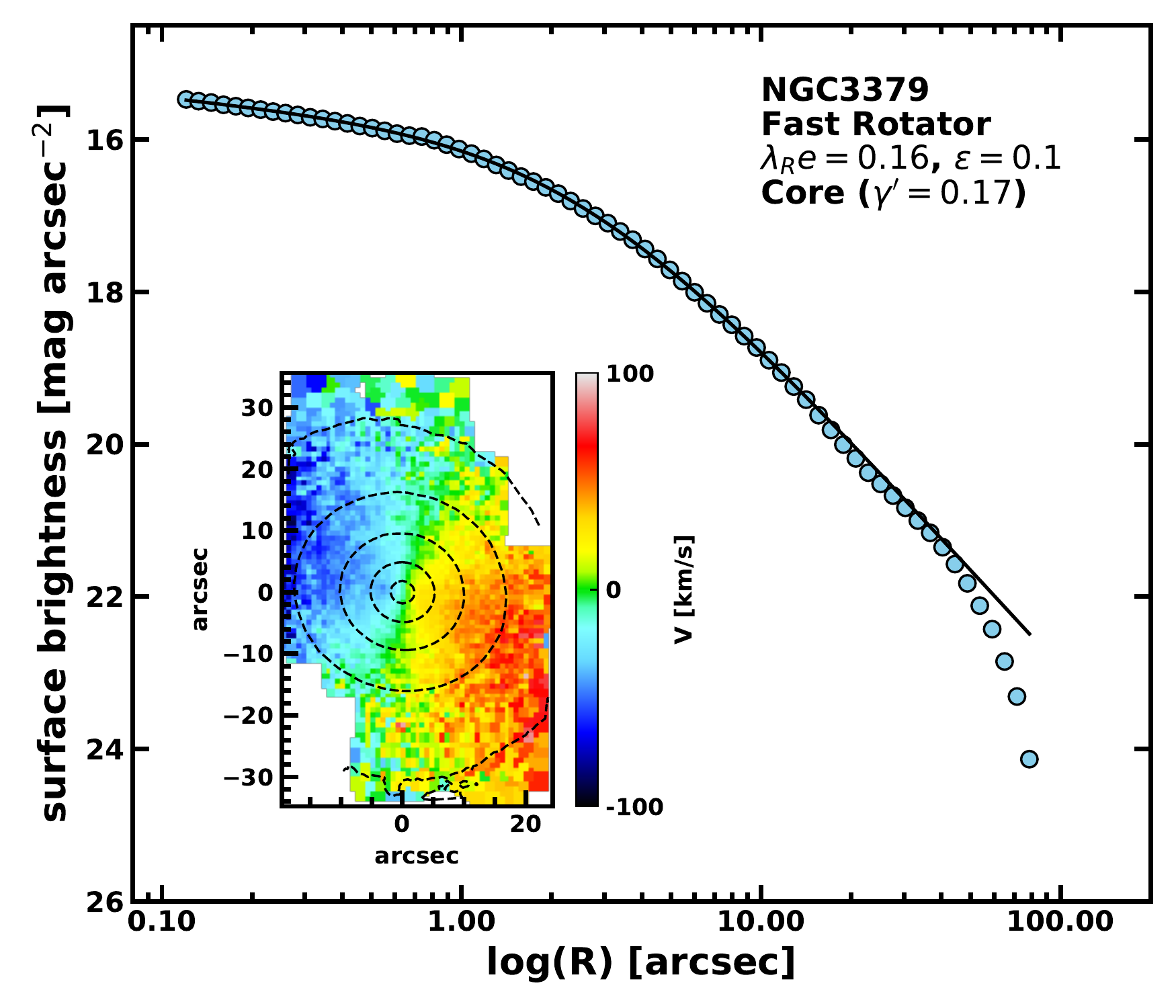}
\includegraphics[width=0.45\textwidth]{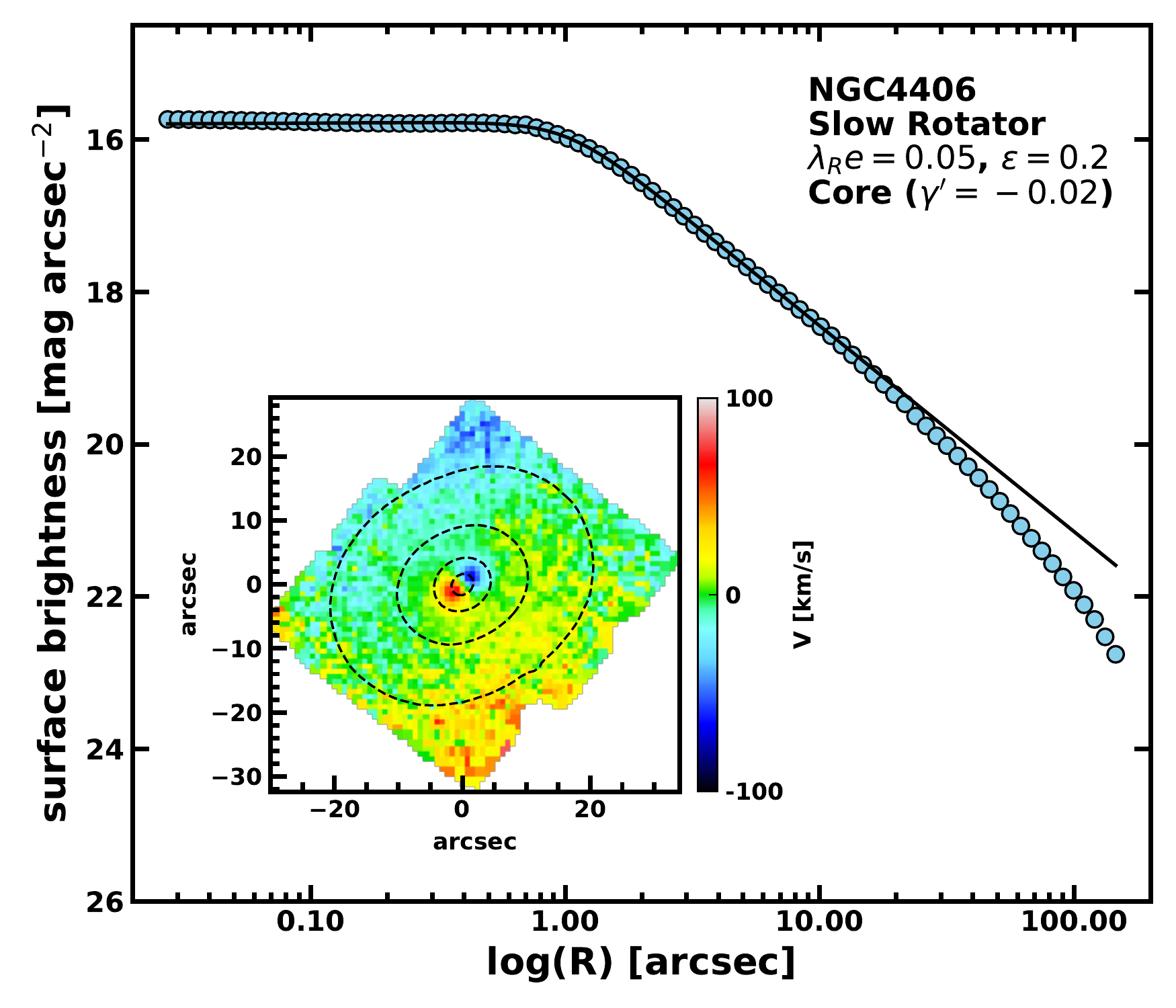}
\includegraphics[width=0.45\textwidth]{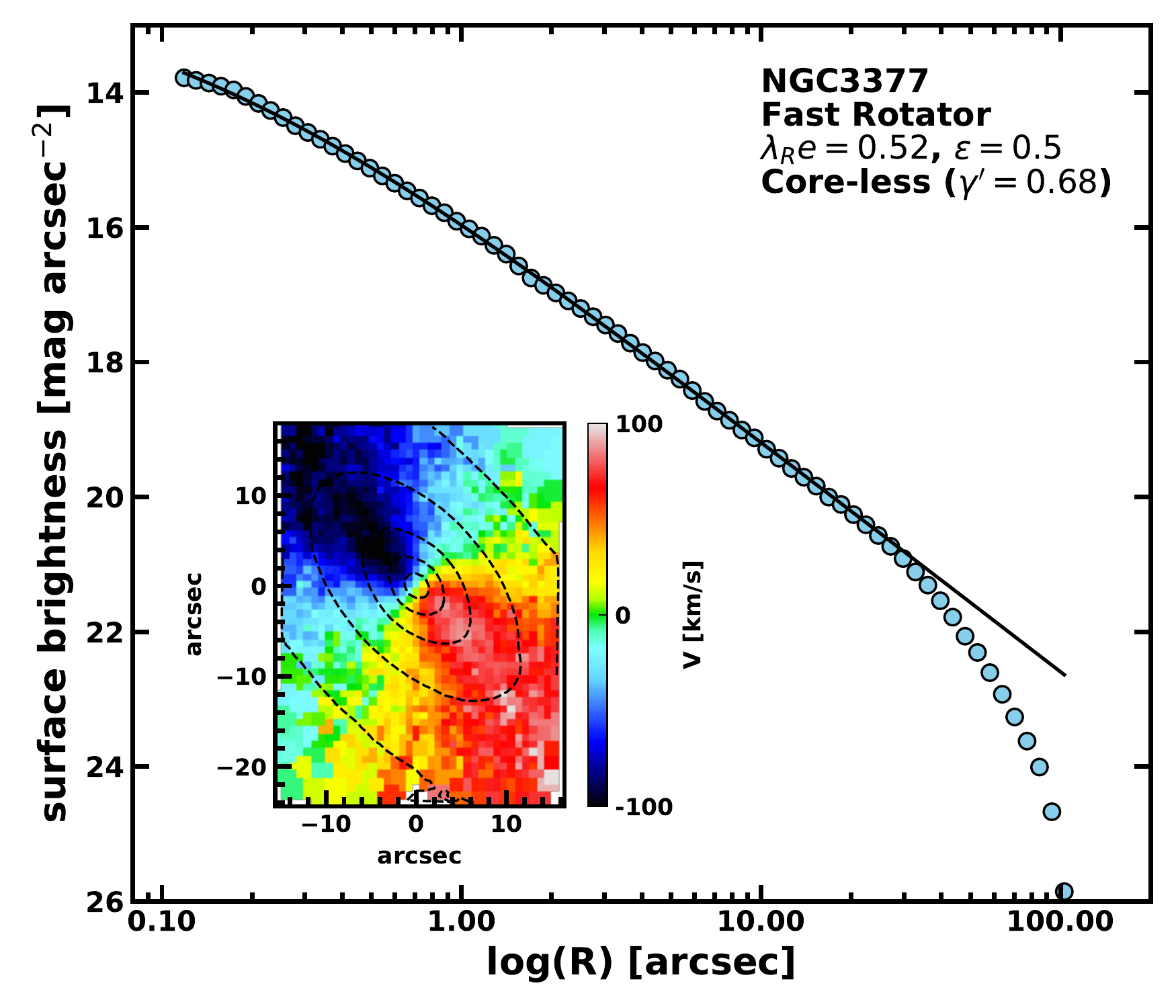}
\includegraphics[width=0.45\textwidth]{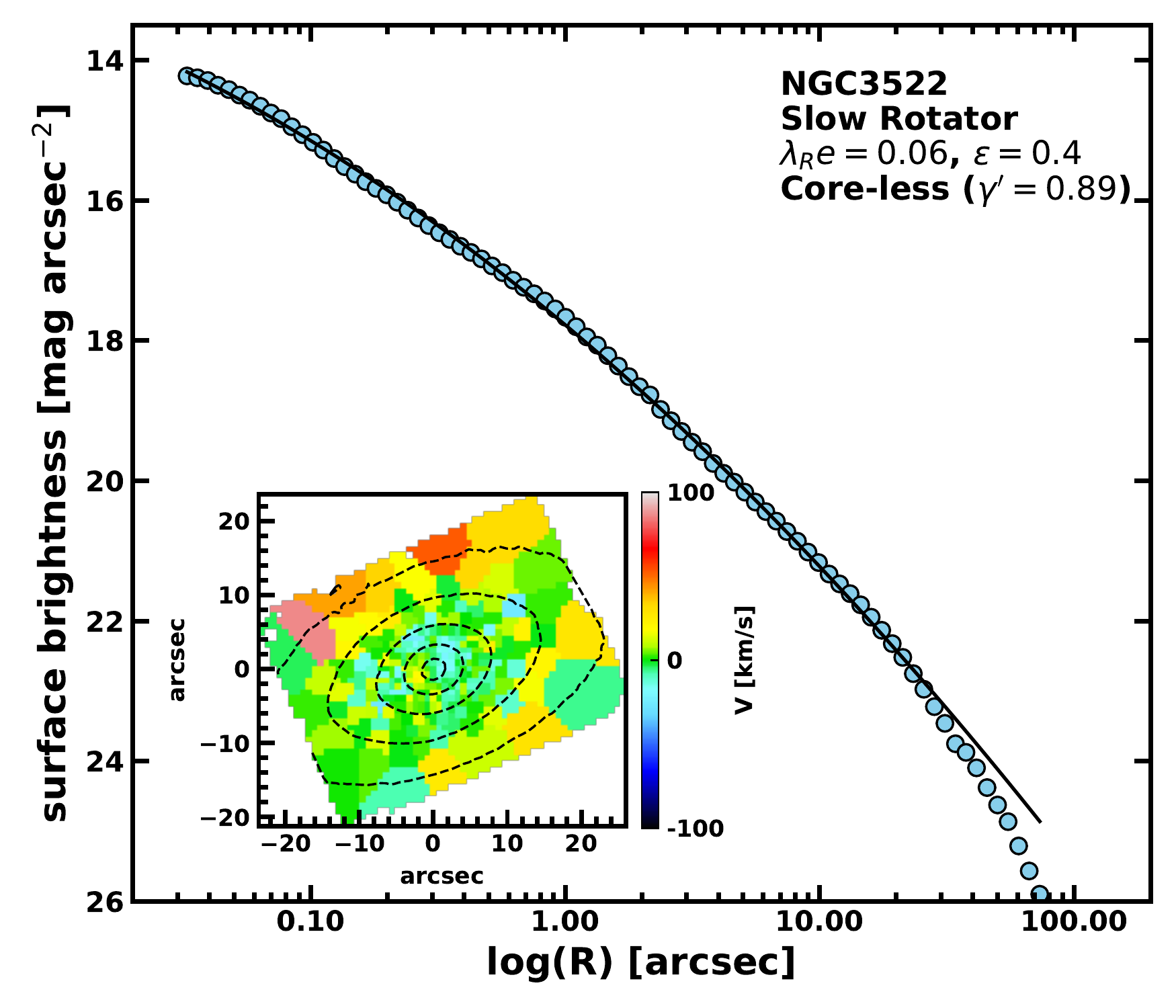}
\caption{Examples of four categories of galaxies with varied stellar kinematics and nuclear surface brightness profiles. In each panel, circular symbols show the nuclear surface brightness obtained in the same way as those in Figs.~\ref{f:sb1} and~\ref{f:sb2}. The solid lines represent the Nuker fits with the gradient $\gamma^\prime$ as given in the legend. Profiles and the Nuker fits for NGC\,3379, NGC\,3377, and NGC\,4406 \citep[data for this galaxy are from][]{2006ApJS..164..334F} are taken from \citet{2013MNRAS.433.2812K}, while data for NGC\,3522 are from the present paper. The insets show respective velocity maps, demonstrating the level of regularity and some specific kinematic structures. The colour bars show the range of velocities, which are kept the same for all galaxies for comparison. Legends also provide details on the projected ellipticity and $\lambda_{Re}$, taken from \citet{2011MNRAS.414..888E}. The two fast rotators (left column) are characterised with regular rotation. The two slow rotators (right column) exhibit a KDC (top) and a CRC (bottom). The CRC in NGC\,3522 has a peak rotation of only about 20 km/s and is located within the central 5\arcsec \citep{2011MNRAS.414.2923K}. }
\label{f:examp}
\end{figure*}

This section starts with reviews of the literature pertaining to the formation of stellar cores and flat metallicity gradient. Subsequently, we apply these ideas and the results presented above to outline possible scenarios for the formation of slow or fast  and core or core-less galaxies. As a visual guide we show in Fig.~\ref{f:examp} examples of surface brightness profiles and velocity maps of galaxy categories that we discuss here: a {\it core slow rotator}, a {\it core fast rotator}, a {\it core-less slow rotator,} and a {\it core-less fast rotator}.

\subsection{Formation of stellar cores}
\label{ss:cores}

There are several possible scenarios for the creation of nuclear stellar cores, of which the most attention has been received by those involving the interaction of two supermassive black holes (SMBH) \citep{1980Natur.287..307B, 1991Natur.354..212E, 1996ApJ...465..527M, 1996NewA....1...35Q, 1997NewA....2..533Q, 2001ApJ...563...34M, 2002MNRAS.331L..51M, 2007ApJ...671...53M, 2012ApJ...744...74G,2012MNRAS.422.1306K, 2018ApJ...864..113R}. The notion that the SMBH binaries are responsible for flat surface brightness profiles observed in bright ellipticals was first proposed by \citet{1997AJ....114.1771F}. Following the work in numerical simulations \citep{1991ApJ...370L..65B, 1994ApJ...437L..47M, 1996ApJ...471..115B}, \citet{1997AJ....114.1771F} showed that core-less galaxies are consistent with being formed through dissipative, gas-rich mergers, but they also showed that pure dissipation-less, gas-poor mergers may not be able to form and maintain cores. Exploring an alternative model for core formation, Faber and collaborators proposed that SMBH binary mergers that remove stars from the nuclei could indeed explain the diversity of surface brightness profiles. Subsequent studies discovered that the SMBH mass correlates with the mass absent from the nuclei \citep{2004ApJ...613L..33G,2006ApJS..164..334F,2008MNRAS.391.1559H,2009ApJ...691L.142K,2013AJ....146..160R,2014MNRAS.444.2700D}, while \citet{2009ApJS..182..216K} summarised a myriad of evidence that core and core-less (extra light in their terminology) galaxies are products of dissipation-less and dissipative mergers, respectively, where SMBH binaries form cores.

The processes of core formation are based on the requirement that for an SMBH binary to become more tightly bound, it has to loose its angular momentum. The angular momentum is transferred to ejected stars that crossed the in-spiraling binary, which forms a depleted nuclear region. This results in the formation of a {\it core} in the light profile \citep{2002MNRAS.335..965Y, 2003ApJ...596..860M, 2004ApJ...602...93M, 2005LRR.....8....8M, 2006RPPh...69.2513M}. Mergers without SMBHs show that the nucleus of the remnant is simply the denser of the progenitor nuclei \citep{1999ApJ...517...92H,2018ApJ...864..113R}. For the core formation in galactic mergers, two prerequisites are therefore necessary: two SMBHs, and an environment lacking cold gas. SMBHs will eject the stars depleting the core, but the paucity of gas will ensure the absence of a nuclear starburst that should refill the depleted region \citep[e.g.][]{1994ApJ...437L..47M, 1996ApJ...471..115B, 2004AJ....128.2098R,2006AJ....131..185R}.

The reduction of the number of stars occurs before the SMBH merge, but it can continue even after the merger event. The anisotropic gravitational radiation can impart a recoil on the remnant SMBH of several hundred km/s, which then leaves the nucleus pulling along stars that are gravitationally bound to it. If the recoil velocity is not higher than the escape velocity, the SMBH will return to the nucleus as a result of dynamical friction, which will create an even larger core \citep{2004ApJ...613L..37B, 2004ApJ...607L...9M, 2008ApJ...678..780G}. 

Overall, the formation of the core is a rapid process \citep[within 50 Myr of the SMBH interaction,][]{2018ApJ...864..113R}, and simulations indicate two straightforward observational predictions. One is the formation of the core, and the other is a tangentially anisotropic orbital distribution, as the core is depleted of orbits that cross the binary SMBH path \citep{1995ApJ...440..554Q, 1997NewA....2..533Q, 2001ApJ...563...34M,2018ApJ...864..113R}. The velocity anisotropy is indeed observable \citep[e.g.][]{2014ApJ...782...39T, 2016Natur.532..340T}, but only within the region dominated by the SMBH gravitational influence. A more readily observable kinematic consequence is a possibility of KDC formation from the interaction between the SMBHs and the surrounding stellar body. As SMBHs pass and exert torques on each other, there is a switch of the angular momentum between the SMBHs (and stars under their direct gravitational influence) and the rest of the galaxy, possibly producing nested KDCs \citep{2019ApJ...872L..17R}. Because the core formation is enhanced if the SMBHs are massive, and because such SMBHs live in massive galaxies \citep[e.g.][]{2016ApJ...831..134V}, as well as because gas-poor mergers are more likely for massive, quiescent elliptical galaxies, we can expect that cores will be present in the most massive ETGs.

Finally, it should be noted that some of the processes invoked for the formation of cores in dark matter profiles \citep[e.g.][]{2010ApJ...725.1707G,2012MNRAS.424..747L,2012MNRAS.420.2859M,2013MNRAS.432.1947M, 2013MNRAS.429.3068T,2015MNRAS.446.1820N,2016MNRAS.461.1745E}, which depend on some sort of feedback mechanism, could also be at work in galactic nuclei. There is little evidence that their mechanisms and predicted properties are compatible with observed galactic nuclei, however.

\subsection{Formation of flat metallicity gradients}
\label{ss:flat}

Monolithic collapse models predict steep metallicity gradients \citep[e.g.][]{2008A&A...484..679P,2010MNRAS.407.1347P}, while hierarchical merger models anticipate shallower gradients \citep[e.g.][]{2004MNRAS.347..740K,2010ApJ...710L.156R}. The metallicity gradients are strongly influenced by the amount of dissipation during the mass assembly, and, therefore an occurrence of a secondary star formation will steepen the gradients \citep[e.g.][]{2013MNRAS.436.3507N}. The mass ratio of the merger is important because equal-mass gas-free mergers are predicted to produce flatter metallicity gradients, while unequal mass mergers (e.g. with mass ratio of 1:5 or more) are shown to produce significant gradients at large radii \citep[e.g.][]{2012MNRAS.425.3119H, 2013MNRAS.429.2924H,2015MNRAS.449..528H}. In such gas-free mergers, the flattening of the gradients also depends on the initial difference of the population gradients between the progenitors \citep{2009A&A...499..427D}. The scatter is also expected to be large depending on the efficiency of star formation and can be interpreted in terms of mergers of progenitors with various gas richness \citep{2009ApJS..181..135H,2009ApJS..181..486H} or in terms of the monolithic collapse model with differing star formation efficiency at a given mass \citep{2010MNRAS.407.1347P}. Furthermore, a feedback mechanism (e.g. winds or  AGN) is often necessary to align the simulation results with observations \citep[e.g.][]{2013MNRAS.428.2885D, 2015MNRAS.449..528H, 2016ApJ...833..158C}

Notwithstanding the details and the absolute level of predicted metallicity gradients by models, a general conclusion is that galaxies undergoing dissipation-less merging will have flatter gradients than galaxies that experience gas-rich mergers or secondary star formation of any origin. This is important because we can link these predictions with formation scenarios for galaxies with cores (gas-poor major mergers) or for those with core-less light profiles (gas-rich mergers followed by a nuclear starburst). The expectation therefore is that galaxies with cores should have flatter metallicity gradients.

\subsection{Formation of slow rotators with and without cores}
\label{ss:form}

The main implication of our results is that slow rotators can be divided into two sub-categories with different formation paths, which are revealed by their nuclear light profiles. The difference seems to be strongly related to the type of the merger at the most significant formation event. In the present and the following subsection, we build on the conclusions reached by \citet{2009ApJS..182..216K}. We further probe a sample of galaxies focusing on the differences in their surface brightness profiles, adding critical information extracted from modern IFU data. The two-dimensional kinematics and stellar population properties provide us with an upgraded perspective, leading to new insights on core and core-less galaxies.

{\it Core-less slow rotators} have somewhat higher $\lambda_r$, lower $\sigma_e$, are typically flatter, less massive, and younger, with steeper metallicity gradients and a larger dispersion of the gradient values. Age gradients span the full range of slopes from steep positive over flat to steep negative values. The detailed kinematic properties show that they have a more significant anti-correlation between $V/\sigma$  and $h_3$, and exclusively have $2\sigma$ type of kinematics. These last two aspects suggest that at least for some of them (40-45\%\  according to the discussion in Section~\ref{ss:nocores}), the low angular momentum is a consequence of the counter-rotation of flattened structures. These are not necessarily actual discs, but are characterised as having well-defined angular momentum vectors, oriented in opposite directions. Crucially, they have lower masses than the {\it core slow rotators}. They are therefore more closely linked in their evolution to fast rotators \citep[Sect.3.4.3 of][]{2016ARA&A..54..597C}.

Following theoretical predictions and previous observational work \citep[e.g.][]{2009ApJS..182..216K}, we conclude that {\it core-less slow rotators} are mostly products of gas-rich interactions. Some might form by accretion of counter-rotating and star-forming gas \citep[e.g.][]{2014MNRAS.437.3596A,2015A&A...581A..65C,2019arXiv190303627S}, where the number of counter-rotating stars that formed defines whether the galaxy is classified as a slow rotator. Other {\it core-less slow rotators} might form in gas-rich mergers of specific orbital configuration. NGC\,1222 is likely an example of such a merger. This gas-rich merger is in an advanced stage \citep{2018MNRAS.477.2741Y}, but the outcome is not yet fully defined. It can, however, be expected that the remnant will have a low net angular momentum, with possibly a KDC or even a prolate-like rotation, but likely with a core-less light profile \citep[e.g.][]{2004AJ....128.2098R,2006AJ....131..185R}. Finally, equal-mass binary mergers \citep[e.g.][]{2005A&A...437...69B, 2011MNRAS.416.1654B} showed that stellar discs might survive if the orbital orientations and the intrinsic spins of progenitors are finely tuned. This is most likely the formation scenario of the prototypical $2\sigma$ galaxy and a {\it core-less slow rotator} NGC\,4550 \citep{2009MNRAS.393.1255C}. 

\citet{2015MNRAS.448.3484M} showed that galaxies with non-regular kinematics (typically slow rotators) fall below the general trend in a [Z/H] versus $\sigma_e$ (see their fig.~11). The trend is more pronounced for galaxies with lower $\sigma_e$ and lower mass, which effectively selects our {\it core-less slow rotators}. As the authors argued, this can be considered as an offset in metallicity (these galaxies are less metal rich than typical fast rotators), or an offset in velocity dispersion or mass (have larger velocity dispersion or higher mass for a given metallicity). The latter can be understood as a consequence of a merger in which violent relaxation \citep{1967MNRAS.136..101L,1992ApJ...397L..75S} plays a major role in reshaping the internal structure of galaxies \citep{2012MNRAS.425.3119H} as well as morphologically transform galaxies \citep{1993ApJ...409..548H}. This applies to minor collisional and major collision-less mergers \citep{2017ARA&A..55...59N}. 

Mergers producing {\it core-less slow rotators} are between progenitors, with the mass ratios spanning a significant range (e.g. 1:1 - 1:10), but the orbital configurations of the mergers must be such to produce low angular momentum remnants \citep[e.g.][]{2006MNRAS.372..839N, 2009ApJ...705..920H, 2010ApJ...723..818H, 2011MNRAS.416.1654B,2014MNRAS.444.1475M,2014MNRAS.444.3357N,2014MNRAS.445.1065R,2018MNRAS.473.4956L}. In some of these mergers, a binary SMBH pair might form, but it could be of too low mass\footnote{{\it Core-less slow rotators} have masses between $10^{10}$ and $10^{11}$ M$_\odot$, which implies black hole masses of $\sim10^7$ up to $10^8$ M$_\odot$ \citep[e.g.][]{2016ApJ...831..134V}} to excavate an observable core \citep{2018ApJ...864..113R}, or the central starburst is able to refill the nuclei \citep[][see also Section~\ref{ss:frcore}]{1994ApJ...437L..47M,2009ApJS..182..216K}. Finally, radiative feedback from the active galactic nucleus, or a mechanical feedback from winds, can also facilitate the production of slow rotators by preventing the cooling and accretion of new gas \citep{2014MNRAS.444.3388S,2019MNRAS.489.2702F}. 

Steeper metallicity gradients of {\it core-less slow rotators} are compatible with these formation scenarios. Crucially, the fact that the gradients show a large dispersion indicates that there are multiple pathways for the formation of {\it core-less slow rotators}. Even a range of star formation histories is allowed, as long as new stars do not settle in a fast, co-rotating, and flat disc-like structure. The formation channels are characterised by the fact that the progenitors are typically less massive ($\sim10^{10}$ M$_\odot$ or smaller), have different stellar populations (typical of fast rotators), and have various amounts of gas.

{\it Core slow rotators} are more massive galaxies, with the lowest angular momenta, show flat age and metallicity gradients, and are made of old stellar populations. They comprise systems of the highest $\sigma_e$ , and some of them show no net rotation. Such properties are predicted from major and multiple minor dissipation-less mergers \citep[e.g.][]{2005MNRAS.360.1185J, 2006ApJ...636L..81N, 2009ApJ...699L.178N,2014MNRAS.444.1475M,2014MNRAS.445.1065R,2018MNRAS.473.4956L,2018ApJ...864..113R}. Furthermore, these properties also require a very active mass-assembly history, rich in accretion of much smaller galaxies \citep[e.g.][]{2013MNRAS.429.2924H}, but also experiencing a few major (equal mass) mergers \citep[e.g.][]{2007MNRAS.375....2D}. The latter are required in order to produce complex kinematics and also to create cores in the nuclei through binary supermassive black hole evolution \citep{1997AJ....114.1771F, 2009ApJS..182..216K,2018ApJ...864..113R}. The location of {\it core slow rotators} in the mass - size diagram is consistent with the predicted location of remnants of equal-mass dissipation-less mergers \citep{2006MNRAS.369.1081B,2009ApJ...697.1290B,2009ApJ...699L.178N,2009ApJ...691.1424H}, which indicates a doubling growth in mass and size, but no change in the effective velocity dispersion. This remains true for galaxies with masses higher than $10^{12}$ M$_\odot$ \citep{2018MNRAS.477.5327K,2018MNRAS.477.4711G}. 

Numerical simulations placed in the cosmological context \citep[see the review by][]{2017ARA&A..55...59N} produce consistent results with simulations of binary mergers. Massive galaxies are gas-poor systems, and their late evolution (z $<2$) is dominated by accretion of stars formed elsewhere \citep[e.g.][]{2007MNRAS.375....2D,2010ApJ...725.2312O,2015MNRAS.449..361W, 2016ApJ...833..158C,2018MNRAS.473.4956L}. The current picture is that low-mass galaxies typically grow through star formation and are more likely to experience dissipative mergers, while massive galaxies accrete stars formed elsewhere and grow through dissipation-less mergers \citep[e.g.][]{2016MNRAS.458.2371R,2017MNRAS.464.1659Q}. Empirical models support this picture \citep{2013MNRAS.428.3121M,2013ApJ...770...57B,2018MNRAS.477.1822M}. The implication is that the SMBH masses will also increase proportionally to the SMBH masses of progenitors and the number of subsequent gas-free mergers. This gives rise to the relation of the SMBH mass with the core size, as well as with the missing light (or stellar mass) compared to the extrapolation of an external light profile \citep{2004ApJ...613L..33G, 2007ApJ...662..808L, 2009ApJS..182..216K, 2013AJ....146..160R}. It also supports the expectation that the SMBH mass will not correlate well only with the velocity dispersion in the most massive systems \citep{2018MNRAS.473.5237K}.

{\it Core slow rotators} are consistent with being merger remnants of galaxies that resemble present-epoch massive {\it core-less fast rotators} or {\it core slow rotators}. This is supported by the overlap in [$\alpha$/Fe] values (Fig.~\ref{f:age}) for some {\it core-less fast rotators} and {\it core slow rotators} and by the substantial population of {\it core-less fast rotators} with large $\sigma_e$ (and masses, Fig~\ref{f:ms}). Furthermore, there are also {\it core fast rotators} that overlap in $\sigma_e$ and [$\alpha$/Fe] (as well as mass) with {\it core slow rotators}. This is important because during the dissipation-less major merging, the velocity dispersion typically remains unchanged. On the other hand, {\it core-less slow rotators} do not seem to be able to contribute to the formation of {\it core slow rotators} and the most massive galaxies in general. Their $\sigma_e$ and [$\alpha$/Fe] are too low. This conclusion is similar to that reached by \citet{2009ApJS..182..216K}, but our spectroscopic data provide stronger constraints on types of possible progenitors of {\it core slow rotators}. 

Galaxy mergers, the lack of cold gas, and the interaction of massive SMBHs seems to be crucial for the creation of cores. This is typical for the most massive galaxies, as is the low angular momentum that is expected to result from such interactions. Moreover, the flat metallicity gradients and the small variation between them imply a unique channel of formation or a few different formation paths that have a dominant physical process in common: dynamical mixing of stellar populations is efficient and likely dependent on mass. Furthermore, maintaining the core is necessary, and this is probably why cores are mostly found in massive slow rotators: they are able to retain their halo of hot X-ray emitting gas \citep{2009ApJS..182..216K,2013MNRAS.433.2812K}. 

\subsection{Formation of fast rotators with cores}
\label{ss:frcore}

{\it Core fast rotators}, while comprising only a small fraction of fast rotators (9 of 111 ATLAS$^{\rm 3D}$ fast rotators with HST imaging), seem to contradict the favoured scenario of core formation. \citet{2013MNRAS.433.2812K} showed that {\it core fast rotators} often have regular kinematics and a significant $V/\sigma - h_3$ anti-correlation. These are indicative of embedded disc-like structures. {\it Core fast rotators} have somewhat lower masses and $\sigma_e$, but a significant overlap also exists with the properties of {\it core slow rotators} (e.g. see Fig.~\ref{f:ms}). Judging by many other galaxy properties, such as stellar populations, age, metallicity gradients, $\alpha$-element abundances (as presented in this paper), molecular gas content, mass of supermassive black holes, or derived core parameters \citep[as presented in][]{2013MNRAS.433.2812K}, {\it core fast rotators} are similar to {\it core slow rotators}. Two more notable differences with respect to {\it core slow rotators} deserve to be highlighted. Firstly, {\it core fast rotators} show a large dispersion in metallicity and age gradients, indicating that they can be formed in different ways (Sections~\ref{ss:metG} and ~\ref{ss:age}). Secondly, {\it core fast rotators} have a lower X-ray emission than {\it core slow rotators}, but it is similar to those of core-less galaxies \citep{2013MNRAS.433.2812K}. This means that at least a theoretically, they might accrete cold gas after the last merger s(and the formation of the core) and change their global kinematics.

At this point it is instructive to consider some of these galaxies in more detail. NGC\,0524 and NGC\,4473 have very clear evidence for embedded discs, both morphologically and kinematically. NGC\,0524, has a nuclear dust disc which makes the analysis of the nuclear light profile difficult, but is the only core galaxy with molecular gas \citep{2011MNRAS.414..940Y}, while dynamical models for NGC\,4473 require two counter-rotating discs \citep{2007MNRAS.379..418C}. The cores in these two galaxies seem to be well defined, but the cores might also depend on the actual choice of the fitting range. For example, \citet{2009ApJS..182..216K} showed that in the case of NGC\,4473, it is possible to fit the surface brightness profile such as to recover a core, an extra-light (core-less) or a combination of an extra-light profile with an indication for a core (their Figs. 58 (top), 17, and 58 (bottom), respectively). This suggests that some of {\it core fast rotators} could be considered as just somewhat unusual fast rotators. 

Alternative examples are provided by NGC\,4649 and NGC\,5485. The former is the third most luminous galaxy in the Virgo cluster and the latter has a prolate-like (around the major axis) rotation \citep{2011MNRAS.414.2923K}. These two galaxies are massive and have relatively high $\sigma_e$. This is especially the case for NGC\,4649, which is one of the most massive galaxies in the ATLAS$^{\rm 3D}$ sample, the most massive fast rotator (see Fig~\ref{f:ms}), and also has one of highest $\sigma_e$ in the sample. The prolate-like rotation of NGC\,5485 requires a special merging configuration, possibly favouring a dissipation-less type of merger as it also has a nuclear gas disc in polar configuration \citep{2017A&A...606A..62T,2017ApJ...850..144E,2018MNRAS.473.1489L}. These galaxies are much more similar to {\it core slow rotators}, and it is possible that their classification depends on the definition of fast rotators as much as their angular momenta depend on the chance outcome of the formation process. 

Possible scenarios for the formation of {\it core fast rotators} are limited by two constraints: they show cores, and they show increased angular momentum and often evidence for discs. On the other hand, as has been argued by \citet{1997AJ....114.1771F} and \citet{2009ApJS..182..216K}, given that almost all bright galaxies contain SMBHs, the question is not why some galaxies have cores, but why most of galaxies are core-less. Their explanation, supported by simulations and other observations (as cited above), is that in a major merger a central starburst will fill any core excavated by the interacting black holes. This might not be the complete picture because we do not yet have full understanding of the specific conditions relevant to ejection of stars by the binary SMBHs, or starburst physics. In particular, the duration of a nuclear starburst can be as short as 30-50 Myr \citep[e.g.][]{2014MNRAS.442L..33R,2017MNRAS.465.1934F}, while the core excavation is completed on a comparable timescale \citep[$\sim50$ Myr,][]{2018ApJ...864..113R}, and both processes are likely to be strongly dependant on the initial conditions \citep[i.e. gas mass, SMBH mass, merger orbital set-up, and feedback ][]{2017ARA&A..55...59N}.

Furthermore, it is useful to recall that some {\it core slow rotators} could also be created through dissipative mergers.  NGC\,5557 is one such case \citep{2011MNRAS.417..863D}, a relatively massive ($~2\times10^{11}$ M$_\odot$) and round {\it core slow rotator}. This galaxy, very typical of slow rotators in its central regions, has a long and narrow tidal tail featuring gas-rich and star-forming objects, which are likely tidal dwarf galaxies. The long-lived tidal structures in NGC\,5557 are likely related to the event that formed the present galaxy, as the luminosity-weighted stellar age in the central regions is only 1-2 Gy old, even though no evidence of a starburst remains and the surface brightness core is detectable. NGC\,5557 is likely a special case and could have ended as a {\it core fast rotator}. Nevertheless, this case, as well as many of the {\it core fast rotators}, should add caution to our discussion because our understanding of the assembly process through dissipative and dissipation-less processes is evidently incomplete. 

We reiterate previous suggestions that {\it core fast rotators} are the galaxies that experienced a shorter central starburst compared to the time of the binary SMBH evolution. They are characterised with (typically) lower masses, diverse stellar population parameters, and lower X-ray luminosities. This makes them candidate remnants of dissipative mergers, where the final outcome (i.e. core or core-less, fast or slow rotator) could depend on initial conditions such as the gas mass, black hole mass, feedback, or general details of the merger. Furthermore, as shown by \citet{2011MNRAS.416.1654B}, dissipation-less mergers can also result in fast rotators \citep[see also][]{2014MNRAS.444.1475M,2014MNRAS.444.3357N} and {\it core fast rotators} might just be special cases, curious products of the vast parameter space of merger orbits.

%
%

\section{Conclusions}
\label{s:con}

We used SAURON IFU observations and HST imaging of the ATLAS$^{\rm 3D}$ sample to investigate the formation of early-type galaxies. In particular, we presented the metallicity and age gradients for the full ATLAS$^{\rm 3D}$  sample. We also link this paper with a public release of the age, metallicity, and $\alpha$-element abundance profiles, SAURON maps of the SSP equivalent stellar age, metallicity, and $\alpha$-element abundances, which can be obtained from the ATLAS$^{\rm 3D}$ website. Furthermore, this paper presents the analysis of the HST imaging of 12 ATLAS$^{\rm 3D}$ Survey slow rotators. We derived their nuclear stellar surface brightness profiles except for one galaxy that was too dusty. Combining these with results of \citet{2013MNRAS.433.2812K} and the kinematic properties of ATLAS$^{\rm 3D}$ galaxies, we divided the ATLAS$^{\rm 3D}$ galaxies into {\it core slow rotators}, {\it core-less slow rotators}, {\it core fast rotators,} and {\it core-less fast rotators}. The last category is the most numerous, but it is also the most incomplete with respect to the HST imaging. Nevertheless, this study presents the first complete volume-limited sample of slow rotators with IFU kinematics and high spatial resolution photometry. 

Using the Nuker profile parametrisation, we showed that all newly analysed slow rotators have rising light profiles and can be classified as core-less. We placed an upper limit on the core sizes of $\sim10$ pc. Because there are no known galaxies with smaller cores, our observations are consistent with expectations and show that lower mass slow rotators do not harbour unexpected nuclear structures. This allowed us to show that more than half (about 55\% ) of the slow rotators in the ATLAS$^{\rm 3D}$ sample have core-less light profiles. 

The metallicity gradients of the ATLAS$^{\rm 3D}$ galaxies are consistent with gradients expected for galaxies in the same mass range. We detect a trend where the most massive galaxies ($>10^{11}$ M$_\odot$) show significantly flatter gradients ($\overline{\Delta [Z/H]} =-0.28$) than the less massive galaxies ($\overline{\Delta [Z/H]} =-0.37$). This is closely reflected for galaxies with core and core-less light profiles ($\overline{\Delta [Z/H]} =-0.23$ and $-0.36$, respectively). {\it Core slow rotators} and {\it core fast rotators} have similarly flat gradients (-0.23 and -0.19, respectively), but the dispersion of gradient values is much higher for {\it core fast rotators} (0.19 compared to 0.07).  The mean gradient and gradient dispersion values of {\it core-less slow rotators} are similar to those of {\it core-less fast rotators}. There is also an indication that metallicity gradients become flatter with increasing mass. Age gradients of {\it core slow rotators} are close to zero, while those of {\it core-less slow rotators} or {\it core fast rotators} can take essentially any value as the underlying ETG population. 

The presence (or absence) of cores correlates with a number of other properties of slow rotators. We find that {\it core-less slow rotators}, compared to {\it core slow rotators}  

\begin{itemize}
\item are less massive (typically $<10^{11}$ M$_\odot$), have lower $\sigma_e$ and overlap with fast rotators in the mass - size diagram, 

\item are flatter, have somewhat higher $\lambda_R$ , and often show evidence of counter-rotation, and

\item have steeper metallicity gradients ($\overline{\Delta [Z/H]} =-0.42$ compared to $\overline{\Delta [Z/H]} =-0.23$), and show a large dispersion of gradient values ($\delta(\Delta [Z/H]) =0.18$ compared to $\delta(\Delta [Z/H]) =0.07$) as well as a range of stellar ages. 

\end{itemize}

{\it Core slow rotators} are extreme galaxies. They are the brightest and most massive, the largest, and have the highest $\sigma_e$. They have the lowest angular momenta. Their stellar populations (within one half-light radius) are always old, and the metallicity gradients are shallow and show little dispersion. There is an indication that the metallicity gradients flatten with increasing mass. 

{\it Core-less slow rotators} and {\it core slow rotators} are equally likely to have KDCs or CRCs, but velocity maps with no net rotation are obtained only for {\it core slow rotators}, while $2\sigma$ velocity dispersion maps are exclusive for {\it core-less slow rotators}. From investigating also the higher order moments of the LOSVD, we conclude that {\it core-less slow rotators} form due to accretion of counter-rotating star-forming gas or stars, and the level of the counter-rotation defines whether they are classified as slow rotators. A fraction of {\it core-less slow rotators} are remnants of dissipational processes, such as gas-rich mergers of various mass ratios with specific orbital configurations, which can also decrease the net angular momentum of the remnant. Violent relaxation can also influence the formation of {\it core-less slow rotators}. 

{\it Core fast rotators} share some properties with {\it core slow rotators} (i.e. presence of cores, [$\alpha$/Fe], mass, $\sigma_e$, mean values of metallicity gradients, and old ages) and some properties with {\it core-less slow rotator} and other fast rotators (i.e. kinematics, counter-rotation, disc-like components, higher $\lambda_R$, larger spread in metallicity and age gradients, and lower X-ray luminosities). In this respect, {\it core fast rotators} are the most diverse class in terms of possible formation channels. They can be products of gas-poor mergers that result in fast rotators (but allow core creation), gas-rich mergers with a central starburst of shorter duration than the binary SMBH evolution, or are able to subsequently increase angular momentum through further interactions without destroying cores. Most likely, every such galaxy will have a unique way of mass assembly, core formation, and subsequent protection. 

In contrast to all this, the formation of {\it core slow rotators} requires at least one and probably a few dissipation-less major mergers in the presence of massive black holes in order to produce the core, flatten their metallicity gradients, and create the variety of observed kinematics. Their progenitors could be galaxies resembling present-day {\it core fast rotators} or most massive {\it core-less fast rotators}, but the assembly of the most massive galaxies is only possible by {\it core slow rotators} themselves.

Our results support the approach of only considering {\it core slow rotators} when searching for massive dry merger relics in observations \citep{2012ApJ...759...64L,2013ApJ...778L...2C} or when describing their evolution as a class \citep{2016ARA&A..54..597C}. Finally, the fact that {\it core slow rotators} only dominate the characteristic mass of M$_{crit}\sim2\times10^{11}$ M$_\odot$ and are completely absent below $\sim10^{11}$ M$_\odot$ indicates that when the characteristics (e.g. their shapes or environments) of massive dry merger relics are to be studied but high-resolution imaging is not available, a further removal of spurious objects is easily achieved by removing slow rotators below M$_{crit}$ \citep[e.g.][]{2018ApJ...863L..19L}.

\begin{acknowledgements}
Michael Wolfe deceased during the final stages of this work, due to complications resulting from a stroke. He was very friendly to everyone, dedicated to his work, and always eager to learn new things. We offer sincere condolences to his family and friends. DK acknowledges support from the grant GZ: KR 4548/2-1 of the Deutsche Forschungsgemeinschaft. UU acknowledges support from the grant 50 OR 1412 of the Deutsches Zentrum f\"ur Luft- und Raumfahrt. MC acknowledges support from a Royal Society University Research Fellowship. RMcD is the recipient of an Australian Research Council Future Fellowship (project number FT150100333). RLD acknowledges travel and computer grants from Christ Church, Oxford, and support from the Oxford Hintze Centre for Astrophysical Surveys, which is funded through generous support from the Hintze Family Charitable Foundation. Support for this project was provided by NASA through grant HST-GO-13324 from the Space Telescope Science Institute, which is operated by the Association of Universities for Research in Astronomy, Inc., under NASA contract NAS5-26555. 
\end{acknowledgements}


\bibliographystyle{aa}


\begin{appendix} 

\section{Comparison between deconvolution methods}
\label{a:deco_comp}

In order to verify the robustness of our surface brightness profiles, we applied the Richardson-Lucy deconvolution method \citep{1972JOSA...62...55R,1974AJ.....79..745L} directly on our images. Among various deconvolution methods available for the comparison, we selected the Richardson-Lucy deconvolution because it was proposed as a direct solution to the known effect that the Burger-van Cittert method can lead to unphysical (negative) values in the deconvolved profiles \citep[e.g.][]{1974AJ.....79..745L}. Richardson-Lucy deconvolution was also extensively used in the analysis of the nuclear surface brightness profiles \citep[]{1995AJ....110.2622L,1998AJ....116.2263L,2001AJ....121.2431R,2005AJ....129.2138L}. An alternative approach is to convolve the models (e.g. Nuker or S\'ersic functions) with the PSF as in \citet{2001AJ....122..653R} and \citet{2006ApJS..164..334F}, for instance.

\begin{figure*}
\includegraphics[width=0.32\textwidth]{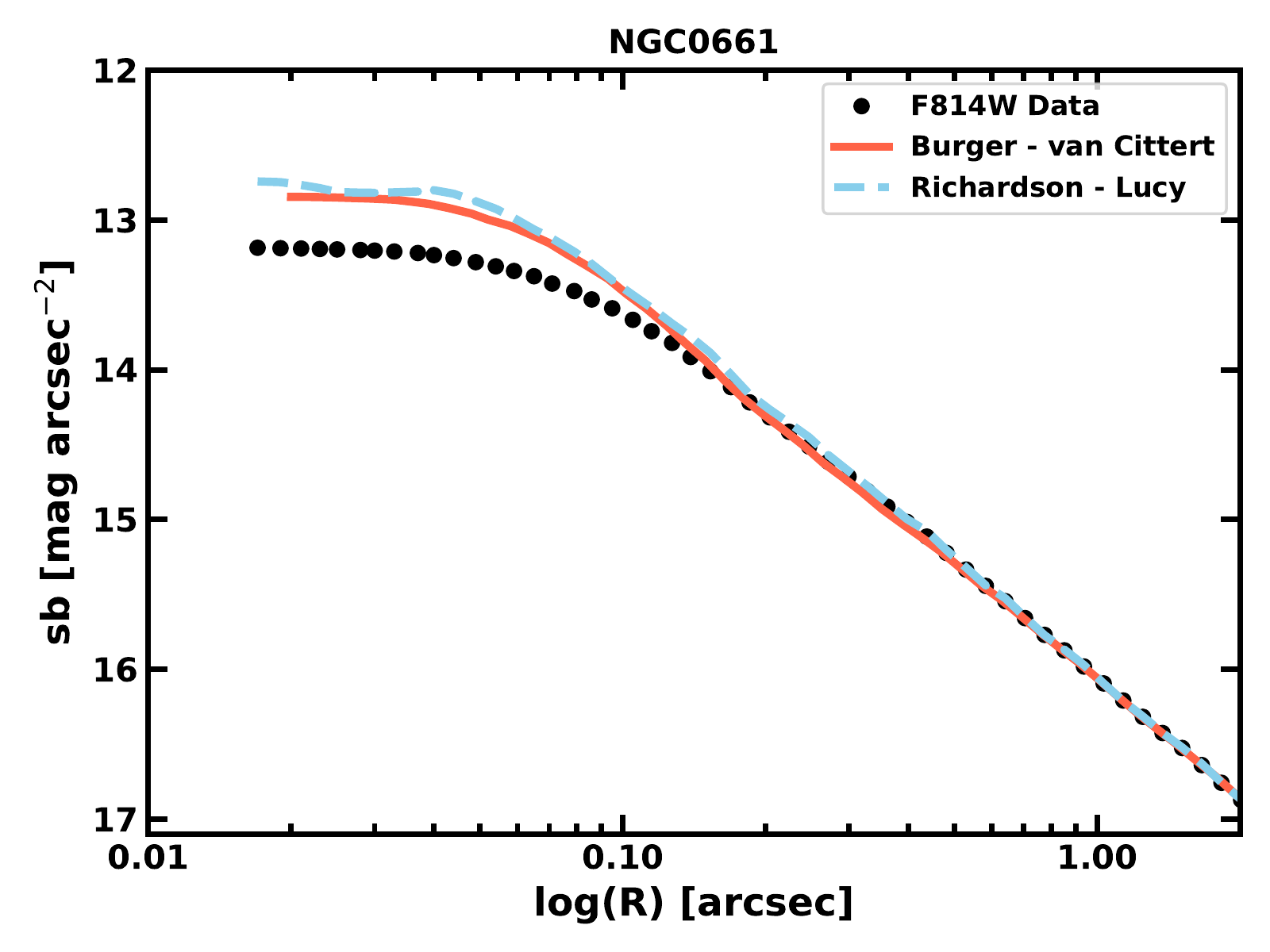}
\includegraphics[width=0.32\textwidth]{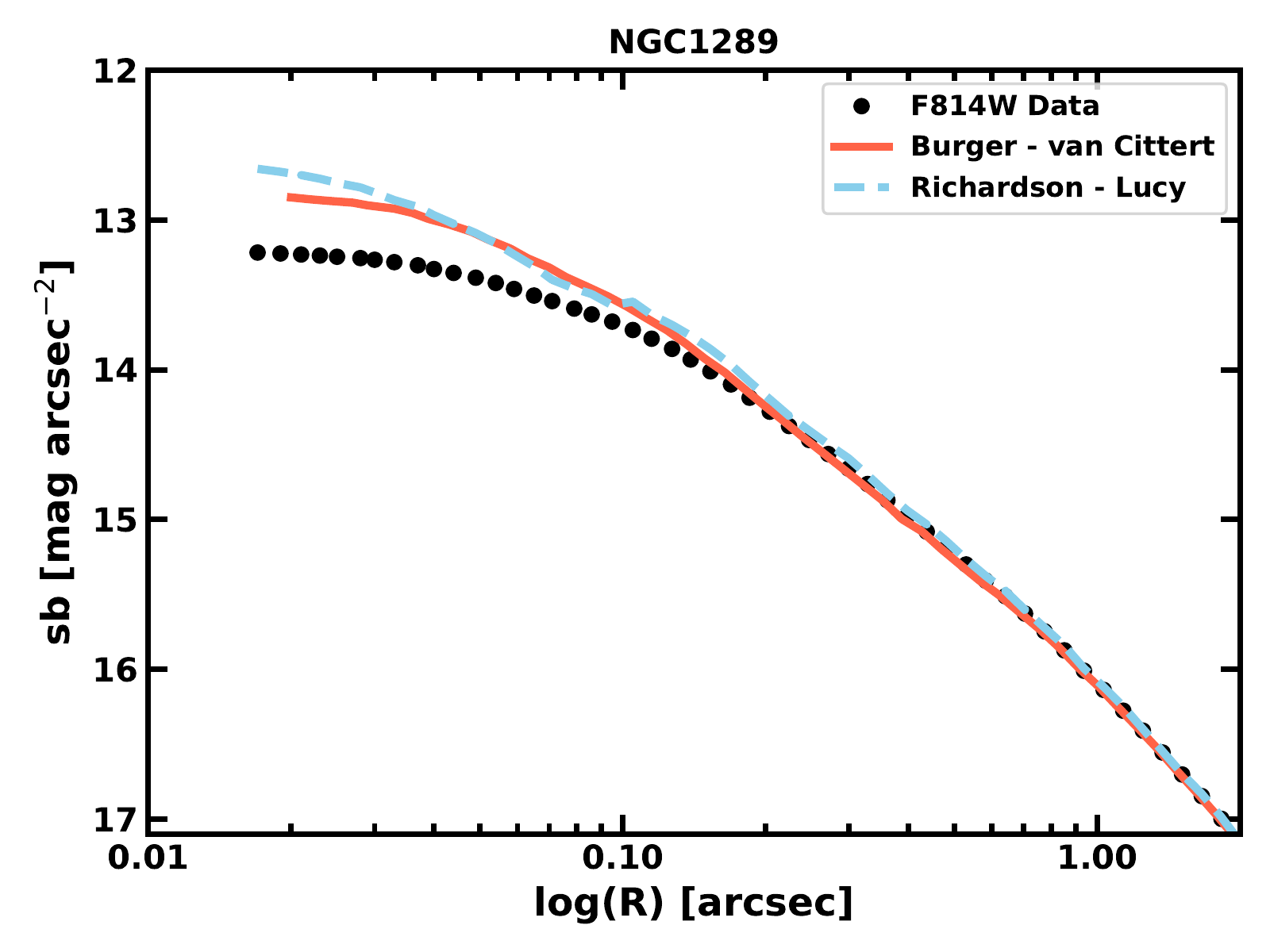}
\includegraphics[width=0.32\textwidth]{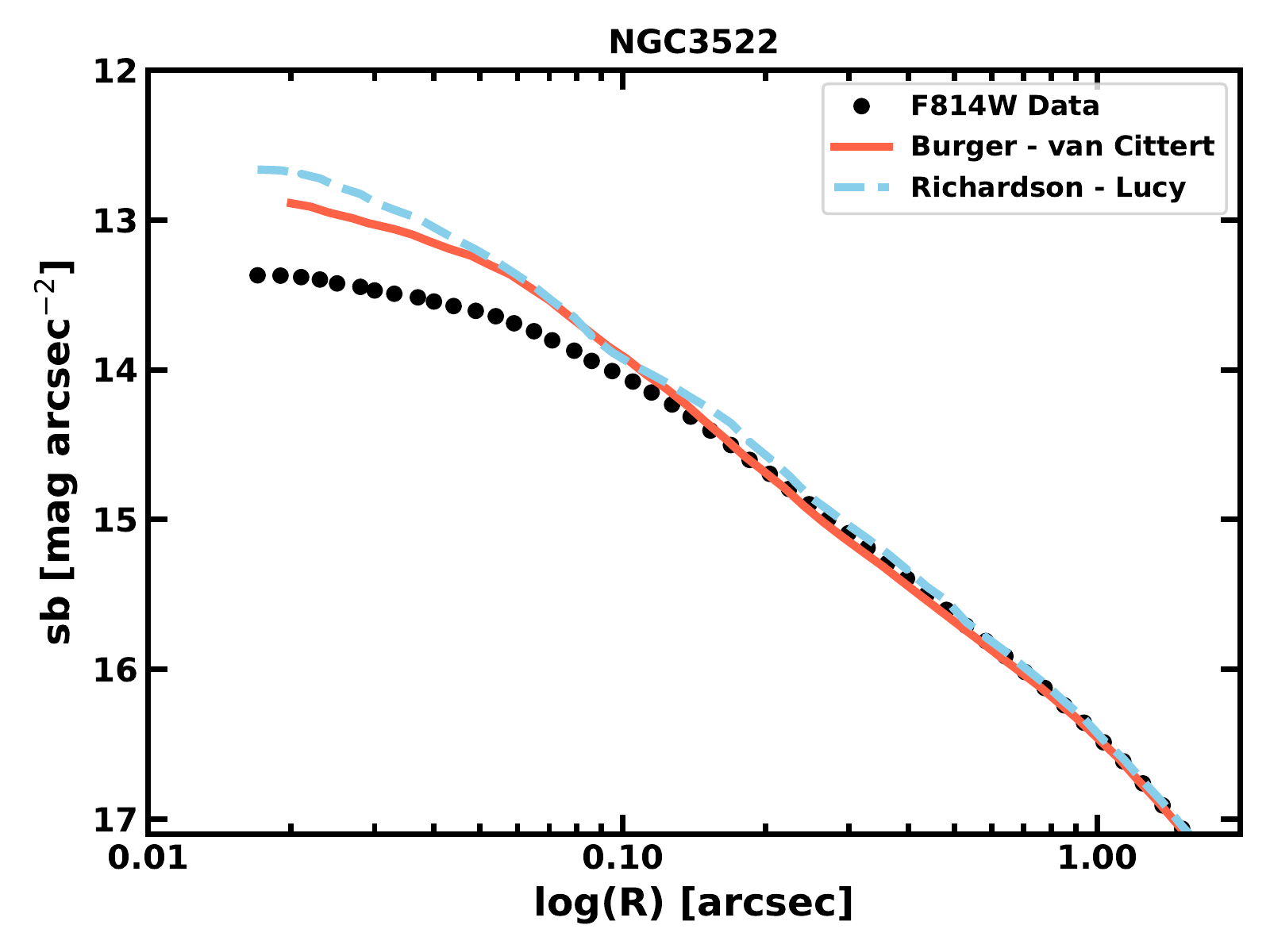}
\includegraphics[width=0.32\textwidth]{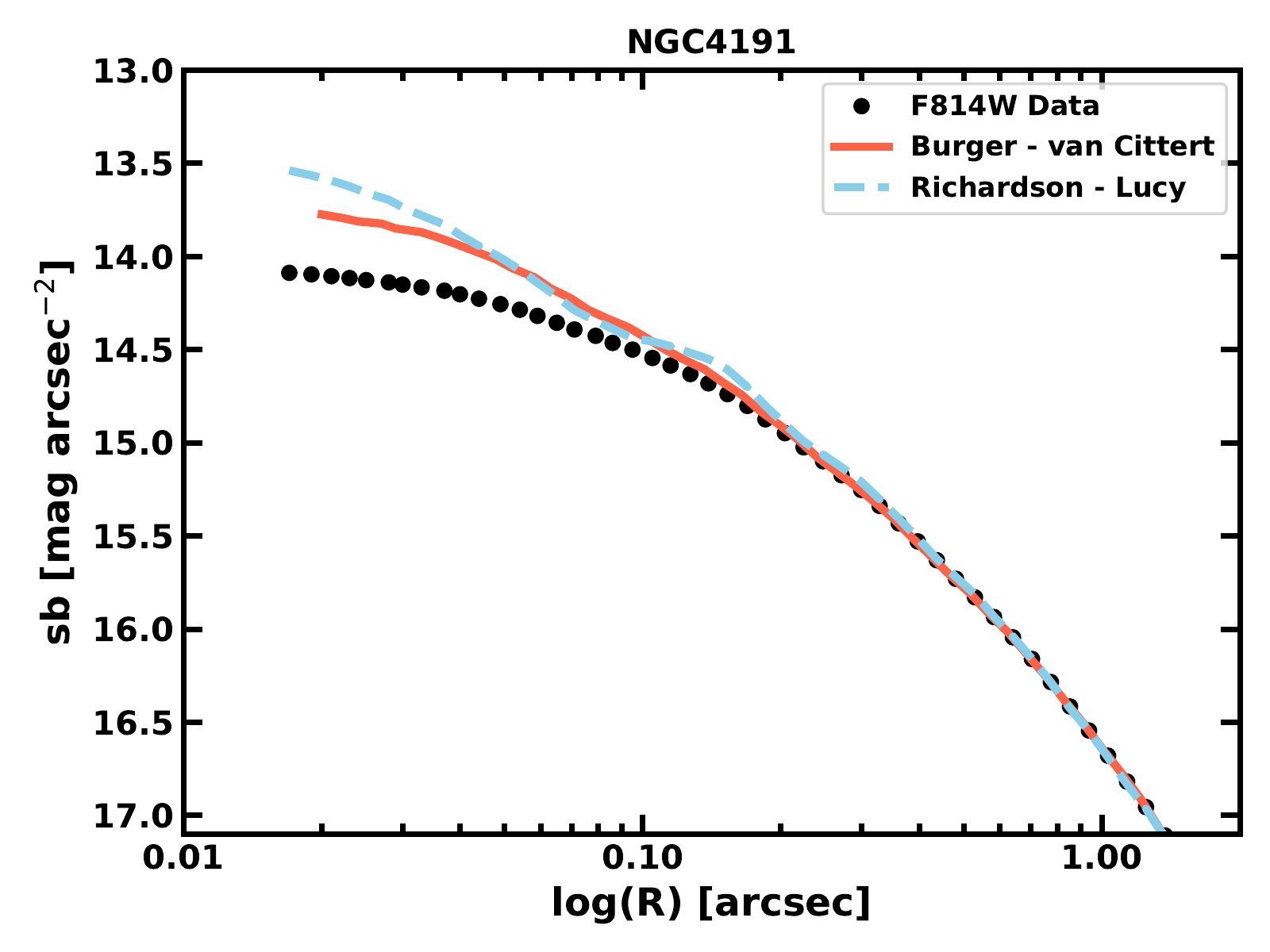}
\includegraphics[width=0.32\textwidth]{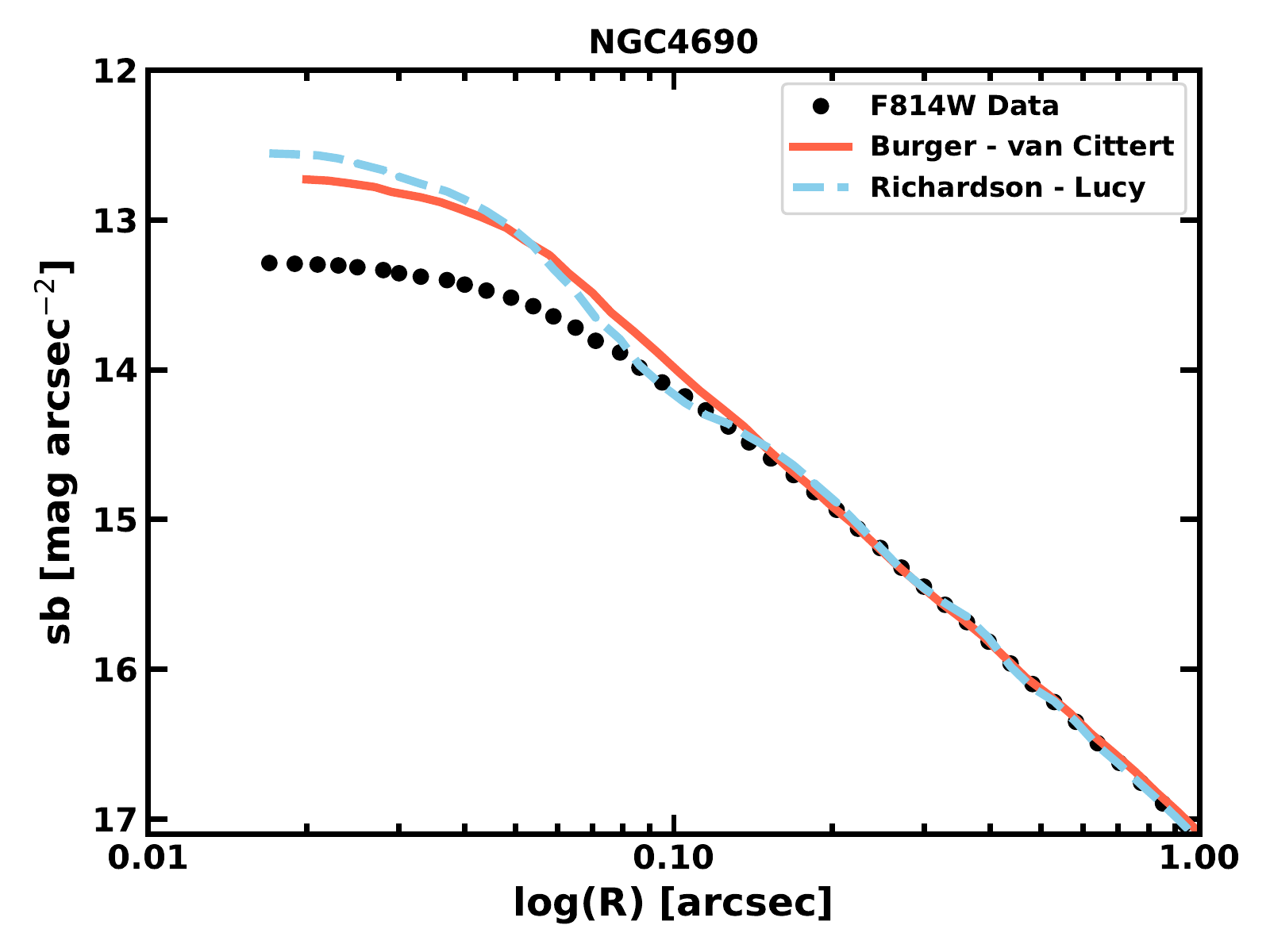}
\includegraphics[width=0.32\textwidth]{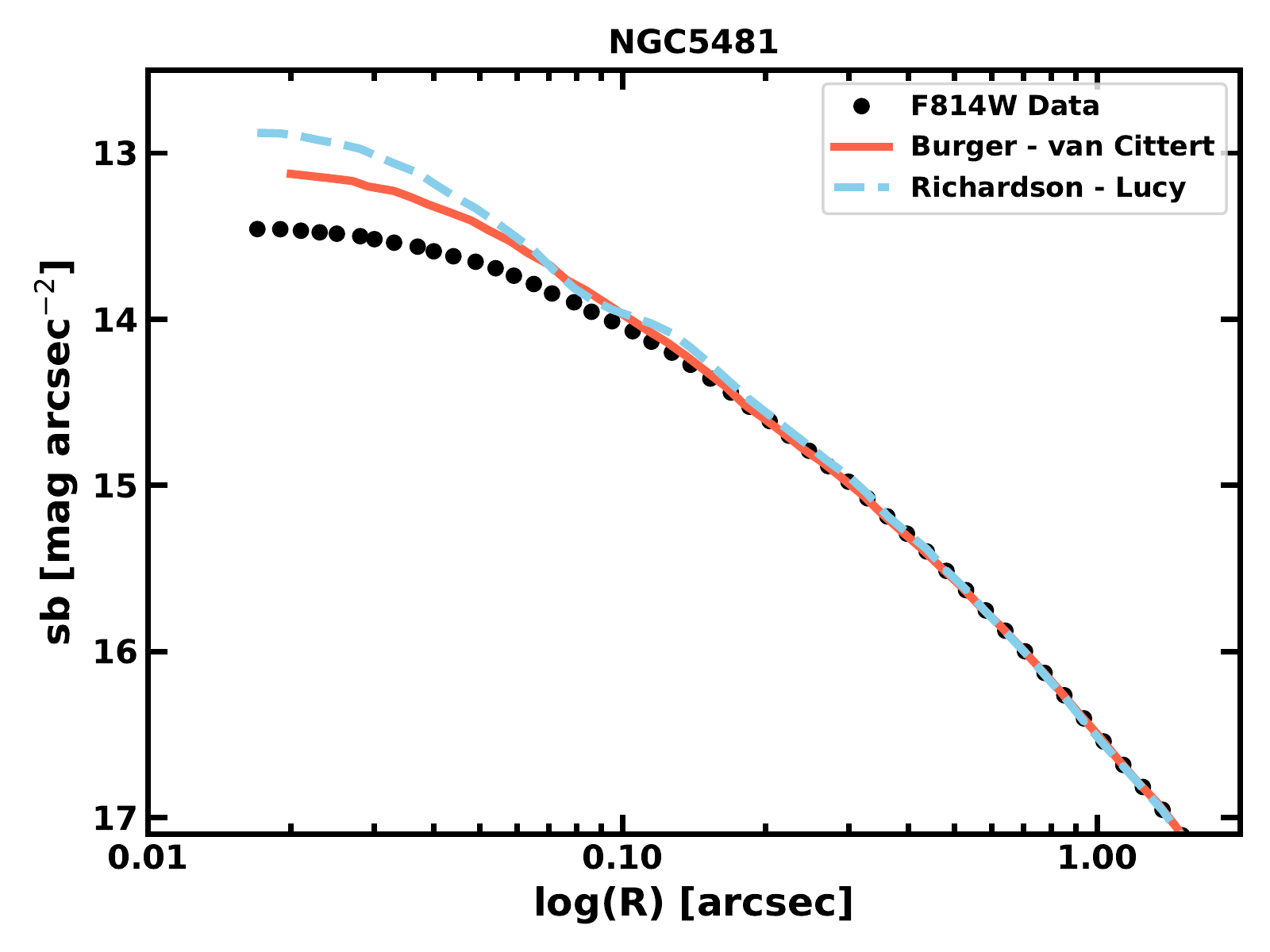}
\includegraphics[width=0.32\textwidth]{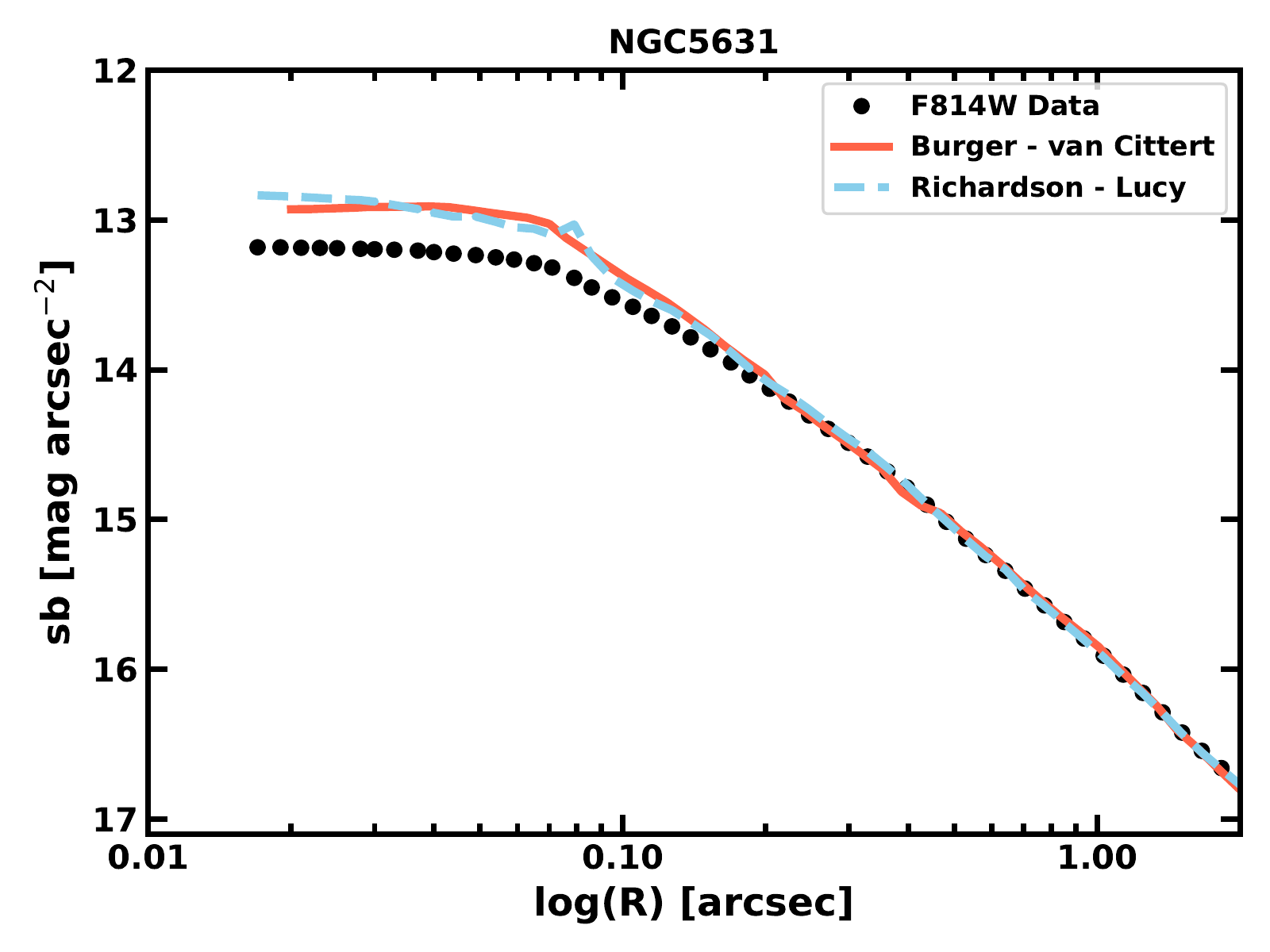}
\includegraphics[width=0.32\textwidth]{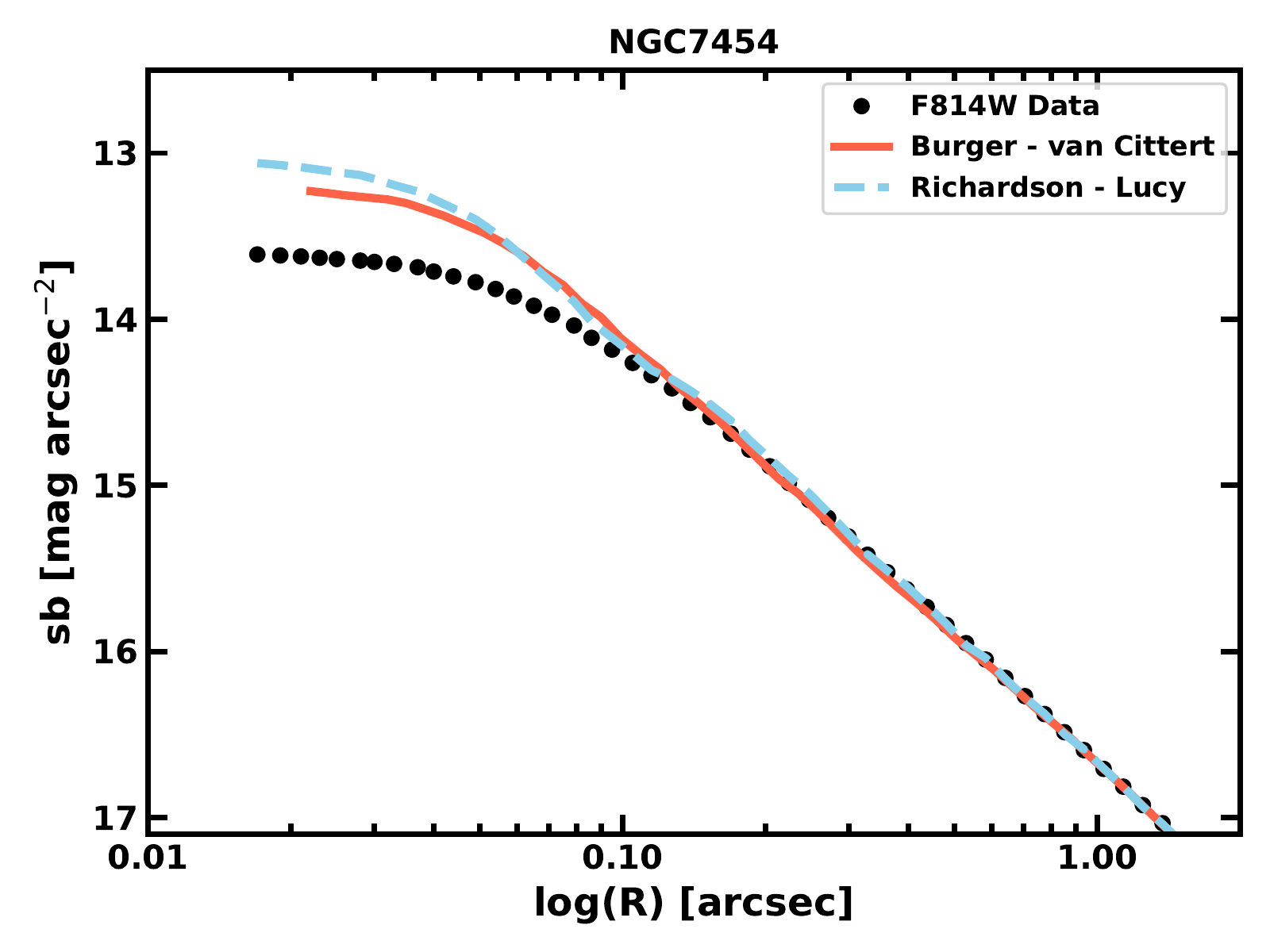}
\includegraphics[width=0.32\textwidth]{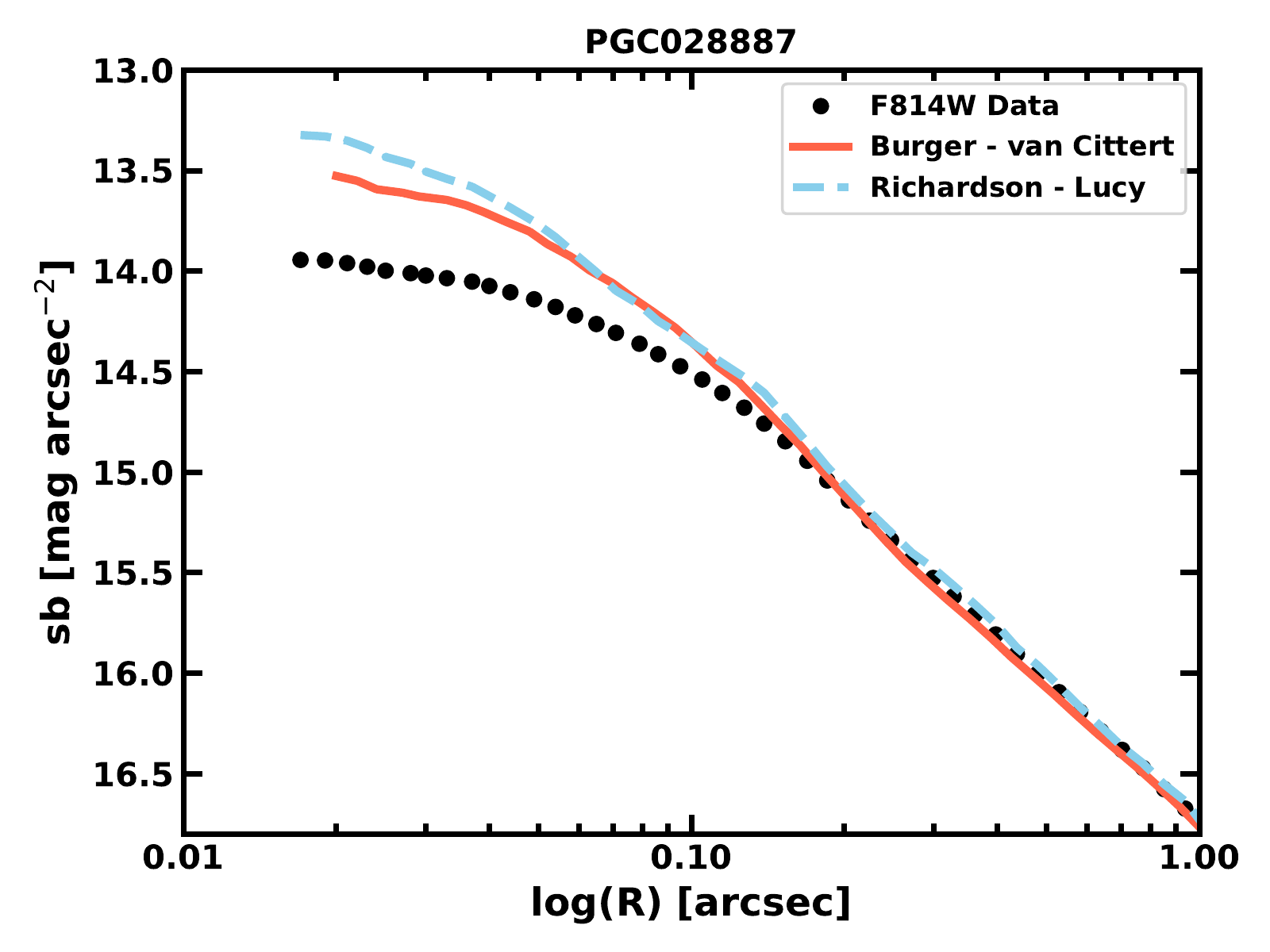}
\includegraphics[width=0.32\textwidth]{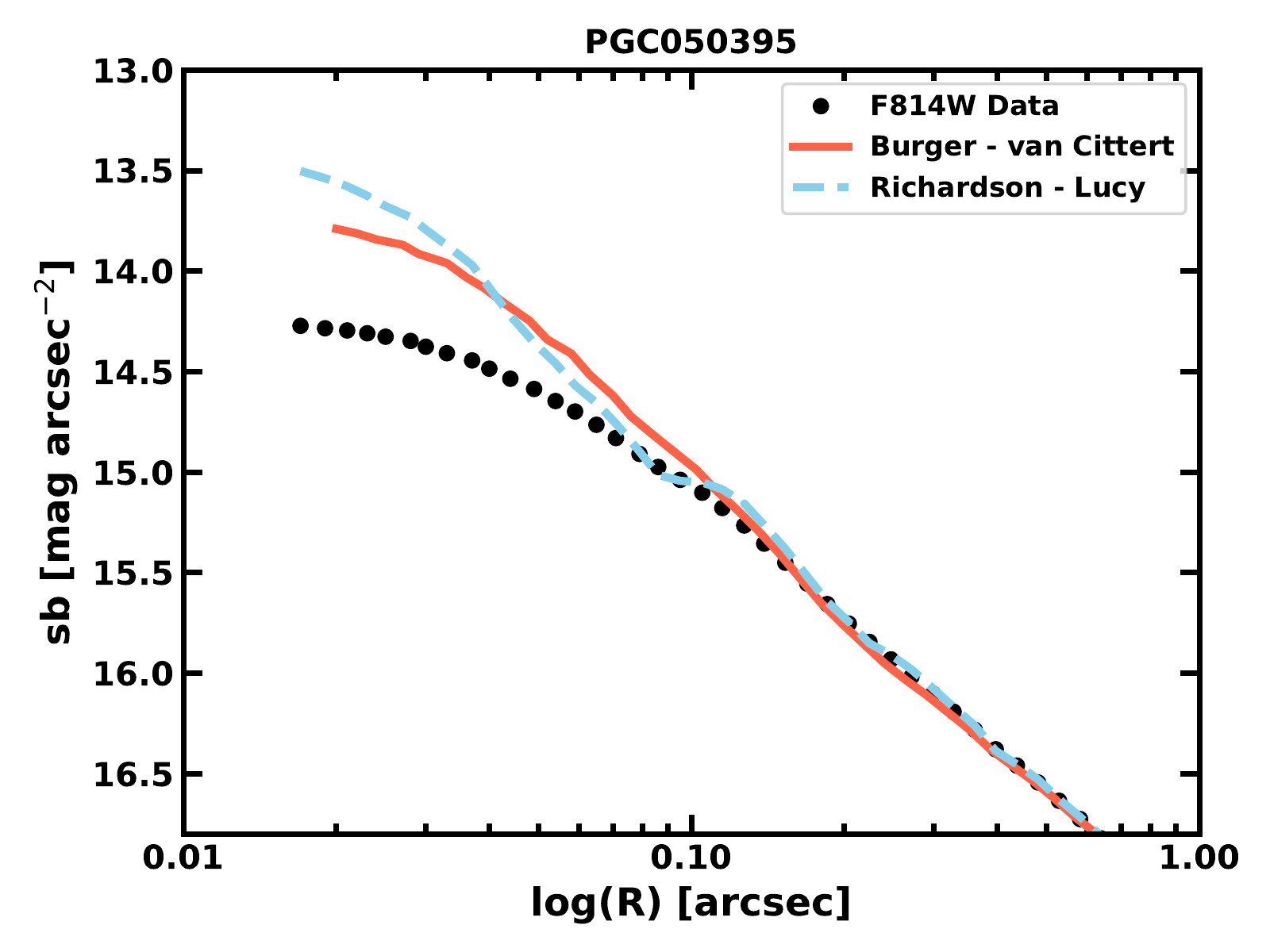}
\includegraphics[width=0.32\textwidth]{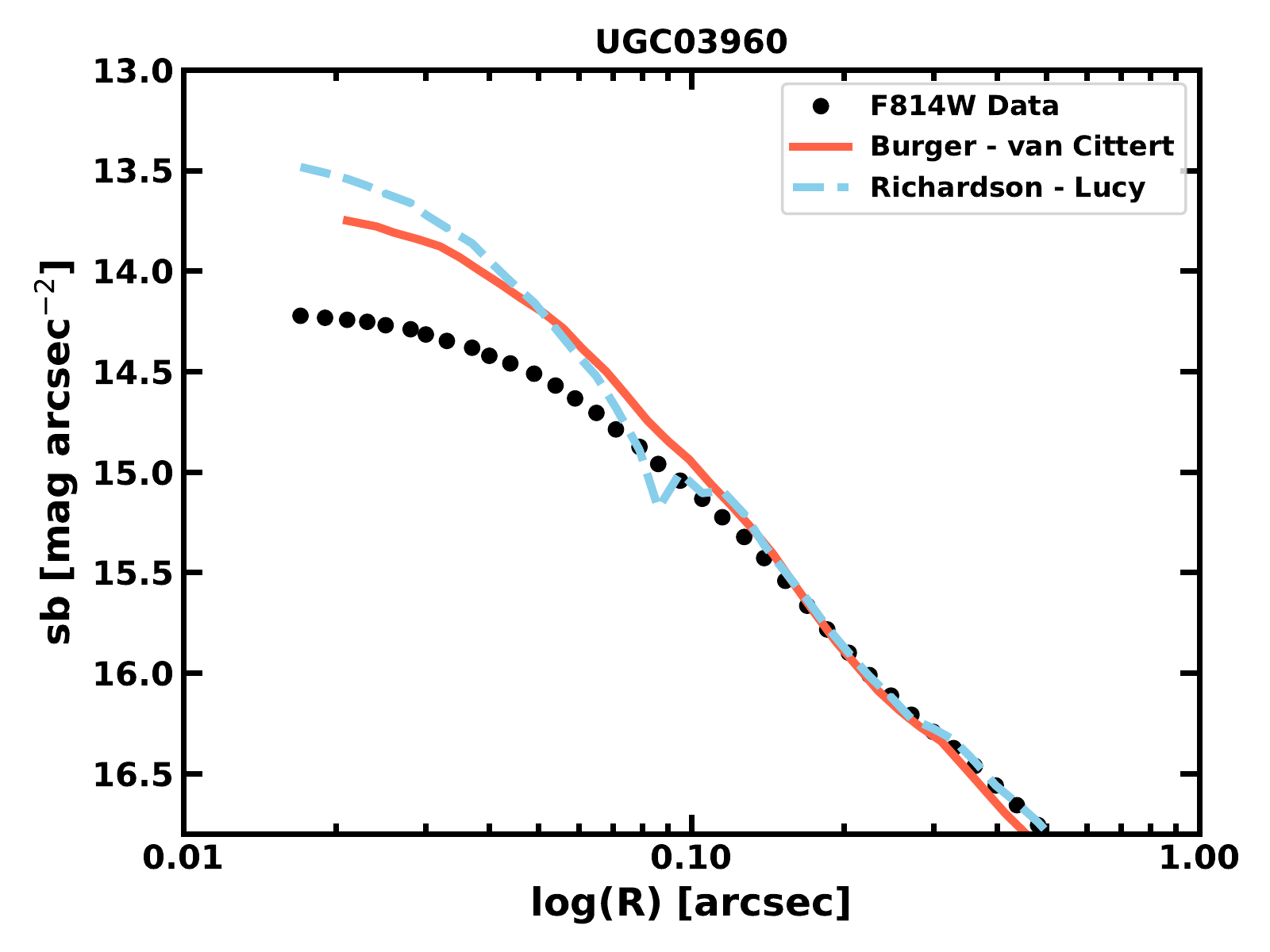}
\caption{Comparison between surface brightness profiles deconvolved using the Burger - van Cittert (red solid line) and Richardson - Lucy (light blue dashed line) deconvolution techniques. Black circles show the original profile in the F814W band. The sampling (black circles) of a light profile does not correspond to the pixel scale of the WFC3 camera, but it is defined by the tool for the isophote analysis. We do not show NGC\,1222. In all galaxies, the two deconvolution methods result in very comparable profiles. The differences are only seen at radii smaller than 0\farcs03-0\farcs04, which correspond to the central pixel.   }
\label{fapp:deco_com}
\end{figure*}

We used the WFC3 PSF estimates obtained using the TinyTim software as before. We followed the suggestion by \citet{1995AJ....110.2622L} and ran 80 iterations for the final decomposition, but we tested the stability of the process with fewer iterations. We used the Richardson-Lucy deconvolution software provided in the python {\tt skimage} package. The surface brightness profiles were extracted from the original and the deconvovled images in the same way as described in Section~\ref{s:ext}. The only difference is that we now used the python package {\tt photutils}, which implements the same \citet{1987MNRAS.226..747J} method as in {\tt STDS IRAF} task {\tt ellipse}.  Figure~\ref{fapp:deco_com}  shows the comparison between three surface brightness profiles for our galaxies, except for NGC\,1222, which is too dusty to allow any reliable comparison. We show the observed profile, the Richardson-Lucy deconvolved profile, and the Burger-van Cittert deconvovled profile (as in Figs.~\ref{f:sb1} and \ref{f:sb2} and Section~\ref{s:nsb}) for each galaxy. The figure zooms on the central 1-2\arcsec to facilitate the comparison. 

The deconvolved light profiles are essentially identical on all scales. The only departure is visible in the innermost 0\farcs03-0\farcs04, which correspond to the central pixel of the WFC3 camera. Similar accuracy is claimed by \citet{1998AJ....116.2263L}, who used Richardson-Lucy deconvolution on WFPC2 images. Profiles deconvolved using the Richardson-Lucy method seem to achieve somewhat higher flux in the centre, but the differences are marginal, and in some cases, the deconvolution results in rougher profiles than the original data. These are dependent on the number of iterations, suggesting that this number needs to be optimised individually for each case. Nevertheless, the overall comparison clearly suggests that the Burger-van Cittert deconvolution method based on high signal-to-noise ratio HST data (and stable HST PSF), while approximate, converges, and provides sufficiently robust profiles comparable to other deconvolution techniques. 

We can trust our deconvolved profiles to $\sim$0\farcs04, therefore the question arises whether we might not also use a smaller radius to determine the slope of the inner profile. As mentioned in the paper, we wish to be consistent with the literature. Nevertheless, we calculated $\gamma^\prime$ values for $r^\prime=0.05\arcsec$, based on our Nuker profiles fits (Table~\ref{t:fits}). The changes in $\gamma^\prime$ were minimal, except for NGC\,0661, where the value changed from 0.96 to 0.33, and NGC\,5631 where $\gamma^\prime$ changed from 0.70 to 0.42. This means that in both cases, the classification of the light profile changed from a power-law to an intermediate, but it still remains core-less in our definition. When the same analysis is performed on the F475W images, the change for both galaxies is smaller, going to $\gamma^\prime_{0.05}=0.52$ and $\gamma^\prime_{0.05}=0.59$ for NGC\,0661 and NGC\,5631, respectively. The deconvolved F814W profile of NGC\,5631 is worth an additional note because it seems rather flat within 0\farcs7. This is equally reproduced in the two deconvolution methods. When the F475W image is used, this is not that case and the deconvolved profiles are smoother. There is no evidence that the F814W image is saturated because our data reduction method used the shallow F814W image to substitute any saturated pixels. Even though the curvature increases for the F814W deconvolved profile, the Nuker fit returns core-less profiles, consistent with F475W profile.

\section{Stellar population gradients}
\label{a:pop}

\begin{table}
   \caption{Metallicity and age gradients for ATLAS$^{\rm 3D}$ sample. The full table is available in the online version of the paper and at the \href{http://purl.org/atlas3d}{ATLAS$^{\rm 3D}$} survey website. \label{t:pop} }
   \label{t:pop}
$$
  \begin{array}{c c c c c c}
    \hline \hline
    \noalign{\smallskip}

$Galaxy$      & \Delta $[Z/H]$  & \delta \Delta $[Z/H]$  & \Delta $Age$& \delta \Delta $Age$ & $HST$ \\
(1)      & (2) & (3) & (4) & (5) & (6) \\
    \hline
   \noalign{\smallskip}
$IC0560$ & -0.454 & 0.068 & 0.411 & 0.054 & 0 \\
$IC0598$ & -0.278 & 0.081 & 0.371 & 0.061 & 0 \\
$IC0676$ & 0.187 & 0.085 & 0.221 & 0.115 & 0 \\
$IC0719$ & -0.56 & 0.05 & -0.08 & 0.106 & 0 \\
$IC0782$ & -0.234 & 0.066 & 0.222 & 0.059 & 0 \\
$IC1024$ & 0.151 & 0.052 & 0.099 & 0.037 & 0 \\
$IC3631$ & -0.519 & 0.071 & 0.744 & 0.128 & 0 \\
$NGC0448$ & -0.403 & 0.04 & -0.058 & 0.053 & 0 \\
$NGC0474$ & -0.257 & 0.041 & -0.295 & 0.061 & 1 \\
$NGC0502$ & -0.305 & 0.059 & -0.03 & 0.076 & 0 \\
$NGC0509$ & -0.488 & 0.073 & 0.781 & 0.067 & 0 \\
$NGC0516$ & -0.141 & 0.125 & -0.169 & 0.164 & 0 \\
$NGC0524$ & 0.042 & 0.033 & -0.264 & 0.033 & 1 \\
$NGC0525$ & -0.287 & 0.102 & 0.097 & 0.164 & 0 \\
$NGC0661$ & -0.33 & 0.062 & 0.02 & 0.075 & 1 \\
       \noalign{\smallskip}
    \hline
  \end{array}
$$ 
{Column (1): Name of the galaxy.
Column (2): Metallicity gradient as defined by Eq.~(\ref{eq:metgrad}).
Column (3): Uncertainty on the gradient obtained from weighted linear regression fits to the metallicity profiles.
Column (4): Age gradient as defined by Eq.~(\ref{eq:agegrad}).
Column (5): Uncertainty on the gradient obtained from weighted linear regression fits to the age profiles.
Column (6): Galaxies that were observed with HST and for which we were able to extract surface brightness profiles are incidated by 1, those with HST data and uncertain profiles are indicated by 2, and those with no HST data are indicated by 0.
This table is also available in the online journal and from the ATLAS$^{\rm 3D}$ project website \url{http://purl.org/atlas3d}.
}
\end{table}


\end{appendix}
\end{document}